\documentclass[aps,rmp,twocolumn,preprintnumbers,amsmath,amssymb,floatfix,groupedaddress,longbibliography]{revtex4-1}

\usepackage{graphicx}
\usepackage{hyperref}
\usepackage{color}

\usepackage{dcolumn}
\usepackage{bm}
\usepackage{amssymb}
\usepackage{xr}
\usepackage{epstopdf}

\newcommand{\ket}[1]{\left | #1 \right \rangle}
\newcommand{\bra}[1]{\left \langle #1 \right |}
\newcommand{\proj}[1]{\ket{#1} \bra{#1}}

 \newcommand{\tr}{{\rm \, Tr }\, }

\newcommand{\beq}{\begin{equation}}
\newcommand{\eeq}{\end{equation}}
\newcommand{\beqa}{\begin{eqnarray}}
\newcommand{\eeqa}{\end{eqnarray}}
\newcommand{\beqan}{\begin{eqnarray*}}
\newcommand{\eeqan}{\end{eqnarray*}}

\newcommand{\expect}[1]{\langle #1 \rangle} \renewcommand{\a}{\hat{a}}  
\newcommand{\e}{\mathrm{e}}

\newcommand{\ee} {{\textrm e}}
\newcommand{\dd} {\hbox{\textrm d}}
\newcommand{\Real}{\textrm{Re}}

\newcommand{\braket}[2]{\left\langle #1\middle|#2\right\rangle}

\newcommand{\ketbra}[2]{|#2\left\rangle\right\langle #1|}

\def\be{\begin{equation}}
\def\ee{\end{equation}}

\renewcommand{\a}{\hat{a}}
\newcommand{\adag}{\hat{a}^\dag}
\newcommand{\Tr}{\mathrm{Tr}}

\graphicspath{{../figures/}}

\begin{document}

\title{Production and applications of non-Gaussian quantum states of light}

\author{A. I. Lvovsky$^{1}$}
\author{Philippe Grangier$^2$}
\author{Alexei Ourjoumtsev$^3$}
\author{Valentina Parigi$^4$}
\author{Masahide Sasaki$^5$}
\author{Rosa Tualle-Brouri$^2$}
\affiliation{$^{1}$ Clarendon Laboratory; University of Oxford; Oxford OX1 3PU UK and Russian Quantum Center; Moscow 143025 Russia}
\affiliation{$^{2}$ Laboratoire Charles Fabry;~IOGS;~CNRS;~Universit\'e~Paris-Saclay; 2 avenue Fresnel 91127 Palaiseau France}
\affiliation{$^{3}$ JEIP;  USR 3573 CNRS;~Coll\`ege de France;~PSL University; 11 place Marcelin Berthelot 75231 Paris Cedex 05 France}
\affiliation{$^{4}$ Laboratoire Kastler Brossel;~Sorbonne Universit\'e;~CNRS;~ENS-PSL~Research~University;~Coll\`ege~de~France; 4 place Jussieu 75252 Paris Cedex 05 France}
\affiliation{$^{5}$ National Institute of Information and Communications Technology (NICT);~4-2-1 Nukuikita~Koganei 184-8795~Japan}

\date{\today}


\begin{abstract}
This review covers recent theoretical and experimental efforts to extend the application of the continuous-variable quantum technology of light beyond ``Gaussian" quantum states,  such as coherent and squeezed states, into the domain of ``non-Gaussian" states with negative Wigner  functions. Starting with basic Gaussian nonclassicality associated with single- and two-mode vacuum states produced by means of parametric down-conversion and applying a set of standard tools, such as linear interferometry, coherent state injection, and conditional homodyne and photon number measurements, one can implement a large variety of optical states and processes that are relevant in fundamental quantum physics as well as quantum optical information processing. We present a systematic review of these methods, paying attention to both fundamental and practical aspects of their implementation, as well as a comprehensive overview of the results achieved therewith.
\\
\\
\\
\\
\\
\\
\end{abstract}

\maketitle

\tableofcontents

\pagebreak

\section*{General introduction}

A mode of electromagnetic radiation, formally analogous to a harmonic oscillator, can be described in phase space by its quadratures. For most quantum states occurring in nature, such as vacuum, coherent or thermal states, the quadrature probability distributions have Gaussian shapes. Second-order nonlinear processes may ``squeeze'' their width below the vacuum level but they preserve their Gaussian character. Quantum optical states that lie outside the Gaussian domain --- the single-photon state being the simplest example --- are more challenging to handle but irreplaceable in any nontrivial quantum information processing and communication. 

The scope of non-Gaussian quantum optics --- and of the present review --- is defined not only by the set of states it covers, but also by their description. It treats each optical mode as an independent electromagnetic harmonic oscillator with an associated Hilbert space. The experimental state characterization method consistent with this approach involves homodyne detectors to measure the correlated quantum statistics of the (continuous) electromagnetic field quadratures in all the relevant modes \cite{Leonhardt1997}.
This method of state description and characterization is referred to as continuous-variable (CV). 

In contrast, the discrete-variable (DV) approach is applied to optical states with a well-defined number of individually manipulated photons known \emph{a priori} to occupy a particular set of modes --- for example, a qubit encoded in the polarization of a single photon. In DV, the Hilbert spaces are associated with individual photons, and the states are defined by the modes which these photons occupy (see \textcite{Kok2007} for a review). Measuring such states experimentally involves detecting single photons in various individual modes. While the DV approach also addresses non-Gaussian states, it is touched by this review only insofar it is relevant to its main subject.

The difference between the DV and CV approaches is related to that between first and second quantization in quantum physics and can be illustrated by the following example. A photon in the superposition of the horizontal ($H$) and vertical ($V$) polarization states (i.e., a $45^\circ$ polarized photon) would be represented as $(\ket H+\ket V)/\sqrt2$ in the DV language, and $(\ket 1_H\ket 0_V+\ket 0_H\ket 1_V)$ in CV. In other words, while in the DV treatment the photon is a quantum \emph{particle}, in CV it is a quantum \emph{state} of an electromagnetic oscillator \cite{Fabre2019}.

Historically, the CV and DV approaches evolved in parallel, with their own sets of theoretical and experimental tools, and reached important milestones such as demonstration of entanglement [\textcite{Ou1992} for CV \emph{vs.} \textcite{Kwiat1995} for DV] and teleportation [\textcite{Bouwmeester1997} and \textcite{Boschi1998} for DV \emph{vs.} \textcite{Furusawa1998} for CV] around the same time. Until about the end of the 20th century, the CV approach was limited to Gaussian states, because such states are available ``on demand'' either directly from lasers (coherent states) and thermal sources (thermal states) or via parametric amplification in nonlinear optical crystals (single- and two-mode squeezed states). Today,  despite the progress discussed in section \ref{section:Outlook}, deterministically preparing a given set of modes in a non-Gaussian state remains a challenge.

In the present review, we will focus on the most common method for generating non-Gaussian states, which provided many significant results during the last 20 years: it is to start from an on-demand Gaussian state produced by nonlinear optics, and to use conditional measurements to bring it outside of the Gaussian domain, typically by photon counting. Combining non-Gaussianity with CV characterization methods adds a further layer of complexity, because the optical modes of these states must be very well defined and match that of the local oscillator used for homodyne detection. In counterpart, this guarantees that the states prepared by two different sources are pure and indistinguishable, so that they can be used as top-quality resources for quantum engineering. 

This approach was first implemented in \cite{Lvovsky2001}, where the single photons were produced and characterized using CV methods. Combining DV and CV toolboxes made non-Gaussian quantum optics arise as a research field reaching across the narrow boundaries of these two domains, giving us access to practically the entire optical Hilbert space and enabling capabilities that exceed the boundaries of each of these two domains taken separately. Unveiling these capabilities is one of the main goals of this review. 

While the field under review has many achievements, it is important to understand its limitations. In particular,  conditioning  cannot produce non-Gaussian light ``on demand". As a result, virtually all existing experiments on non-Gaussian state engineering resort to the ``crutch" of probabilistic conditional state preparation via non-Gaussian measurements on squeezed states. 
This limitation can be traced back to the harmonic oscillator nature  of light. Its equidistant energy structure implies that any attempt to move the oscillator out of the ground state will necessarily populate not only the first excited level (the single-photon state) but also higher levels, leading to the Gaussian character of the resulting excitation. Possible ways to shun this trap in the optical domain are either operating in a highly nonlinear medium in which the light wave would lose its harmonic characteristic, or coupling the light to a non-harmonic physical system. A few platforms on which this can be achieved are discussed in the last section of this manuscript.

This work is not the first review of non-Gaussian quantum optics. The article by \textcite{Zavatta2005} summarized the first results in the field: tomography of the single-photon Fock state and photon added states, as well as technology of the state preparation and homodyne detection. A subsequent development of experiments on ``Schr\"odinger's cats'' (superpositions of opposite-amplitudes coherent states) has triggered a study on this subject by \textcite{Glancy2008}. A detailed review by \textcite{Lvovsky2009} paid significant attention to state reconstruction algorithms and surveyed further experimental developments, in particular single-rail optical qubits and ``cat'' states. A more recent, albeit brief, progress article \cite{Andersen2015} summarized the progress of hybrid CV-DV quantum technology not only in optics, but also in superconducting circuits. 

In addition, we would like to mention a few review articles that are tangentially related to our field of interest. \textcite{Kurizki2015} survey the rapidly developing research on technology of interfacing among physical systems of different nature, which often requires hybridization of the CV and DV approaches. The paper by \textcite{Pirandola2015} covers recent advances in quantum teleportation, again in both the discrete and continuous bases. The theory, technology and application of squeezed light is surveyed in \cite{Lvovsky2015} and \textcite{Andersen2016}, the latter of which concentrating more on the historic aspect of this phenomenon, and the former being more pedagogical. Additional educational resources that might be useful to newcomers to the field is the book by \textcite{Lvovsky2008}, which covers DV quantum optics, presents detailed analysis of the harmonic oscillator, squeezing and the Heisenberg picture, and the book by \textcite{Leonhardt1997}, detailing the basics of quantum optics with the emphasis on phase-space representation of states, and homodyne tomography.

This article is organized into two main parts. The first part is entitled ``Methods \& techniques", and introduces the theoretical and experimental tools which are generic to this domain of research (sections I to V). The second part  is entitled ``Results and analysis", and describes more specific experiments, which either have been realized, or are still open challenges (sections VI to VIII). Sections IX and X describe respectively applications to quantum communications and quantum computing, and section XI  concludes with some future perspectives.

\part{Methods \& techniques}
\label{part.methods}

\section{Continuous variables and quasi-probabilities functions.}

In classical mechanics it is always possible, in principle, to access complete information on a system. In particular,  its statistical properties can be described by introducing a joint distribution for the position $x$ and the momentum $p$ in the phase-space picture. In an analogous way, in classical optics the field can be described by the statistics of the complex amplitude $\alpha$; in particular, the expectation value for all the physical quantities can be calculated by knowing the joint probability distribution $P(x,p,t)$ for the real and the imaginary components of $\sqrt{2}\, \alpha = x + i p$, named quadratures, at a given time $t$.  The function $P(x,p,t)$ provides the joint probability of finding
$x$ and $p$ in a simultaneous measurement. 

In quantum mechanics, on the contrary, proper probability densities in phase-space cannot be defined since the Heisenberg uncertainty principle forbids the simultaneous knowledge of two canonical variables with infinite precision. 
However, it is still possible to define particular
functions in phase-space, named quasi-probability distributions, useful to calculate averages of physical quantities \citep{Gardiner2000,Leonhardt1997}. In contrast to their classical counterparts, and may become negative or highly singular.

In the quantum description the quadratures correspond to the position and momentum operators of the harmonic oscillator which describes the  field in a well defined mode.  They can be expressed by the following  combinations of
the bosonic ladder operators $\hat{a}$ and $\hat{a}^{\dag}$
\begin{equation}\label{quad1}\begin{cases}\hat{x}=\frac{1}{\sqrt{2}}(\hat{a}^{\dag}+\hat{a});
\\ \hat{p}=\frac{i}{\sqrt{2}}(\hat{a}^{\dag}-\hat{a}),
\end{cases}
\end{equation}
which obey the commutation rule
$[\hat{x},\hat{p}]=i$. More generally the phase-dependent quadrature is defined as
\begin{equation}\label{phasequad}
 \hat{x}_{\theta}=\hat{x}\cos(\theta)+\hat{p}\sin(\theta)=\frac{1}{\sqrt{2}}(\hat{a}^{\dagger} e^{i \theta}+\hat{a}e^{-i \theta}),
 \end{equation}
 where $\theta=0$ ($\pi/2$) gives $\hat{x}$ ($\hat{p}$). Any couple of two orthogonal quadratures ($\hat{x}_{\theta}$ and $\hat{x}_{\theta+\pi/2}$ for any $\theta$) can be chosen as the two axes which define the phase-space coordinates.

The most famous quasi-probability distributions, the Wigner, the P- and the Q-function, can be classified according to the type of expectation values that are suitable to calculate.
The P-function diagonalizes the density operator in terms of coherent states:
\begin{equation}\label{pfunc}
\hat{\rho}=  \int P(\alpha) \; |\alpha \rangle\langle \alpha |\; \dd^{2}\alpha.
\end{equation}
It corresponds to the normally ordered averaging, which means that the mean values of expressions with powers of $\hat{a}^{\dag}$ operator on left can be calculated in a classical-like manner by:
\begin{equation}
\tr\{\hat{\rho}\; \hat{a}^{\dag \mu}\hat{a}^{\nu} \}=
\int_{-\infty}^{\infty}\int_{-\infty}^{\infty}P(\alpha)\; \alpha^{ \ast
\mu} \alpha^{\nu} \; \dd^{2} \alpha.
\end{equation}
The P-function has a central role in the theory of photodetection \citep{Glauber1963}.

The Q-function, defined as $Q(\alpha)=
\frac{1}{2\pi}\langle \alpha |\hat{\rho}|  \alpha \rangle$,  is  useful to express anti-normally ordered expectation values
 \begin{equation}
 Tr\{\hat{\rho}\; \hat{a}^{\nu}\hat{a}^{\dag \mu} \}=
\int_{-\infty}^{\infty}\int_{-\infty}^{\infty}Q(\alpha) \; \alpha^{\nu}\alpha^{
\ast \mu } \; \dd^{2} \alpha.
\end{equation}

The distribution which will be extensively used in this manuscript is the Wigner function: introduced in 1932 \citep{Wigner1932}, it corresponds to the symmetric order averaging.
 A symmetric ordered expression is one in which every possible ordering is equally weighted, for example $S\{\hat{a}^{2}\hat{a}^{\dag}\}= 1/3(\hat{a}^{\dag}\hat{a}^{2}+\hat{a}^{2}\hat{a}^{\dag} +\hat{a}\hat{a}^{\dag}\hat{a})$. Note that an expression that is symmetric in $\hat{a}$ and $\hat{a}^{\dag}$ is also symmetric in position $\hat{x}$ and momentum $\hat{p}$. The mean value of this kind of expressions is given by 
 \begin{equation}
\tr\{\hat{\rho} \; S\{\hat{a}^{\dag \mu}\hat{a}^{\nu}\}\}=
\int_{-\infty}^{\infty}\int_{-\infty}^{\infty}W(\alpha) \; \alpha^{ \ast
\mu} \alpha^{\nu} \; \dd^{2} \alpha
\end{equation}

The Wigner function of a state can be expressed through its density matrix according to
\begin{equation}\label{wigner}W(x,p)=\frac{1}{2\pi}\int^{+\infty}_{-\infty}\exp[i p x']\langle x-\dfrac{x'}{2}| \; \hat{\rho} \; |x+\dfrac{x'}{2}\rangle \dd x',
\end{equation}
where, again, $x=\sqrt 2\,{\rm Re}\alpha$ and $p=\sqrt 2\,{\rm Im}\alpha$. 

 The main property of the Wigner function is that its marginal distributions correspond to the  probability distributions of quadratures $ \langle x_{\theta} \vert \hat{\rho} \vert x_{\theta} \rangle = {\rm pr}_{\theta}(x_{\theta})$:
 \begin{eqnarray} \label{ptheta}
&& {\rm pr}_{\theta}(x_{\theta}) =    \\ 
 &&\int^{+\infty}_{-\infty} W(x_{\theta}\cos(\theta)-p_{\theta} \sin(\theta),x_{\theta} \sin(\theta)+p_{\theta} \cos(\theta)) \; \dd p_{\theta} \nonumber
 \end{eqnarray}
 This gives the possibility of reconstructing  the Wigner function of the experimental states by tomographic techniques applied to a complete set of quadrature measurements \cite{Leonhardt1997,Lvovsky2009}. As the Wigner function has a univocal correspondence with the density operator of the state,
it is therefore possible to get a complete information on the quantum state by measuring the quadrature observables.

Since the quasi-probability distributions are used to calculate quantum averages in a classical-like way, any deviation from the behavior of a classical joint probability distribution is considered as a signature of the non-classical properties of the corresponding quantum state. In particular, for a state with a non-negative P-function, the density operator \eqref{pfunc} can be viewed as a statistical mixture of coherent states, which are considered the most closely related to the classical behavior of an electromagnetic wave. In contrast,  states with an irregular or negative P-function are considered non-classical. It is the general definition of non-classicality  \citep{Glauber1963} widely accepted in the quantum optics community. 
Unfortunately a direct test of non-classicality by reconstructing the P-function generated state is not always possible because it often exists only in the form of a generalized function and cannot be connected with any measurable quantity, unless a regularized version of the function is used \citep{Kiesel2010}.

The Wigner function of a state is a convolution of its P-function with the Gaussian Wigner function of the vacuum state. Hence a negativity in the Wigner function implies that the P-function is also not non-negative definite, and is hence as sufficient witness of a state's non-classicality. Because the Wigner function can always be reconstructed from quadrature measurements, is a more convenient criterion,  although there do exist states that are non-classical, yet have non-negative Wigner functions, such as the squeezed  states.

A pure state has a positive Wigner function if and only if it is Gaussian \citep{Hudson1974}. For mixed states, however, one cannot affirm in general that a non-Gaussian state exhibits a negative Wigner function. There have been several attempts to extend the Hudson theorem or to find equivalent criteria in the more general case of mixed states \citep{Mandilara2009,Genoni2013}.

It must be added that, in addition to non-classicality criteria based on phase-space description, there exist other criteria, arising, for example, from quantum information theory \citep{Ferraro2012}.

\section{Measurement-induced non-Gaussian state generation } 

\label{NonGauss}

\subsection{Principle}
\label{NonGauss:Principle}

An important class of transformations in quantum optics, is called \emph{linear unitary Bogoliubov operations (LUBO)}. In the Heisenberg picture, LUBO $M$ applied to a multimode quadrature vector $\vec{R}=[x_1,p_1,...,x_N,p_N]^T$, transforms it into $M\vec R$, or, equivalently, transforms the Wigner function $W(\vec{R})$ of the initial state into $W(M^{-1}\vec{R})$. Any arbitrary LUBO can be implemented by combining the following standard experimental tools.
\begin{itemize}
	\item{Beam splitter (BS)} transformations
	\begin{equation}
	\label{EqUnitBS}
	\hat{B}_{ij}{(\tau)} =\exp\left[\tau\left(\hat{a}_i\hat{a}_j^\dag-\hat{a}_i^\dag\hat{a}_j\right)\right]
	\end{equation}
	mix the annihilation operators $\a_i$ and $\a_j$ of two modes $i$ and $j$ according to
	\begin{equation}
	\label{EqAAdagBS}
	\left[\begin{array}{c}\a_i\\\a_j\end{array}\right] \rightarrow \left[\begin{array}{cc}\cos(\tau)&\sin(\tau)\\-\sin(\tau)&\cos(\tau)\end{array}\right] \left[\begin{array}{c}\a_i\\a_j\end{array}\right]
	\end{equation}
	where $t=\cos\tau$ and $r=\sin\tau$ are the BS's amplidude transmission and reflection coefficients. In this review, we shall also use $T=t^2$ and $R=r^2$ for the intensity transmission and reflection coefficients of the BS, respectively.
	\item{Squeezing} transformations
	\begin{equation}
	\label{EqUnitOPA}
	\hat{S}_{ij}{(r)}=\exp\left[ \frac{r}{1+\delta_{ij}}\left(\hat{a}_i^\dag\hat{a}_j^\dag-\hat{a}_i\hat{a}_j\right)\right],
	\end{equation}
	usually implemented using optical parametric amplifiers (OPAs), result in the linear combination
	\begin{equation}
	\label{EqAAdagOPA}
	\left[\begin{array}{c}\a_i\\\a_j^\dag\end{array}\right] \rightarrow \left[\begin{array}{cc}\cosh(r)&\sinh(r)\\\sinh(r)&\cosh(r)\end{array}\right] \left[\begin{array}{c}\a_i\\a_j^\dag\end{array}\right]
	\end{equation}
	where $G=\cosh^2(r)$ is the amplifier's gain. Here the case
	$\delta_{i\neq j}=0$ corresponds to two-mode squeezing, and $\delta_{i=j}=1$ to single-mode squeezing.
\end{itemize}
For instance, using coherent states as a resource, one can prepare any Gaussian state, with a Wigner function
\begin{equation}
\label{EqWigGauss}
W\left(\vec{R}\right)=\frac{\exp\left[-(\vec{R}-\vec{\mu})^T K^{-1} (\vec{R}-\vec{\mu})\right]}{\pi^N|K|}
\end{equation}
completely determined by the mean value $\vec{\mu}=\expect{\vec{R}}$ and the covariance matrix $K=\expect{(\vec{R}-\vec{\mu})(\vec{R}-\vec{\mu})^T}$ \citep{Olivares2012}. Importantly, the equation above shows that if the initial state is Gaussian, after any LUBO the final state will be Gaussian as well. Making it non-Gaussian requires a non-linear transformation on the quadrature operators, such as one effected by a third-order optical non-linearity. Unfortunately, in standard optical materials, not only such non-linearities are too weak to be used in practice, but they have been shown to be useless in theory \cite{Shapiro2006,Gea-Banacloche2010}. Getting around these no-go theorems requires highly non-linear systems with mode-selectivity and dynamical control. Their realization, while not impossible (see section \ref{section:Outlook}), remains very challenging.

This problem can be circumvented by using LUBOs to entangle Gaussian states with each other and performing \textit{measurements} on some of the output modes. Then, conditioned on certain results of these measurements, other output modes may collapse onto non-Gaussian states. Such a ``conditional measurement" technique typically involves photon detectors because of their inherent non-Gaussian nature and is capable to effectively \emph{simulate} a non-linear optical transformation \cite{Lapaire2003}. The price to pay is that the evolution is no longer deterministic: the operation succeeds only when the measurement produces the desired outcome.

Non-Gaussian state engineering by means of conditional measurements is impossible if all of the initial states are classical, i.e. are characterized by positive definite P-functions. Indeed, such a state can be expressed as a mixture of coherent states (or their products) as per Eq.~(\ref{pfunc}). The BS operation (\ref{EqAAdagBS}) transforms a product of coherent states into a product of coherent states \cite{Sanders2012} and is  hence incapable of producing entanglement from a classical input. The state prepared by any partial measurement on the BS output will therefore remain a classical mixture of coherent states.

Hence at least one of the initial states must be nonclassical. The states that are most commonly used in this context are the single- and two-mode squeezed vacua, which we study next.

\subsection{Squeezing via parametric fluorescence} 
\begin{figure}[tb]
	\includegraphics[width=0.8\columnwidth]{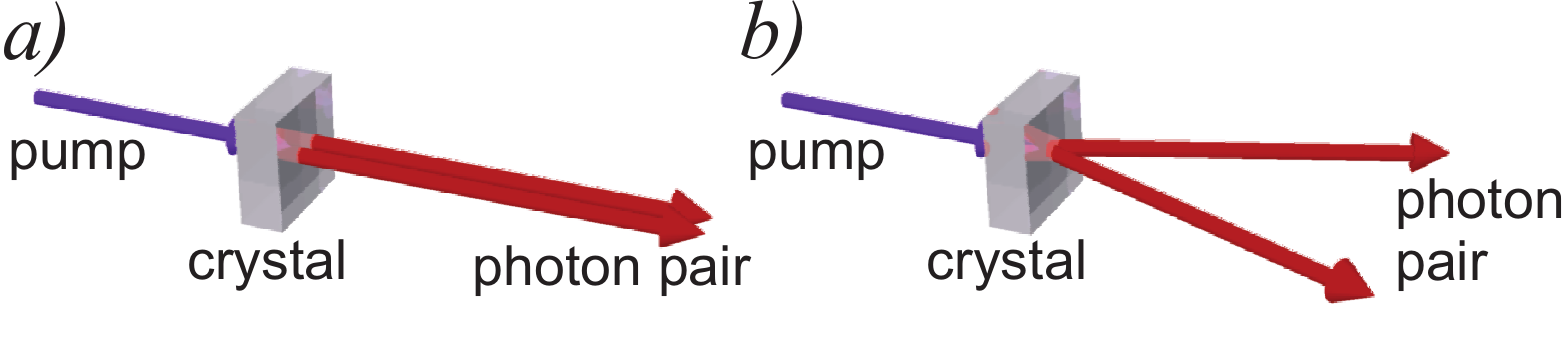}
	\caption{Optical parametric amplifier. Degenerate (a) and non-degenerate (b) configurations, leading to single- and two-mode squeezed vacuum states, respectively.\label{PDCFig}}
\end{figure}

One of the primary methods of producing squeezing is spontaneous parametric down-conversion (SPDC), also known as optical parametric amplification: a nonlinear process where the annihilation of a high frequency pump photon in a nonlinear crystal leads to the simultaneous creation of two lower-frequency photons, known as  ``signal'' and ``idler''. The frequencies, wavevectors and polarizations of the generated photons are governed by phase-matching conditions. If the phase-matching configuration is such that  the two generated photons are indistinguishable, SPDC is \emph{degenerate} and gives rise to single-mode squeezing [Fig.~\ref{PDCFig}(a)]. \emph{Non-degenerate} SPDC, on the other hand, leads to entanglement in the form of the two-mode squeezed vacuum also known as twin-beam state or the Einstein-Podolsky-Rosen (EPR) state [Fig.~\ref{PDCFig}(b)]. 

The Heisenberg picture evolution (\ref{EqUnitOPA}) allows us to find the expression for both types of squeezed vacuum in the Fock basis in the Schr\"odinger picture \citep{Walls2008,Lvovsky2015,Lvovsky2008}:
\begin{align}
\label{EqStateSqz}
\ket{\psi_{\rm sqz}[r]}&=\hat S{(r)}\ket0\\ \nonumber&=\frac{1}{\sqrt{\cosh(r)}}\sum_{k=0}^\infty \sqrt{\left(\begin{array}{c}2k\\k\end{array}\right)}\left[-\frac{\tanh(r)}{2}\right]^k\ket{2k}
\end{align}
for single-mode squeezing and 
\begin{equation}
	\label{EqStateEPR}
	\ket{\psi_{\rm EPR}}=\hat S_{12}{(r)}\ket{0,0}=\frac{1}{\cosh(r)}\sum_{k=0}^\infty \tanh^k(r)\ket{k,k},
	\end{equation}
for the EPR state. The entanglement $E$ of this pure state, determined by the Von Neumann entropy of the reduced density matrix for each of the modes
\begin{equation}
	\hat{\rho}_1=\hat{\rho}_2=\frac{1}{\cosh^2(r)}\sum_{k=0}^\infty \tanh^{2k}(r)\proj{k},
\end{equation}
increases with the parametric gain $G=\cosh^2(r)$ as
\begin{equation}
	\label{EqEntEPR}
	E=-\Tr\left(\hat{\rho_1}\log({\rho_1})\right)=G\log(G)-(G-1)\log(G-1)
\end{equation}

In the quadrature picture, the squeezed states are best visualized in terms of their Wigner functions:
\begin{equation}
\label{EqWigSqz}
W_{\rm sqz}=\frac{1}{\pi}\exp\left[-\e^{2r}x^2-\e^{-2r}p^2\right],\\
\end{equation}
\begin{align}
\label{EqWigEPR}
W_{\rm EPR}=\frac{1}{\pi^2}\exp&\left\{-\e^{2r}[(x_1+x_2)^2+(p_1-p_2)^2]\right.\\
&\left.-\e^{-2r}[(x_1-x_2)^2+(p_1+p_2)^2]\right\}.\nonumber
\end{align}
A single-mode vacuum state is squeezed by a factor $\e^{-2r}$ in variance along the $x$ quadrature and antisqueezed  by $\e^{2r}$ along $p$ if $r$ is positive. For the EPR state,  quantum
fluctuations of the position difference $(x_1-x_2)/\sqrt{2}$ and of the momentum sum $(p_1+p_2)/\sqrt{2}$ become reduced below the vacuum level by a factor $\e^{2r}$ in variance, if $r$ is negative. 

The EPR state and  two orthogonally squeezed single-mode vacuum states produced by degenerate OPAs can be interconverted into one another by mixing on a symmetric BS. This is a common method for producing the EPR state, starting with  \cite{Furusawa1998}. In practice, it can be implemented by two degenerate OPAs, which produce squeezing in orthogonal polarizations, placed in sequence; then the EPR state will be generated in the $\pm45^\circ$ polarization modes \cite{Chen2014,Fedorov2017}. Such a configuration ensures phase stability of the two single-mode squeezed states which is essential for high-quality two-mode squeezing.

Both for single- and two-mode squeezing, we can distinguish two regimes. In the low-gain regime ($r,G\ll 1$), the probability to generate a pair is small, so each subsequent term in Eqs.~(\ref{EqStateSqz}) and (\ref{EqStateEPR}) is much lower in amplitude than the previous one. This regime is appropriate, for example, for heralded single-photon production: the detection of a photon in the idler mode of the EPR state heralds a single-photon state in the signal mode. Because the probability of multi-pair events is very low, to herald a state with a high fidelity it is unnecessary to use photon-number-discriminating detectors (see section \ref{NonGauss:PhotonCount}). The high-gain regime, on the other hand, is required to achieve significant quadrature squeezing. In the limit $r,G\to \infty$, the squeezing is infinite, so the twin-beam state will approach the state described in the original EPR Gedankenexperiment \cite{Einstein1935}: by choosing to measure either the position or momentum in the idler mode, once can remotely prepare a state with either a certain position or a certain momentum in the signal mode, thereby violating local realism.

Since their first experimental observation by \textcite{Slusher1985}, squeezed states found many applications for quantum information processing where, for example, they enabled the first demonstration of quantum teleportation for quadrature observables \citep{Furusawa1998}, or for precision measurements where they are currently used in gravitational-wave detectors to improve their sensitivity beyond the vacuum-noise limit \citep{LIGO2019,VIRGO2019}. A high squeezing is desirable in most cases, as it enhances the non-classical features of these states and hence the performance of the protocols where they are used. Much progress has been made in this direction, by developing materials with stronger non-linearities and by increasing the available peak powers. A detailed review to this effect can be found in \cite{Lvovsky2015,Andersen2016}.

The highest squeezing levels have so far been observed using OPAs in the continuous-wave (cw) regime: 15 dB for single-mode squeezing \citep{Vahlbruch2015}; 8.3--8.4 dB for two-mode squeezing\cite{Zhou2015,Feng2018}. Optical powers available in this regime are too low to obtain significant parametric gains in a single pass through a non-linear crystal. Therefore, the crystals are usually placed inside resonant optical cavities, forming optical parametric oscillators (OPOs). The cavity defines a well-controlled spatial mode for the prepared state and reduces its bandwidth to a few MHz. As a result, the temporal profiles of the prepared states are long enough to be resolved by detection electronics which offers an extra degree of freedom for their manipulation.

An alternative approach is to use ultrashort pulses with high peak powers \citep{Slusher1987} which nowadays allow one to observe squeezing levels of $\sim 6$ dB with a single pass through a non-linear crystal \citep{Kim1994} or through an optical fiber \citep{Dong2008}. In this case the optical power enhancement is obtained not at the crystal's but at the laser's level:
compared to  mode-locked lasers used to generate parametric photon pairs, lasers used in squeezing experiments often operate with pulse pickers, cavity dumpers \citep{Wenger2004b} or amplifiers \citep{Dantan2007}. Pulsed lasers provide an intrinsic timing source for the photon detection, but as the ultrashort pump pulses are very sensitive to optical dispersion, the control of spatial and temporal modes is more involved than in cw experiments.

In some applications, the amount of squeezing is the only quantity of interest, and the antisqueezing of the orthogonal quadrature plays no role. In the experiments considered here the situation is quite different. While stronger squeezing can indeed help preparing more complex non-Gaussian states or increasing the preparation success rate, the antisqueezing of the orthogonal quadrature must be kept at the minimum allowed by the Heisenberg uncertainty principle $\Delta X\Delta P\geq 1/2$, otherwise the squeezed state becomes impure and, as a consequence, the measurements preparing the non-Gaussian state become unreliable. Unfortunately, excess noise becomes more important as the gain of the amplifier increases. It is a known problem which limits the maximal squeezing level measurable in a given setup \citep{Dong2008}, but even before this level is reached the excess noise makes the squeezed states impossible to use as a resource for non-Gaussian state generation. 

The purity of the measured state is also degraded if it becomes multimode, or if its mode becomes different from the one defined by the local oscillator used for the homodyne measurement. In practical experimental situations, especially in the pulsed regime, this may occur for many reasons, e.g.: gain-induced diffraction of the parametric photons \citep{Kim1994b,Anderson1995}, group velocity dispersion \citep{Raymer1991}, spatial walk-off or group velocity mismatch between the pump pulse and the downconverted light, thermal or photorefractive effects \citep{Goda2005}.  As a result, if the mode of interest is defined by a fixed, independently prepared local oscillator, the squeezing measured in pulsed systems usually do not exceed $3$ dB \citep{Wenger2005,Gerrits2010}. To measure higher squeezing levels the measured mode must be properly modified by shaping the local oscillator \citep{Kim1994,Eto2008}. Unfortunately, to transform these squeezed states into non-Gaussian states, the modes measured by photon counters should be shaped accordingly, which is difficult to do in practice (see Sec.~\ref{condSPDC}). Therefore, a lot of efforts are focused on preparing very pure Gaussian resource states and controlling their spatial, spectral and temporal properties \citep{Banaszek2001,Wasilewski2006,URen2006,Hendrych2007,Garay-Palmett2007,Eckstein2011,Branczyk2011,Edamatsu2011,Cui2012}.

\subsection{Photon counting projective measurements}
\label{NonGauss:PhotonCount}

Photon detection is certainly the simplest non-Gaussian measurement in quantum optics. Formally, an ideal photon counter (able to determine the exact number of photons) destructively projects the measured mode $i$ on a non-Gaussian $n$-photon Fock state $\ket{n}_i$. In order to do this in practice, however, the detector must be photon-number-resolving, highly efficient, and present a low probability of false positive detections (dark count rate). 

Different practical implementations of such devices are described in a review paper by \textcite{Hadfield2009}. To date, most experiments have been performed with commercial semiconductor avalanche  photodiodes (APD) operating in the Geiger mode: the absorption of one or several photons starts a quenched electronic cascade leading to a detectable, macroscopic electronic signal. These devices cannot resolve the number of photons but only detect their presence or absence. 
The efficiency of commercially-available models can reach $80\%$. 

In recent years, a new generation of detectors became broadly available. The primary element in such a detector is a metal wire cooled down to a superconducting state. When a photon is absorbed, it heats a region of this wire, lifting superconductivity and giving rise to measurable resistance. Such devices can reach efficiencies up to $98\%$ \citep{Lita2008} and discriminate the number of incident photons up to $\sim 29$ \citep{Lolli2012}.

\medskip

Photon detection can be used to implement two essential operators in quantum optics: photon subtraction and photon addition.

\begin{figure}[tb]
\begin{center}
\includegraphics[width=6.5cm]{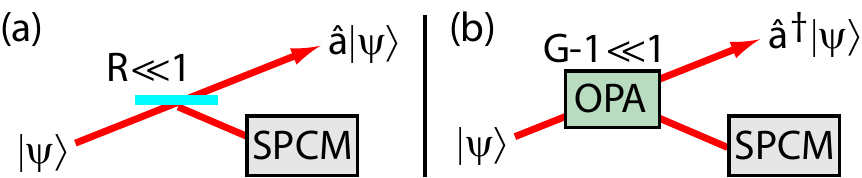}
\end{center}
\caption{(Color online) 
Heralded subtraction and addition of a photon. (a) Heralded subtraction of a photon from the state $\ket{\psi}$, by reflection a small fraction $R\ll1$ of the beam towards a single photon counting module (SPCM). (b) Heralded addition of a single photon, by sending the state $\ket{\psi}$ through a low-gain non-degenerate OPA and detecting a photon in the idler mode.} \label{FigAlexei:PrincipePhotSubtractAdd}
\end{figure}

The photon subtraction operator $\a$ can be approximately realized by reflecting a very small fraction  of a beam towards a photon detector [Fig.~\ref{FigAlexei:PrincipePhotSubtractAdd}(a)]. Indeed, by applying the BS transformation (\ref{EqUnitBS}) to the input modes in state $\ket\psi$ and $\ket 0$, we find
\begin{equation}\label{BSsub}
\hat B{(\tau)}\ket\psi\ket0=\ket\psi\ket0+\tau\hat a\ket\psi\ket 1+O(\tau^2). 
\end{equation}
A detection event in the second input mode heralds the subtraction of at least one photon from the beam, and, if the BS's reflectivity is sufficiently low ($\tau\approx \sqrt R \ll1$), the probability to subtract more than one photon becomes negligible. In practice, to keep a sufficient photon detection rate, the reflected fraction must be finite, which mixes the prepared state with the vacuum mode entering through the other input port of the BS, and degrades its non-Gaussian structure. 

An ideal photon subtraction from the mode $k$ transforms the density matrix $\hat{\rho}$ into
\begin{equation}
\label{EqRhoPhotSubtract}
\hat{\rho}'=\frac{\a_k \; \hat{\rho} \; \a_k^\dag}{\Tr(\a_k\; \hat{\rho} \; \a_k^\dag)}.
\end{equation}
For a single-mode state, its expression in the Fock basis
\begin{equation}
\hat{\rho}'=\frac{\sum_{m,n=1}^\infty \rho_{m,n}\sqrt{mn}\ket{m-1}\bra{n-1}}{\sum_{n=1}^\infty \rho_{n,n}n}
\end{equation}
easily allows one to calculate the average number of photons $\bar{n}'$ remaining after subtraction, related to their average number $\bar{n}$ in the initial state by $\bar{n}'=\bar{n}-1+F$
where the Fano factor $F\equiv\langle(n-\bar{n})^{2}\rangle /\bar{n}$ is equal to, greater or lower than 1 for a state with Poissonian, super-Poissonian or sub-Poissonian statistics respectively. Therefore, although a photon was removed from the initial state, $\bar{n}'$ is not necessarily smaller than $\bar{n}$: for Poissonian states it remains identical, and for super-Poissonian states it is actually larger.

An interesting feature of the photon-subtraction operation implemented in this way is that the mode detected by the photon detector (we will denote its annihilation operator as $\hat b$) does not necessarily have to match the signal mode $\hat a$. Indeed, we can decompose $\hat b=\alpha\hat a+\beta\hat a_\perp$, where $\hat a_\perp$ is a mode that is orthogonal to $\hat a$, and is in the vacuum state. We then find 
$$\hat b\ket\psi_{\hat a}\ket 0_{\hat a_\perp}
=\alpha(\hat a\ket\psi_{\hat a})\ket 0_{\hat a_\perp} +\beta \psi_{\hat a}(\hat a_\perp\ket 0_{\hat a_\perp}).$$
The second term in the above equality is zero, so any mismatch between the modes $\hat a$ and $\hat b$ manifests itself only in the probability of the photon detection event.

While mathematically simple, this observation may appear intriguing, as the following example shows. Suppose a cloud of weakly absorbing atoms, initially in the ground state, is placed into a pulsed mode $\hat a$ in such way that they are occupying only a part of that mode's geometric cross-section. If these atoms are observed to spontaneously emit a photon after the end of the pulse, this means that the photon has been removed from the optical mode. While one would expect the photon subtraction to have affected only the part of mode $\hat a$'s cross-section where the atoms are located (mode $\hat b$), in fact it affects the entire mode $\hat a$. This effect was named ``quantum vampire" in reference to the folkloric property of vampires to cast no shadow. It was observed in application to Fock \cite{Fedorov2015b} and thermal \cite{Katamadze2018} states.

\begin{figure}[tb]
	\begin{center}
		\includegraphics[width=\columnwidth]{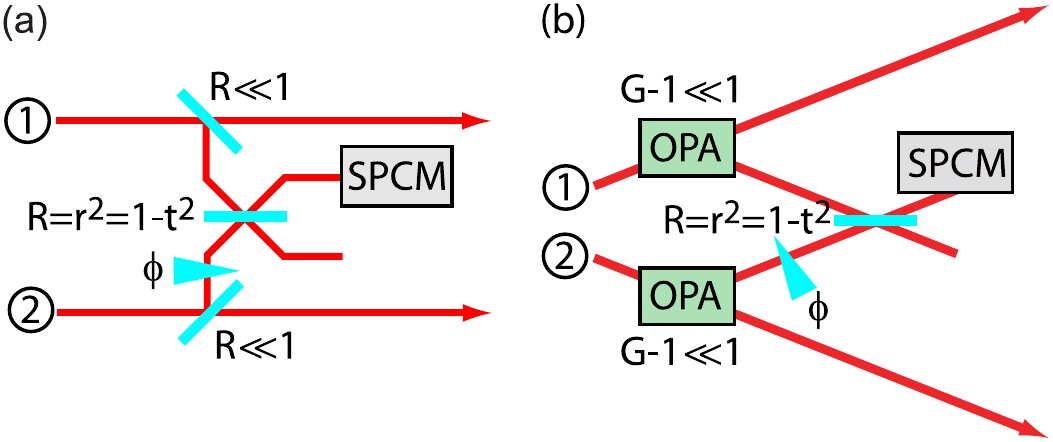}
	\end{center}
	\caption{(Color online) 
		Non-local photon subtraction (a) and addition (b). The heralding modes of the subtraction and addition operations (Fig.~\ref{FigAlexei:PrincipePhotSubtractAdd}) applied to modes 1 and 2 interfere with a phase $\phi$ on a BS with a reflectivity $R=r^2=1-t^2$. Detecting a photon in one of the output ports corresponds to applying the operator $r\a_1+t\e^{i\phi}\a_2$ (a) or $r\a_1^\dag+t\e^{i\phi}\a_2^\dag$ (b) to the incoming 2-mode state.}
	\label{fig:AddSub2mode}
\end{figure}

Recently an alternative method based on up-conversion process for implementing single-photon subtraction \cite{Averchenko2014} has been experimentally demonstrated \cite{Ra2020}. The signal field and a strong
coherent gate field interact in a non-linear crystal, and when a photon at the sum-frequency of the two impinging beams  is detected, a single photon has been conditionally subtracted from the signal field. Compared to the use of a weakly-reflective beamsplitter, this approach offers, via the shaping of the gate beam, the control over the subtracted photon's mode and also allows for subtraction from a coherent superposition of modes.

Photon addition $\a^\dag$ can be approximated in a similar way by replacing the BS with a low-gain non-degenerate OPA [Fig.~\ref{FigAlexei:PrincipePhotSubtractAdd}(b)]. We can see this from Eq.~(\ref{EqUnitOPA}), applied, again, to the input modes in states $\ket\psi$ and $\ket 0$:
\begin{equation}\label{BSsub}
\hat S_{12}{(r)}\ket\psi\ket0=\ket\psi\ket0+r\hat a^\dag\ket\psi\ket 1+O(\tau^2). 
\end{equation}
The higher-order terms can be neglected for $r^2\approx G-1\ll1$. Detecting a photon in the idler mode heralds the creation of an additional photon in the signal beam, and, if the gain is low enough, the creation of several photons is unlikely. Adding a single photon transforms a density matrix $\hat{\rho}$ into
\begin{equation}
\label{EqRhoPhotAdd}
\hat{\rho}'=\frac{\a_k^\dag \; \hat{\rho} \; \a_k}{\Tr(\a_k^\dag\; \hat{\rho} \; \a_k)}.
\end{equation}
For a single-mode state, this can be written in the Fock basis as
\begin{equation} \label{add}
\hat{\rho}'=\frac{\sum_{m,n=1}^\infty \rho_{m-1,n-1}\sqrt{mn}\ket{m}\bra{n}}{\sum_{n=1}^\infty \rho_{n-1,n-1}n}:
\end{equation}
the mean photon number $\bar{n}'=\bar{n}+1+F$ increases by at least 1, and the vacuum term vanishes. As shown by \cite{Lee1995} such a state is necessarily nonclassical. In reality, however, the finite OPA gain required for a sufficient success rate mixes the state with the amplified vacuum mode entering through the other port of the OPA. 

Successfully performing addition and subtraction operations in the desired modes requires good understanding of the spatial and temporal modes of the resource states and of the modes selected by the detectors \cite{Aichele2002,Rohde2007,Sasaki2006,Tualle-Brouri2009}. Experimentally, selecting the desired modes is a non-trivial task which usually requires narrow filtering of the light reaching the photon counters and, on the homodyne detection side, an optimization of the local oscillator's mode (see Sec.~\ref{condSPDC}).

\begin{figure}[tb]
	\begin{center}
		\includegraphics[width=\columnwidth]{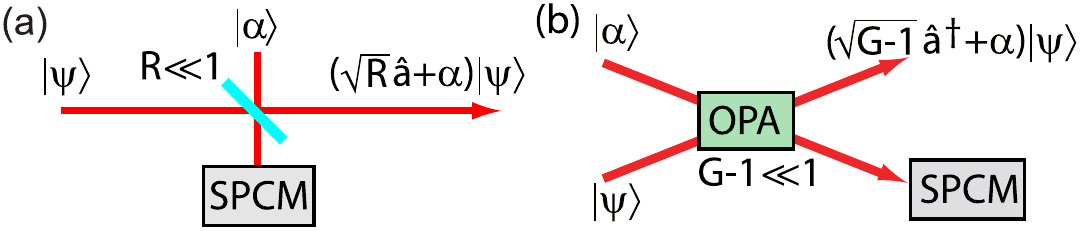}
	\end{center}
	\caption{(Color online) 
 Injecting a weak coherent state in the second input port of the BS used for photon subtraction (a) or the OPA used for photon addition (b) implements, respectively, the superpositions of the photon subtraction or addition operators with the identity operator.}
	\label{fig:AddSubGen}
\end{figure}

These two elementary ``building blocks'' can be combined to implement more complex operations using the following important idea. Realization of both operators $\hat a $ and $\hat a^\dag$ is heralded by photon detection in an ancillary mode. If one combines these ancillary modes from two sources on a BS, and performs photon detection in one of its outputs [e.g. as shown in Fig.~\ref{fig:AddSub2mode}], the detector ``will not know" which of the two sources the photon came from. As a result, the detection event will herald a \emph{superposition} of the processes associated with the photon detection in each of the individual modes --- the photon subtraction operators in modes 1 and 2 in the case of  Fig.~\ref{fig:AddSub2mode}(a) or addition for Fig.~\ref{fig:AddSub2mode}(b). The BS parameters and the relative phases of the ancillary modes determine the coefficients of this superposition. Experimentally, the implementation of this scheme requires high quality matching of the BS inputs in order to ensure that the heralding photons are truly indistinguishable.


In principle, combining photon additions \citep{Dakna1999a,Dakna1999b} or subtractions \citep{Fiurasek2005} with linear operations allows one to prepare any quantum state of light. However, the heralding photon-counting operations usually present a low success rate, and using other available tools such as homodyne measurements may offer practical advantage.

Of particular interest is the operation of photon subtraction from single-mode squeezed vacuum, which permits one to generate the optical analogue of ``Schr\"odinger's cat''. This is our next subject.

\subsection{``Schr\"odinger's cat'' and photon subtraction}\label{ScCatTheoSec}
Since Schr\"odinger's seminal paper \citep{Schroedinger1935}, the transition between the ``microscopic'' (quantum) and ``macroscopic'' (classical) worlds became a central question in quantum physics. Quantum information processing, and the quest for larger and more complex quantum devices, made this subject even more important. In general, a ``Schr\"odinger's cat'' is defined as a classical object being in a quantum superposition of classical, distinguishable states, like a cat being dead and alive at the same time. In optics, the closest analogy to a classical light wave with a given amplitude and phase is a coherent state. Therefore, in the following we will call ``Schr\"odinger's cat'' a quantum superposition of coherent states $\ket{\alpha_k}$, written in its general form as $\sum_k a_k\ket{\alpha_k}$. In most cases it is restricted to only two ``dead'' and ``alive'' states with equal probabilities, which reduces its expression to $c(\ket{\alpha_1}+\e^{i\phi}\ket{\alpha_2})$. Quantitatively, the ``size'' of such a superposition can be measured in different ways, but all of them are related to the phase-space distance between the two coherent states $|\alpha_1-\alpha_2|$ which must be large compared to the shot noise for these states to be macroscopically distinguishable: $|\alpha_1-\alpha_2| \gg 1$. Therefore, phase-space displacements and rotations do not change the key features of this state, and without loss of generality we can consider that it can be expressed as
\begin{equation}\label{EqStateCatIdeal}
\ket{\psi_{{\rm cat}}[\alpha]}=\frac{1}{\sqrt{2c}} \; (\ket{\alpha}+\e^{i\theta} \ket{-\alpha}),
\end{equation}
with $\alpha\in \mathbb{R}_+$ and $c=1+\cos(\theta)\e^{-2|\alpha|^2}$. The associated Wigner function is
\begin{multline}
\label{EqWigCatIdeal}
W_{{\rm cat}}[\alpha]=\frac{1}{\pi c}\left[\e^{-(x-\sqrt{2}\alpha)^2-p^2}+\e^{-(x+\sqrt{2}\alpha)^2-p^2}\right.\\
\left.+\e^{-x^2-p^2}\cos(2\sqrt{2}\alpha p-\theta)\right],
\end{multline}
where the first two terms correspond to the two coherent states and the last one produces the phase-space fringe pattern arising from their quantum superposition. The phase $\theta$ determines the phase of the fringes. This ``interference", where the Wigner function takes negative values, is the characteristic signature of the quantum nature of the cat. 

In the microwave domain, where the photons can be strongly confined and efficiently coupled to superconducting or Rydberg qubits, quantum superpositions of two or more coherent states have been created deterministically over a decade ago \citep{Brune1996,Deleglise2008,Hofheinz2009}. In the optical domain this has proven to be more difficult \cite{Hacker2019}, and several measurement-based non-deterministic protocols have been developed meanwhile \citep{Glancy2008}.

\begin{figure}[tb]
	\begin{center}
		\includegraphics[width=7.5cm]{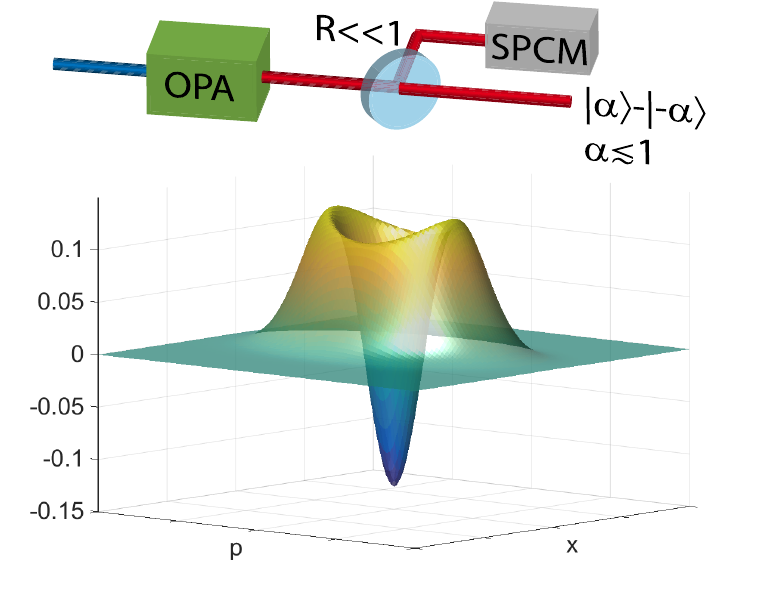}
	\end{center}
	\caption{(Color online) A small fraction of a squeezed vacuum state, created by a degenerate optical parametric amplifier (OPA), is ``tapped": reflected towards a single photon counter which heralds the subtraction of a single photon. The resulting state resembles a ``Schr\"odinger's kitten'', a quantum superposition of coherent states $\ket{\alpha}$ and $\ket{-\alpha}$ with opposite phases and a small amplitude $\alpha\lesssim 1$. Bottom: Wigner function of experimentally generated states in the pulsed regime for $\alpha=0.9$ \citep{Ourjoumtsev2006a}. See Sec.~\ref{Cats} for details.}
	\label{FigAlexei:PulsedKitten}
\end{figure}

The first proposals to prepare cat states using squeezed light and photon counting \citep{Song1990,Yurke1990,Tombesi1996} were based on an experimental design originally developed for quantum non-demolition measurements \citep{LaPorta1989}. However, the first experimental demonstrations \citep{Ourjoumtsev2006a,Neergaard-Nielsen2006,Wakui2007} used a simpler approach proposed by \textcite{Dakna1997}, who pointed out that a small odd ``Schr\"odinger's  kitten" state $c(\ket{\alpha}-\ket{-\alpha})$ with $\alpha \lesssim 1$ could be produced with a very high fidelity by subtracting a single photon from a single-mode squeezed vacuum beam (Fig.~\ref{FigAlexei:PulsedKitten}, top). 
To see this, we recall the decomposition of the coherent state into the Fock basis: 
\begin{equation}\label{cohdecom}\ket{\alpha}=e^{-|\alpha|^2/2}\sum_{n=0}^\infty\frac{\alpha^n}{\sqrt{n!}}\ket{n}.\end{equation}
The decomposition coefficients for $\ket\alpha$ and $\ket{-\alpha}$ are the same for even Fock states, but opposite for even ones. This means that the Fock decomposition of ``even" cat state (with $\theta=0$) contains only even number terms, whereas that of the ``odd" cat state ($\theta=\pi$) only odd terms. Next, we observe that the single-mode squeezed vacuum $\hat S(r)\ket0$, given by Eq.~\eqref{EqStateSqz} also contains only even number terms. For $r^2=\alpha$, the first two terms of the even cat and $\hat S(r)\ket0$ coincide up to a normalization factor, the difference appearing  in higher-order terms. That is, any weakly squeezed vacuum approximates a kitten state of the corresponding amplitude $\alpha=\sqrt{r}$  with a high fidelity. 

Applying the photon subtraction operator $\hat a$ to the even cat state will produce an odd cat of the same amplitude because the coherent states $\ket{\pm\alpha}$ are eigenstates of that operator with the respective eigenvalues $\pm\alpha$. When we apply that operator to $\hat S(r)\ket0$, we also obtain an approximation of an odd kitten, but the amplitude will increase to $\alpha=\sqrt{3r}$. This difference arises because only the first two terms ($\ket 0$ and $\ket 2$) in the Fock decomposition of $\hat S(r)\ket0$ coincide with those of the even cat, whereas the terms of $\hat S(r)\ket0$ that give rise to the odd kitten after the photon subtraction are $\ket 2$ and $\ket 4$. A detailed calculation to this effect can be found in the supplementary information to \cite{Ulanov2017a}. Subtracting more than one photon increases the cat amplitude even further, by the same token.

The photon subtracted squeezed state is equivalent to a squeezed single photon up to a normalization factor. Indeed, from Eq.~\eqref{EqAAdagOPA} we have $\hat S^\dag(r)\hat a\hat S(r)\ket 0=(\hat a\cosh r+\hat a^\dag \sinh r)\ket 0=\sinh r \ket 1$, and hence $\hat a\hat S(r)\ket 0=\sinh r\hat S(r) \ket 1$. This means that, if a deterministic photon source is available, kittens can be prepared in a deterministic fashion (Fig.~\ref{FigAlexei:PulsedKitten}, bottom). It also follows that whatever the squeezing level, the Wigner function of $\hat a \hat S(r)\ket0$ will  become negative only around $x=p=0$ (single photon squeezed along one quadrature), whereas the Wigner function of a larger cat state will present several negative ``interference fringes'' between the ``dead'' and the ``alive'' coherent states. Therefore, a single photon subtraction only allows one to prepare small ``kitten'' states, and merely increasing the squeezing of the initial state does not transform them into large ``cats''.


%


\subsection{Homodyne projective measurements}
\label{NonGauss:HomodyneProject}

An ideal homodyne measurement (discussed in detail in the next section) of the quadrature
$\hat{x}_\theta$
in the mode $k$ giving the outcome $y$ corresponds to the projection on the infinitely squeezed state $\ket{x_\theta=y}_k$. These states have non-trivial expressions in the Fock state basis: for example, $\ket{x_\theta=0}$ corresponds to a quantum superposition containing only even photon numbers. 
A homodyne measurement performed on mode $k$ of a multimode state $\hat\rho$ yields
\begin{equation}
\hat{\rho}'= \frac{_k\!\bra{x_\theta)=y}\hat{\rho}\ket{x_\theta=y}_k}{ \tr\left(_k\!\bra{x_\theta=y}\hat{\rho}\ket{x_\theta=y}_k\right)}.
\end{equation}
In terms of Wigner functions (\ref{wigner}), the  single-mode  infinitely squeezed state along $x(0)=y$ corresponds to  $W(x,p)=\delta(x-y)$ for the corresponding Wigner function, which becomes $W({\vec R},{\vec R}^*)=\delta(\Real({\vec R}\e^{-i\theta})-y)$ for any measurement phase $\theta$.
A homodyne projection onto  $\ket{x(\theta)=y}_k$ in mode $k$ transforms the Wigner function of the initial multimode state into
\begin{multline}
W'({\vec R},{\vec R}^*)\\
=\frac{\int \dd {\vec R}_k\dd {\vec R}_k^* \; \delta\left(\Real({\vec R}\e^{-i\theta})-y\right) W({\vec R},{\vec R}^*)} {\int \dd{\vec R}^{2N} \delta\left(\Real({\vec R}\e^{-i\theta})-y\right) W({\vec R},{\vec R}^*)}
\end{multline}
Unlike photon counting, homodyne detection is a Gaussian process: using the equation above, one can easily see that if the initial state is Gaussian, the final state will be Gaussian as well. However, such measurements may be used as a convenient way to transform one non-Gaussian state into another, and to create complex non-Gaussian states from relatively simple ones.

Since quadrature eigenstates form a continuous basis, one must allow a finite tolerance $y\pm\epsilon$ around the required value $y$ to prepare the desired state with a finite success rate.
In addition, detection losses play a particularly detrimental effect: unlike for photon counting they don't necessarily decrease the heralding event rate, but they modify the structure of the remotely prepared state.

\medskip

Linear optics, parametric amplifiers, photon counting and homodyne measurements allow one to generate a variety of non-classical light states. These states are generally analyzed by quantum homodyne tomography, presented in the next section.

\section{Quantum tomography}

Quantum tomography aims at characterizing some quantum objects
from experimental data: a quantum state, a quantum process, or a
detector \citep{Lundeen2009,Brida2012,Zhang2012,Lvovsky2008}. There is already a wide
literature on this subject, and
our purpose here is mainly to focus on the problematic raised by
the use of continuous variables: measurement space of high
dimensionality, use of hybrid non-trace-preserving processes, or encoding of discrete qubits with continuous variables.

\subsection{Quantum state tomography}

\subsubsection{Homodyne detection}\label{Homodynedet}
A primary method for measuring optical states, in addition to photon detection, is balanced homodyne detection \citep{Yuen1978}. This method enables one to measure the phase-dependent electromagnetic field quadrature $\hat{x}_{\theta}=(\hat{a}^{\dag}e^{+i\theta}
+\hat{a}e^{-i\theta})/\sqrt{2}$. Acquiring the statistics of such measurements at various phases enables one to reconstruct the density matrix and the Wigner function of the state, thereby enabling quantum state tomography \cite{Leonhardt1997,Lvovsky2009}. In this section, we concentrate on homodyne detection as a measurement technique, deferring the discussion of quantum state tomography to Sec.~\ref{section:MaxLik}.

\begin{figure}[tb]
	\begin{center}
		\includegraphics[scale=0.33]{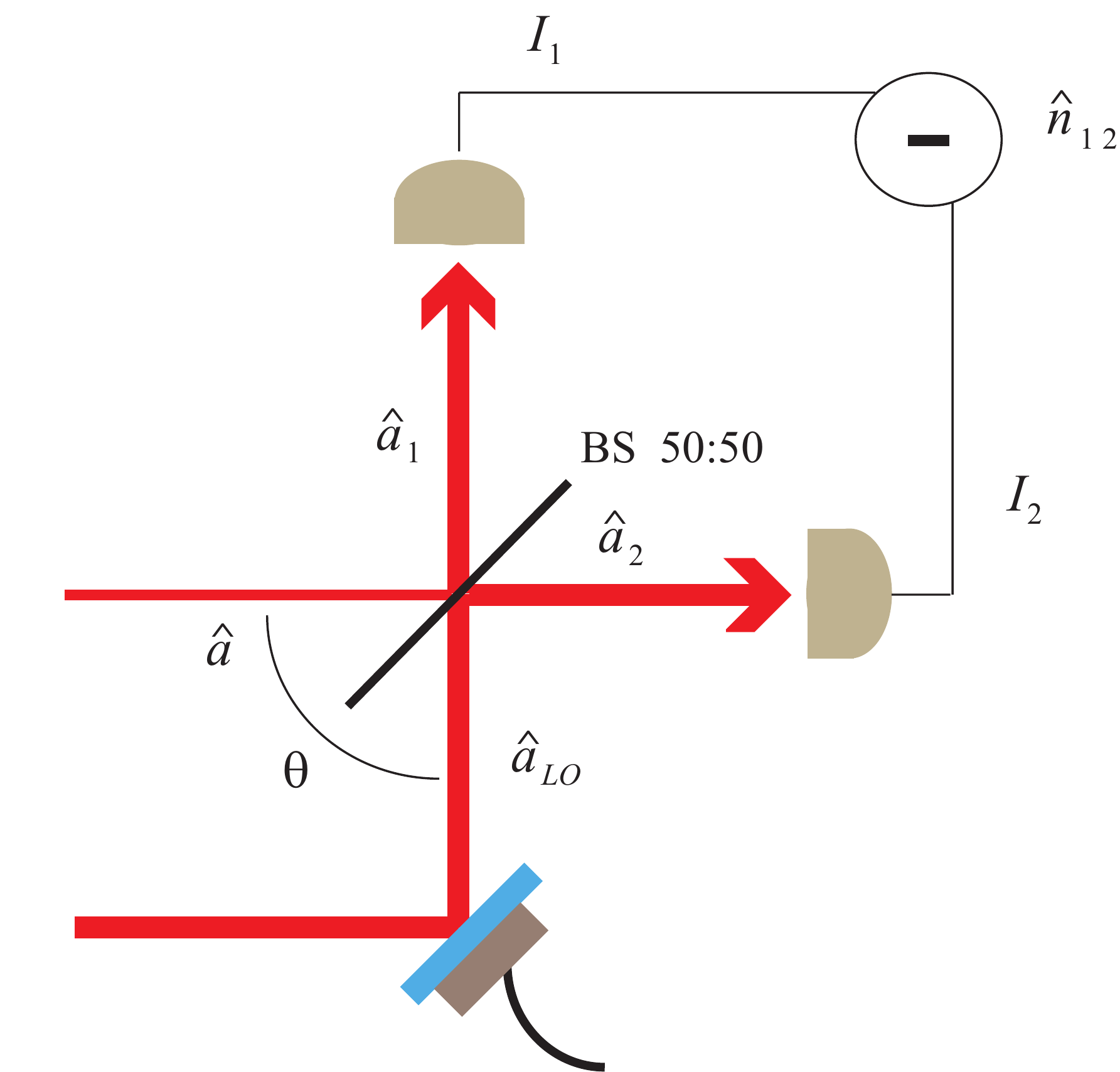}
	\end{center}
	\caption{(Color online) 
		Homodyne detection: The signal and LO field are mixed on the 50:50 BS and the two outputs are detected by two photodiodes, the integral of the difference photocurrent is proportional to the quadrature operator.} \label{Homodynefig}
\end{figure}

The general scheme of this technique involves a classical reference beam, named local oscillator (LO),  impinging on a 50:50 beam splitter, while the signal to be analyzed enters in  the second input port. The two outputs are detected by two efficient linear photodetectors, generally PIN photodiodes (Fig.~\ref{Homodynefig}). The quantity of interest is the difference of the corresponding photocurrents. The time integrated photocurrent measured by each detector is proportional to the photon numbers $\hat n_1$ and $\hat n_2$ in the beam splitter output ports, and hence the subtraction photocurrent is proportional to $\hat n_2-\hat n_1$.

A simple mathematical description is based on the BS transformation (\ref{EqAAdagBS}) of the  $\hat{a}$ and $\hat{a}_{LO}$ annihilation operators describing the signal and LO fields. For $T=1/2$, the two output fields are 
\begin{equation}\begin{cases}\hat{a}_{1}=\frac{1}{\sqrt{2}}(\hat{a}+\hat{a}_{LO}) \cr
\cr
\hat{a}_{2}=\frac{1}{\sqrt{2}}(\hat{a}-\hat{a}_{LO})\end{cases}
\end{equation}
The photon-number operators at the output are then
\begin{equation}\label{n1n2}\begin{cases}\hat{n}_{1}=\hat{a}_{1}^{\dag}\hat{a}_{1}=\frac{1}{2}(\hat{a}^{\dag}\hat{a}+\hat{a}_{LO}^{\dag}\hat{a}_{LO}+\hat{a}^{\dag}\hat{a}_{LO}+\hat{a}_{LO}^{\dag}\hat{a}) \cr
\cr
\hat{n}_{2}=\hat{a}_{2}^{\dag}\hat{a}_{2}=\frac{1}{2}(\hat{a}^{\dag}\hat{a}+\hat{a}_{LO}^{\dag}\hat{a}_{LO}-\hat{a}^{\dag}\hat{a}_{LO}-\hat{a}_{LO}^{\dag}\hat{a})\end{cases}
\end{equation}
and the difference is
\begin{equation}\label{n12}\hat{n}_2-\hat{n}_1=\hat{a}^\dagger \hat{a}_{LO}+ \hat{a}_{LO}^\dagger \hat{a}.
\end{equation}

Because the above transformation applies to operators, i.e., is performed in the Heisenberg picture, the measurement of the operator $\hat n_{12}$ is performed on the BS input state. That is, the operators $\hat a$ and $\hat a^\dag$ are measured on the signal state, and the operators $\hat a_{LO}$ and $\hat a_{LO}^\dag$ - on the bright coherent state $\ket{\alpha_{LO} e^{i\theta}}$ in the local oscillator mode. Considering the limit $\alpha_{LO}\gg 1$ for the amplitude of the local oscillator allows us to treat it classically and replace the operators  $\hat a_{LO}$ and $\hat a_{LO}^\dag$ with their expectation values  $\alpha_{LO} e^{i \theta}$ and $\alpha_{LO} e^{-i\theta}$. Eq.~(\ref{n12}) can then be rewritten as follows:
\begin{equation}\hat{n}_2-\hat{n}_1=\alpha_{LO} \; (\hat{a}^{\dag}e^{+i\theta}
+\hat{a}e^{-i\theta})= \sqrt{2}\; |\alpha| \hat{x}_{\theta}.
\end{equation}

The homodyne signal is hence the measurement of the  $\hat{x}_{\theta}$ operator.
The angle $\theta$  in the phase space corresponds to the relative phase between the signal and the reference field, generally adjusted by a piezo-controlled mirror in the LO path.
If the phase is fixed, the histogram obtained from a large set of repeated measurements, giving different outcomes $x_{\theta}$ on identically prepared states, is a good evaluation of the  probability distribution $P_\theta(x_{\theta})$. Importantly, as long as the detection stays balanced, the macroscopic terms $\hat{a}_{LO}^{\dag}\hat{a}_{LO}$ in Eq.~\eqref{n1n2} are canceled, meaning that the effects of any classical excess noise of the reference field are minimized \citep{Bachor2004}.

Understanding many important features of homodyne detection requires a more elaborate model than the above simple single-mode description. In particular it is necessary to include a multimode picture to explain that in homodyne detection, which is an inherently interferometric technique, the strong local oscillator selects the spatial and temporal mode in which the measurement is performed. Accordingly,  it is necessary to have a good mode-matching between the LO and the signal under analysis in order not to lose information about the quantum field to be measured. Further possible imperfections of homodyne detections include optical losses, finite detection efficiency of the photodiodes, and electronic noise in the measuring chain. They can be modeled by additional beam splitters in the signal path, so that the signal is transmitted with a coefficient corresponding to the global efficiency, and is mixed with  either vacuum or thermal states entering in the other input ports \citep{Lvovsky2009,Leonhardt1997}.

The first application of homodyne detection in quantum optics was in the frequency domain \citep{Abbas1983,Slusher1985} for the measurement of squeezed light. Measurements of the quadrature noise in different spectral components of the signal is still performed by homodyne setups \citep{Vahlbruch2008,Mehmet2011}.
However the application where the homodyne detection has been most extensively used, starting from the pioneering work of \citep{Smithey1993},  is the time domain analysis for the tomographic reconstruction of the Wigner function of quantum states localized in a certain temporal mode. In the pulsed regime the homodyne detector analyzes the quadratures contained in the temporal mode of the pulsed LO which normally comes from the same laser source used in  the quantum field generation \citep{Smithey1993,Wenger2004b}. In the conditioned schemes, where the quantum state is announced by a certain event in a correlated heralding field, only the quadratures in the pulses corresponding to the heralded events are processed \citep{Lvovsky2001,Zavatta2004a,Huisman2009,Ourjoumtsev2006b,Zavatta2008}.
In the cw regime the time behavior of the quantum states is often determined
by the detection of the heralding event which is time-correlated with the analyzed quantum field. The homodyne signal which is acquired with a continuous wave LO needs to be filtered using a temporal function corresponding to the right mode shape \citep{Neergaard-Nielsen2007,MelholtNielsen2009}. These schemes will be described in more details below. 

Scaling up optical non-Gaussian quantum engineering motivated specific research efforts to move from the single-mode to the multi-mode regime \citep{Pinel2012,Pysher2011,Janousek2009, Roslund2013,Chen2014, Yokoyama2013,Cai2020,Asavanant2019}. The requirement for arbitrary, selective quadrature measurement of such states stimulated the emergence of adaptive methods which realize a mode-selective and a multimode homodyne detection of quantum light states \citep{Ferrini2013,Cai2020}. Spatial and temporal shaping techniques for femtosecond light can be used to iteratively adjust the LO field shape to the quantum signal by maximizing the homodyne efficiency \citep{Polycarpou2012}.

The main performance characteristics of homodyne detectors are high subtraction efficiency, high clearance of the shot over the electronic noise, high photodetector quantum efficiency and high bandwidth for increasing the temporal resolution. Achieving high performance in all these benchmarks is challenging. In particular, higher-efficiency photodiodes often feature higher capacitances, which, in turn, leads to narrower bandwidths. A higher  bandwidth often compromises the shot-to-electronic-noise clearance, which, in turns, affects the measurement efficiency \citep{Appel2007}. A typical approach to constructing the balanced detector circuitry involves using a low-noise operational amplifier \citep{Zavatta2002,Neergaard-Nielsen2007,Okubo2008,Haderka2009,Kumar2012,Cooper2013a,Qin2016}. Performance characteristics of such circuits are well understood \citep{Masalov2017}. Further improvement in performance can be achieved by using a field-effect transistor instead of an operational amplifier for the first cascade of amplification \citep{Duan2013}, which however requires a good expertise in high-frequency electronics. For some applications, resonant AC-coupled detectors may be advantageous \cite{Serikawa2018}. 
An important recent achievement is the realization of a homodyne detector integrated on a photonic chip \cite{Raffaelli2018}.

\subsubsection{State reconstruction in homodyne tomography}
\label{section:MaxLik}
As already discussed before, the Wigner function $W(x,p)$ is particularly relevant to picture a
quantum state with continuous variables as it is closely connected
to the probability distributions  $P_\theta(x_\theta)$ of quadratures
measured with the homodyne detector. Such distributions indeed simply correspond to the projection of this function
onto the axis of the considered quadrature as per Eq.~\eqref{ptheta}. If one introduces the Fourier transforms 
$\tilde P_\theta(k)$ and $\tilde W(k_x,k_p)$ of $P_\theta(x_\theta)$ and $W(x,p)$, 
respectively, this relation  can be rewritten \citep{Leonhardt1997}:
\begin{equation} \label{Rosa:ExpWigner}
\tilde W(k\cos\theta,k\sin\theta)=\tilde P_\theta(k_\theta)
\end{equation}

The reconstruction of $W$ from its projections is quite similar to
X-ray computed tomography in medical imaging, and this is the task
of the Radon transform, which belongs to the class of state
reconstruction methods via inverse linear transformation
\citep{Lvovsky2009}: by writing the inverse Fourier transform of
Eq.(\ref{Rosa:ExpWigner}) with polar coordinates, it can be
expressed as:
\begin{equation} \label{Rosa:EQURadonInverse}
W(x,p)=\frac{1}{2\pi^{2}}\int_{0}^{\pi}d\theta\int
\tilde P_{\theta}(k_\theta)K(x\cos\theta+p\sin\theta-k)\dd k
\end{equation}
This method however is not
very robust to noise due to high oscillations of the integration kernel, which
could introduce artifacts in the reconstructed Wigner function. It
can even lead to unphysical states: the density matrix $\rho$
deduced from $W$ using pattern functions \citep{Leonhardt1997} is not guaranteed to be positive and of unit trace \citep{Lvovsky2009}. That is the reason why maximum-likelihood reconstruction is usually
preferred \citep{Hradil1997,Hradil2004,Lvovsky2004}. 

The likelihood function can be defined
as the probability to observe a set of measurement outcomes in the state with the density matrix $\hat\rho$
\begin{equation} \label{Rosa:EQUVraisemblance}
L(\hat\rho)=\prod_{j}p_{j}(\hat\rho)
\end{equation}
where $p(\hat\rho)=\bra{x_{\theta_j}}\hat\rho\ket{x_{\theta_j}}$ is the probability to observe the quadrature value $x_\theta$ in the $j$th measurement with the local oscillator phase $\theta_j$. 
An important  result
\citep{Hradil1997,Lvovsky2004} is that the state $\hat\rho$ for which the likelihood is maximized is obtained by the following iterative procedure:
\begin{equation} \label{Rosa:EQUsuiteR}
\hat\rho^{(k+1)}=\mathcal{N}[\hat  R(\rho^{(k)})\hat\rho^{(k)}\hat R(\rho^{(k)})]
\end{equation}
\noindent where $\hat R$ is the positive operator:
\begin{equation} \label{Rosa:EQUdefR}
\hat R(\hat\rho)=\sum_{j}\frac{\ket{x_{\theta_j}}\bra{x_{\theta_j}}}{p_{j}(\hat \rho)}
\end{equation}
 and $\mathcal{N}$ denotes normalization to a
unitary trace. 

The maximum-likelihood approach offers a
major advantage of preserving trace and positivity, thereby guaranteeing
the reconstruction of a physical state. The iterative sequence usually
converges when starting from $\hat\rho^{(0)}=\mathcal{N}[\hat 1]$, even
if this algorithm can sometimes fail. Other schemes have been
proposed to prevent that risk \citep{Rehacek2007} and guarantee
convergence. Stopping rules have also been derived for such
iterative processes \citep{Glancy2012}.

Importantly, this approach is able to account for a
non-ideal detection efficiency which, in the case of homodyne detectors, mainly stems from optical losses. To model them, we can assume that the signal mode (labeled $s$ in the equation below) is transmitted through a beam splitter (BS) whose transmissivity equals the detection efficiency $\eta$ and whose other input (2) is in a vacuum state.  The loss transforms the state in the signal path into
\begin{equation}
\hat{\rho}_{s}'(\eta)=\tr_{2}\{\hat{B}_{s2}{(\tau)^\dag}\hat{\rho}_{s} \vert 0 \rangle_{2} \, _{2} \langle 0 \vert \hat{B}_{s2}{(\tau)}\},
\end{equation} 
where the operator $\hat{B}^\dag(\eta)$ corresponds to the BS operator defined in Eq.~(\ref{EqUnitBS}) with $\cos\tau=\sqrt{\eta}$.

If the density operator is expressed in the Fock basis as $\hat{\rho}_{s}= \sum \varrho_{n,m}\vert n \rangle  \langle m \vert\ $ the density matrix elements are transformed as
\cite{Leonhardt1997}

\begin{equation}\label{GenBernoulli}
\varrho_{n,m}'(\eta)= \sum_{k=0}^{\infty}[\textbf{B}_{n}^{n+k}(\eta)\textbf{B}_{m}^{m+k}(\eta)] \varrho_{n+k,m+k} 
\end{equation}
where $\textbf{B}^{n+k}_{n}=\sqrt{\binom{n+k}{n} \eta^{n}(1-\eta)^{j}}$. This
formula describes a \textit{generalized Bernoulli transformation} and in the case of a pure single photon state gives a mixture of the single photon and the vacuum state: $\hat{\rho}_{s}'(\eta)= \eta  \vert 1 \rangle \langle 1 \vert + (1-\eta)  \vert 0 \rangle \langle 0 \vert $.
The Wigner function of the resulting state can be directly calculated as a convolution of the entering state and the vacuum by \cite{Leonhardt1997}
\begin{equation}\label{Weff}\begin{split}W'_{s}(x,y,\eta)=&\int \int
W\big(\sqrt{\eta}x - \sqrt{1-\eta}x_{2},\sqrt{\eta}p - \sqrt{1-\eta}p_{2}  \big)\\
& \times W_{0}\big(\sqrt{1-\eta}x - \sqrt{\eta}x_{2},\sqrt{1-\eta}p - \sqrt{\eta}p_{2})\\
& dx_{2} dp_{2} 
\end{split}
\end{equation}
In other words, losses leas to the ``blurring'' of the
Wigner function. 

It is common to correct for the effect of the losses in the numerical procedure of state reconstruction. In maximum-likelihood reconstruction \citep{Lvovsky2004}, this is done by redefining the POVM or, equivalently, by replacing
$\hat\rho$ in Eq.~(\ref{Rosa:EQUdefR}) with its generalized Bernoulli
transform (\ref{GenBernoulli}). In this way, the likelihood function
(\ref{Rosa:EQUVraisemblance}) is estimated with respect to the density operator that represents the state before the
detection losses, so maximizing the likelihood reconstructs the state at that stage.

\subsubsection{Parameter estimation}
\label{section:ParameterEstimation}
A quantum state is characterized by a set of parameters (for example, the elements of the density matrix). Quantum state tomography implies finding the approximate values of these parameters within a certain subspace of the Hilbert space. Even for a moderate size subspace, this parameter set is quite large. Therefore, although the reconstruction methods described above  are quite efficient,  state tomography
requires one to handle a high amount of data, with a
substantial computation time.  A faster way to evaluate a
quantum state in real time (e.g. when aligning a setup) is to
exploit  the physical \textit{a priori} knowledge of the state in order to build its  
parametrized model. Then  the characterization of a quantum state reduces to a much simpler task of determining only
a few parameters.

\begin{figure}[tb]
\begin{center}
\includegraphics[scale=0.6]{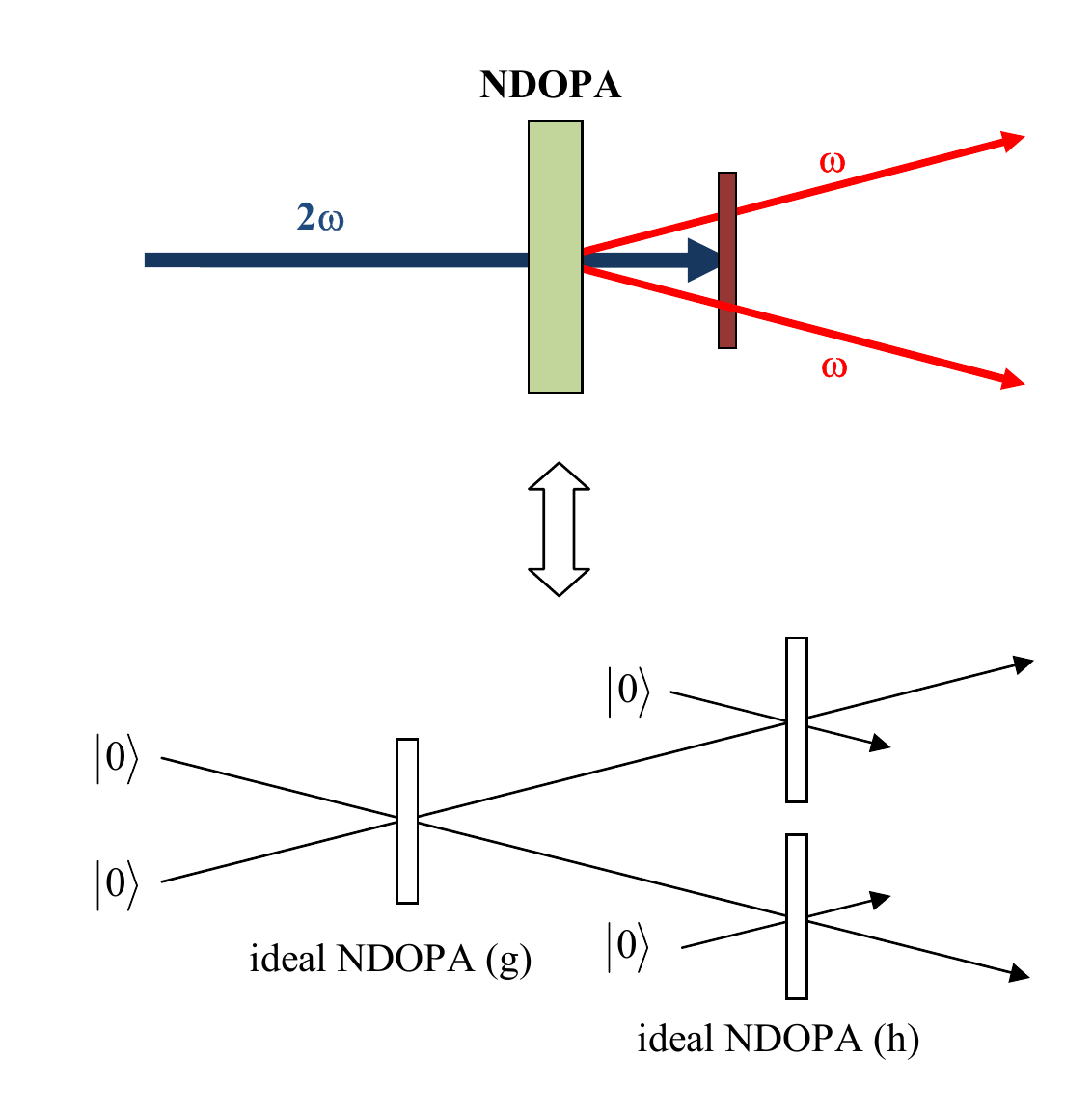}
\end{center}
\caption{(Color online) 
A non-degenerate optical parametric amplifier
(NDOPA) can be modeled by the succession of an ideal NDOPA of
gain $g$ and of two ideal NDOPAs of gain $h$ to account for added
noise.} \label{Rosa:Model_NDOPA}
\end{figure}

For instance, a degenerate OPA
used to produce a squeezed state can be modeled by the succession of
an ideal degenerate OPA and of an ideal
non-degenerate optical parametric amplifier (NDOPA) \citep{Paris2003,Adam1995} or an attenuator \cite{Lvovsky2015} to
account for added noise. In the same way, a NDOPA for the
generation of Gaussian EPR states can be modeled by the
succession of an ideal NDOPA and of two ideal NDOPAs
 (Fig.~\ref{Rosa:Model_NDOPA}).

A $1$-photon Fock state can be conditionally produced through a heralding detection event on one mode of the EPR state of Fig.~\ref{Rosa:Model_NDOPA}. A spatial and spectral filtering system is needed on the heralding mode in order to clearly define the spatio-temporal mode in which the Fock state is emitted, as discussed in detail in Sec.~\ref{condSPDC}. A modal purity parameter $\xi$ can be introduced in order to characterize the quality of this filtering system by quantifying the fraction of detected photon in the desired mode \citep{Lvovsky2001,Wenger2004a,Tualle-Brouri2009}: the signal state will then be a statistical mixture of the ideal heralded state and the state that is present in the signal mode without heralding. Analytical expressions can then be derived for its Wigner function  \citep{Ourjoumtsev2006b}:
\begin{equation} \label{Rosa:WignerFock1}
W_1(x,p)=\frac{e^{-R^2/\sigma^2}}{\pi\sigma^2}
\left[1-\delta+\frac{\delta R^2}{\sigma^2}\right],
\end{equation}
where $R^2=x^2+p^2$ and  $\sigma^2$ and $\delta$ are functions of parameters $g$, $h$, $\xi$, of the homodyne efficiency $\eta$  and of the electronic noise $e$. 
The Wigner function in this case is therefore supposed to depend on only two parameters which can be easily deduced 
from the first non-zero moments of the measured quadrature distributions:
\begin{eqnarray} \label{Rosa:Fock1Moments}
\mu_2\equiv\langle x^2\rangle = \sigma^2(1+\delta)/2 \\
\mu_4\equiv\langle x^4\rangle = 3\sigma^4(1+2\delta)/4
\end{eqnarray}
This very fast method allows a real-time reconstruction, which is
a major help for optimizing an experiment. It also allows one to
understand which kind of parameters can be deduced from 
experimental data: it is clear that all the model parameters $g$,
$h$, $\xi$, $\eta$ and $e$ cannot be deduced from the only two
parameters $\sigma$ and $\delta$ involved in Eq.~(\ref{Rosa:WignerFock1}). 

More information can be obtained from a homodyne tomography of the photon
subtracted squeezed vacuum discussed in Sec.~\ref{ScCatTheoSec}. In this case, the Wigner function depends on the phase and is of the form
\begin{equation} \label{Rosa:WignerKitten}
W_{c}(x,p)=\frac{e^{-\frac{x^2}{a}-\frac{p^2}{b^2}}}{\pi\sqrt{ab}}
[1-\frac{a'}{a}-\frac{b'}{b}+2x^2\frac{a'}{a^2}+2p^2\frac{b'}{b^2}].
\end{equation}
The four parameters $a$, $a'$, $b$, $b'$ can be readily deduced from quadrature
moments. On the other hand, they can be expressed
\citep{Ourjoumtsev2006a} in terms of the five parameters
$s$, $h$, $\xi$, $\eta$ and $e$ of the model in which the squeezed vacuum is produced by an ideal degenerate OPA of
squeezing $s$ followed by an ideal NDOPA of gain $h$. In this case, independent knowledge of one
model parameter is still needed: the homodyne efficiency
$\eta$ cannot be recovered from experimental data and must be
evaluated from separate measurements. Physical constrains on the parameters can be enforced using a Bayesian approach \cite{Blandino2012a}.

To finish with this short review on parameter estimation, let us
mention the research on state non-classicality witnessing
\citep{Bednorz2011,Kot2012}. A widely recognized criterion for
non-classicality is the negativity of the Wigner function, which  is
directly connected to state tomography. Many
reconstruction methods are however based on quantum laws, as for
instance maximum likelihood reconstruction which needs the
introduction of the density matrix. The Radon transform does not rely
on such knowledge, but presents drawbacks such as parasitic
oscillations which can lead to irrelevant negative values. A way around 
is to note that violating the following inequality implies
negativity of the Wigner function \citep{Bednorz2011}:
\begin{equation} \label{Rosa:NCWitness}
\langle\mathcal{F}\rangle\equiv\int \mathcal{F}(x,p) \; W(x,p) \; \dd x\dd p\ge 0
\end{equation}
\noindent where $\mathcal{F}$ is any positive function. One can
choose \citep{Kot2012} for such a function the square of a polynomial
of $R^2=x^2+p^2$, so that the criterion (\ref{Rosa:NCWitness}) can
be evaluated from quadrature moments using
\begin{equation} \label{Rosa:Moments}
\langle R^{2n}\rangle=\begin{pmatrix}2n\\n\end{pmatrix}^{-1}
\frac{2^{2n}}{2n}\sum_{m=1}^{2n}\langle
x_{\theta=m\pi/(2n)}^{2n}\rangle
\end{equation}
Such a criterion allows one to assess the negativity of the Wigner
function directly from the moments, without reconstructing the state.

Errors on the estimated parameters can be derived from the
statistics of the experimental data through linear approximations
and error propagation. This can be easily performed using the results of the previous section. Let us consider for instance the value $W_{1,0}\equiv W_1(0,0)$ of the Wigner function (\ref{Rosa:WignerFock1}) of the one-photon Fock state, taken at the origin. As the negativity of this value is a criterion for non-classicity, it is of great importance to have an error estimation on it. Following Eq.(\ref{Rosa:WignerFock1}) and inverting Eq.(\ref{Rosa:Fock1Moments}), this parameter can be expressed as a function of the moments $\mu_2$ and $\mu_4$ of the measured quadrature distributions. The error on the parameter is therefore:
\begin{equation} \label{Rosa:ErrorOnW10}
\langle\delta W_{1,0}^2\rangle=\sum_{k,l\in\{2,4\}}\frac{\partial W_{1,0}}{\partial\mu_k}\frac{\partial W_{1,0}}{\partial\mu_l}\langle\delta\mu_k\delta\mu_l\rangle
\end{equation}

Let us note that the error on the moments can be estimated from the moments themselves: using the estimator $N^{-1}\sum_{i=1}^N x_i^k$ for the moment $\mu_k$, and injecting it into the expression $\langle\delta\mu_k\delta\mu_l\rangle\equiv\langle\mu_k\mu_l\rangle-\langle\mu_k\rangle\langle\mu_l\rangle$, one finds
\begin{equation} \label{Rosa:ErrorOnMuk}
\langle\delta\mu_k\delta\mu_l\rangle=\frac{\mu_{k+l}-\mu_{k}\mu_{l}}{N}
\end{equation}

A simple alternative to this approach is bootstrapping \citep{Home2009,Lvovsky2009}, where the distribution of the estimated parameters is simulated from random sets of data drawn from the reconstructed parameters.

\subsubsection{Intermediate approaches}

Quantum tomography that does not rely on any \emph{a priori} information about the state (section \ref{section:MaxLik}) is very general but becomes intractable when the size of the Hilbert space grows. Reconstruction through parameter estimation (section \ref{section:ParameterEstimation}) is very efficient but relies on a detailed model of the physical process used to generate the quantum state. Between these two extremes, several intermediate approaches exist, such as quantum compressed sensing \cite{Gross2010}, permutationally-invariant tomography \cite{Toth2010}, matrix product state tomography \cite{Cramer2010} or machine learning \cite{Carleo2017,Torlai2018,Macarone2019,Rocchetto2019,Yu2019}. Their common point is to describe the state with a model (\emph{Ansatz}) assuming some knowledge about the state's structure and rely on a restricted set of parameters to specify the state within the Ansatz. In particular, such an Ansatz can be provided by neural networks, which is attractive due to their known ability to serve as universal approximators \cite{Hornik1993}, as well as  broad availability and understanding of optimization techniques. Most results on applying neural networks for quantum tomography were so far obtained in the DV context, but \textcite{Tiunov2019} have recently used a restricted Boltzmann machine to efficiently reconstruct optical non-Gaussian states, while \textcite{Cimini2020} have used neural networks to directly test Wigner negativity of multimode non-Gaussian states.
Rapid progress along these lines is expected.

\subsubsection{Unbalanced homodyne detection}

Alongside the dramatic development of quantum tomography  based on balanced homodyne detection, alternative methods for the local sampling of the phase-space distribution
have been proposed  \citep{Wallentowitz1996,Banaszek1996c}. They are based on an unbalanced homodyne setup where the signal and the LO fields are mixed on a BS with low reflectivity $|r|^2=1-|t|^2 \ll 1$. This operation, acting on the transmitted signal, corresponds to the application of the displacement operator   $\hat{D}(\beta)= \exp(\beta\hat{a}^\dagger- \beta^*\hat{a})$ on the entering signal field  $\hat{a}$, which gives $\hat{D}^\dagger(\beta)\hat{a}\hat{D}(\beta)=\hat{a}+\beta$  where $\beta= (r/t)\vert \alpha_{LO}\vert e^{i\theta}$ \cite{Paris1996,Lvovsky2002a}. The probability $P_n(\beta, \eta)$ to detect $n$ photons in the emerging field  contains the information on the Wigner function in the phase-space point $\beta$:
\begin{equation}
W(\beta,\eta)=\frac{2}{\pi}\sum_0^\infty
\Big(\frac{2-\eta}{\eta}\Big)^n P_n(\beta, \eta),
\end{equation}
where $\eta=\vert t\vert^2\eta_{\rm det}$ with $\eta_{\rm det}$ being the quantum efficiency of the photon detector.
The main drawback of this technique is that it requires 
photon-number resolving detectors. So the first  implementations of the quantum sampling method have been realized with on/off detectors in the low-mean-photon number regime \citep{Juarez-Amaro2003,Allevi2009,Bondani2010} or using time-multiplexing detectors \citep{Laiho2010}.
To overcome the  technical problems of poor discrimination in the photon number,  cascaded schemes have been proposed: they combine the unbalanced homodyne scheme for the local reconstruction of phase-space distributions and the phase randomized
balanced homodyne detection for measuring the photon statistics \citep{Kis1999,Munroe1995}. Then schemes for the reconstruction of photon number distributions by mixing the signal with a thermal state has been proposed and realized \citep{Harder2013}. More recently implementations with number-resolving detectors have become available, along with schemes that are less sensitive to non-ideal detectors \citep{Sridhar2014,Harder2016,Olivares2019,Nehra2019,Sperling2020}.

\subsection{Quantum process tomography}

In the framework of this review, quantum process tomography is not as crucial as quantum state tomography
considered in the previous section. However, it does matter for the evaluation of processors, 
such as amplifiers or quantum gates, which will be presented in Part II. Therefore we present here
some relevant features, with special emphasis on heralded (non-deterministic) processes.

\subsubsection{Generalities}

A quantum process, which associates an input state $\hat\rho$ to an
output state $\mathcal{E}(\hat\rho)$, can be viewed as a completely
positive linear map \citep{Kraus1983,Chuang1997}. A positive map
preserves the Hermiticity and the positivity, what is clearly a
mandatory requirement for the transformation of density matrices.
Complete positivity consists in the additional requirement that
$\mathbb{I}\otimes\mathcal{E}$ is also a positive map (where
$\mathbb{I}$ is the identity on any ancillary space), what accounts for the fact that $\mathcal{E}$ can also be applied to parts of entangled states. This last requirement is closely linked to the Jamio{\l}kowsi isomorphism \citep{Jamiolkowski1972}, which
associates to any map $\mathcal{E}$ the quantum state
\begin{equation} \label{Rosa:Jamiolkowsi}
\hat\rho_\mathcal{E}=(\hat{\mathbb{I}}\otimes\mathcal{E})\ket{\Phi}\bra{\Phi}
\end{equation}
\noindent where
$\ket{\Phi}=\frac{1}{\sqrt{d}}\sum_j\ket{j}\ket{j}$ is a maximally
entangled two-mode state described in a $d$-dimensional Hilbert
space. The physical meaning of this isomorphism has to be
underlined: if (for a trace-preserving map) a state $\ket{\psi}$
is measured on the first mode of $\hat\rho_\mathcal{E}$, the second
mode is projected into $\mathcal{E}((\ket{\psi}\bra{\psi})^*)$, where the asterisk denotes the operator whose matrix is complex conjugate to that of $\ket{\psi}\bra{\psi}$. In this way, we describe the the map through a $2$-mode quantum state.


 A quantum map corresponding to a physical operation only needs to be trace-non-increasing 
\citep{Nielsen2000}, and may include non-deterministic processes, which are trace-non-preserving.
In fact, for any non-deterministic (heralded) process $\mathcal{E}$,  a trace-preserving ($\it{i.e.}$
deterministic) quantum map can be constructed by including both success and
failure events as parts of the process. The trace $\tr[\mathcal{E}(\hat\rho)]$ then
corresponds to the heralding success probability $P_{\rm succ}$, and
the final state is
\begin{equation} \label{Rosa:FinalState}
\hat\rho_{\rm out}=\frac{\mathcal{E}(\hat\rho)}{\tr[\mathcal{E}(\hat\rho)]}=\frac{\mathcal{E}(\hat\rho)}{P_{\rm succ}}
\end{equation}

A primary property of any quantum map is its linearity with respect to density operators. That is, if the effect of the process
$\mathcal{E}(\hat\rho_i)$ is known
for a set of density operators $\{\hat\rho_i\}$, its effect on any linear
combination $\hat\rho=\sum \beta_i\hat\rho_i$ is 
\begin{equation}\label{QPTbasis}\mathcal{E}(\hat\rho)=\sum \beta_i \mathcal{E}(\hat\rho_i). 
\end{equation}
Thanks
to this linearity, there exists a compact representation of any map in the form of a matrix multiplication. By applying Eq.~\eqref{QPTbasis} to the density matrix of an arbitrary input state $\hat\rho=\sum_{m,n}\hat\rho_{mn}\ketbra mn$, we find 
\begin{equation} \label{Rosa:MapFockBasis}
\bra{k}\mathcal{E}(\hat\rho)\ket{l}=\sum_{m,n}\mathcal{E}_{k,l}^{m,n},
\end{equation}
where
\begin{equation}\label{ProcessTensor}
\mathcal{E}_{k,l}^{m,n}=\bra k\mathcal{E}(\ketbra mn)\ket l.
\end{equation}
is the so-called process tensor and $\ket k,\ket l,\ket m,\ket n$ denote the elements of any basis, for example, the Fock basis \citep{Lvovsky2008}. 

A quantum map can also be represented using continuous variables: in the same way as
the Wigner function $W(x,p)$ can be viewed as a quasi-probability
distribution for the quadratures of a quantum state, a
Markovian-like process can be associated to a quantum map
\citep{Berry1979}, with:
\begin{equation} \label{Rosa:Ferreyrol}
W_{out}(x,p)=\int \; W(x',p') \; f_\mathcal{E}(x,p,x',p')\dd x'\dd p' 
\end{equation}
The function $f_\mathcal{E}$ can be obtained even for non-deterministic operations, and allows an intuitive picture of quantum maps, with very simple expressions for many elementary processes.  
The corresponding expression  for chained processes is very similar to the
Chapman-Kolmogorov equation for Markovian processes \citep{Ferreyrol2012}.

\subsubsection{Methods}
Quantum process tomography (QPT) is a technique of reconstructing the process tensor \eqref{ProcessTensor} by applying the process to a number of ``probe" states that constitute  a spanning set within the space $\mathcal{L}(\mathcal{H})$ of  linear operators over a particular Hilbert space $\mathcal{H}$. Then the knowledge of $\{ \mathcal{E}(\hat\rho_i)\}$ may be sufficient to extract complete information about the quantum process as per Eq.~\eqref{QPTbasis}. One possibility \citep{Lobino2008} is to use coherent states as probe, taking advantage of Eq.~\ref{pfunc}: any density operator can be written as a linear combination of density operators of coherent states. 
Applying Eq.~\eqref{ProcessTensor} to Eq.~\eqref{pfunc}, we find
\begin{equation}\label{QPTpfunc}
\mathcal{E}_{k,l}^{m,n}= \int d^{2}\alpha \; P_{\ketbra mn}(\alpha, \alpha^{\ast} ) \; \bra k\mathcal{E}(|\alpha \rangle\langle \alpha |)\ket l,
\end{equation}
where $P_{\ketbra mn}(\alpha, \alpha^{\ast} )$ is the P-function of the operator $\ketbra mn$. The advantage of this technique is that any quantum process, however complex, can, in principle, be reconstructed by means of simple laser pulses \citep{Lobino2008}, without nonclassical probe states. 

Coherent-state QPT in the form of Eq.~\eqref{QPTpfunc} is difficult to apply in practice because  $P_{\ketbra mn}(\alpha, \alpha^{\ast} )$ is a highly singular generalized function. This formula has been simplified in  \citep{Rahimi-Keshari2011}:
\begin{equation} \label{Rosa:Rahimi}
\mathcal{E}_{k,l}^{m,n}=\frac{1}{\sqrt{m!n!}}\partial_\alpha^m\partial_{\alpha^*}^n\left[e^{|\alpha|^2}\bra k\mathcal{E}(|\alpha \rangle\langle \alpha |)\ket l\right]_{\alpha=0}
\end{equation}
The derivatives at $\alpha=0$ can be deduced from a finite set of probe
coherent states through a polynomial interpolation. A further simplification is provided by the QPT  maximum-likelihood algorithm \citep{Anis2012} which iteratively calculates the process tensor directly from the quadrature measurements $\mathcal{E}(|\alpha \rangle\langle \alpha |)$. This algorithm utilizes the Jamiolkowski isomorphism to effectively reduce QPT to quantum state tomography.

In the case of a heralded process, the final state
$\hat\rho_{\rm out}(\alpha)$ is different from
$\hat\rho(\alpha)=\mathcal{E}(\ket{\alpha}\bra{\alpha})$ due to the
normalization in Eq. (\ref{Rosa:FinalState}). As underlined by
\cite{Rahimi-Keshari2011}, this problem can be circumvented with the knowledge of the
success probability $P_{\rm succ}$ by setting
$\mathcal{E}(|\alpha \rangle\langle \alpha |)=P_{\rm succ} \; \hat\rho_{\rm out}(\alpha)$ in Eq.
(\ref{Rosa:Rahimi}). In the maximum-likelihood algorithm  \citep{Anis2012}, trace-non-preserving processes are 
 taken into account by treating instances in which the heralding event did not occur as the process output being a special state that is not an element of the original Hilbert space. 
 
 Since its development, the method of coherent-state QPT has been applied to a large variety of processes, such as quantum-optical memory \citep{Lobino2009}, photon addition and subtraction \citep{Kumar2013,Ra2017} as well as other conditional state-engineering quantum processes \citep{Cooper2015}, and optically-controlled kerr non-linearity \citep{Kupchak2015}. It has also been extended to the multimode case \citep{Fedorov2015a}. On the other hand, it has been observed that ``coherent states are very `classical', and provide  exponentially little information about parameters of some  quantum processes" \citep{Rosema2014}. This has motivated a search for a more efficient set of probe states. In particular, squeezed states have been investigated for this role \citep{Fiurasek2015}.


\subsubsection{Physical models for quantum processes}
Generally, QPT
is a task far more difficult than quantum state tomography. The total number of parameters contained in the process tensor scales as the fourth power of the Hilbert space dimension, which means that QPT requires one to acquire and  manipulate a vast amount of data. Under some circumstances, however, certain features of a process are known \emph{a priori}. Then the general QPT, as described above, is not necessary; the problem effectively reduces to parameter estimation.

For example, if the process is known to be of linear optical nature, its characterization is straightforward even in the case of multiple input and output modes. The transformation matrix can be reconstructed by measuring the intensities in the output ports in response to coherent states injected into individual input ports as well as pairs thereof \citep{Rahimi2013}. A similar goal can be achieved by using single-photon states as probes, in which case the photon count rates in individual outputs, as well as pairwise coincidences must be measured \citep{Laing2012,Dhand2016}.

Another instance is the non-deterministic phase gate proposed in \citep{Marek2010} and discussed in detail in Sec.~\ref{QCompSec}. For the phase shift of $\pi$, this gate reduces to a photon-subtraction experiment with a sampling BS and a photon detection heralding event. This setup can be characterized by two parameters: the reflectivity $R$ of the sampling BS and the modal purity parameter $\xi$. These parameters can be determined from only one input state like a squeezed vacuum, and can successfully predict the behaviour of the gate for a squeezed single photon at the input, then allowing a quantitative comparison between the experimental gate and the ideal one \citep{Blandino2012b}. A similar approach was also applied to the non-deterministic Hadamard gate \citep{Tipsmark2011}.

\section{Conditional photon measurements in  parametric down conversion}
\label{condSPDC}
The parametric down conversion process (PDC) in the pulsed low-gain regime is the most widely used scheme
for the conditional implementation of nonclassical states of light with few photons in a localized temporal frame.  Fock states  and (multi-)entangled states, which are the basic ingredients for photonic quantum information and communication protocols, were obtained  in PDC setups \citep{Lvovsky2001,Ourjoumtsev2006b,Bouwmeester1999,Yao2012}.

The basic principle of heralded state preparation by means of photon detection has been discussed in Sec.~\ref{NonGauss:PhotonCount}. Here we analyze this procedure in view of an important practical complication: photon pair produced in PDC are typically entangled spectrally. Therefore photon detection in the heralding channel, unless performed judiciously, will result in the heralded state in a highly impure optical mode that is unsuitable for further applications. The optimal approach to heralding strongly depends  on how the PDC bandwidth compares to the inverse time resolution of the photon detector. Accordingly, we identify two main regimes of operation: pulsed (broadband) and cw (narrowband). 

\subsection{Pulsed low-gain regime}\label{PDC-PLG}

\begin{figure}[tb]
\begin{center}
\includegraphics[width=0.9\columnwidth]{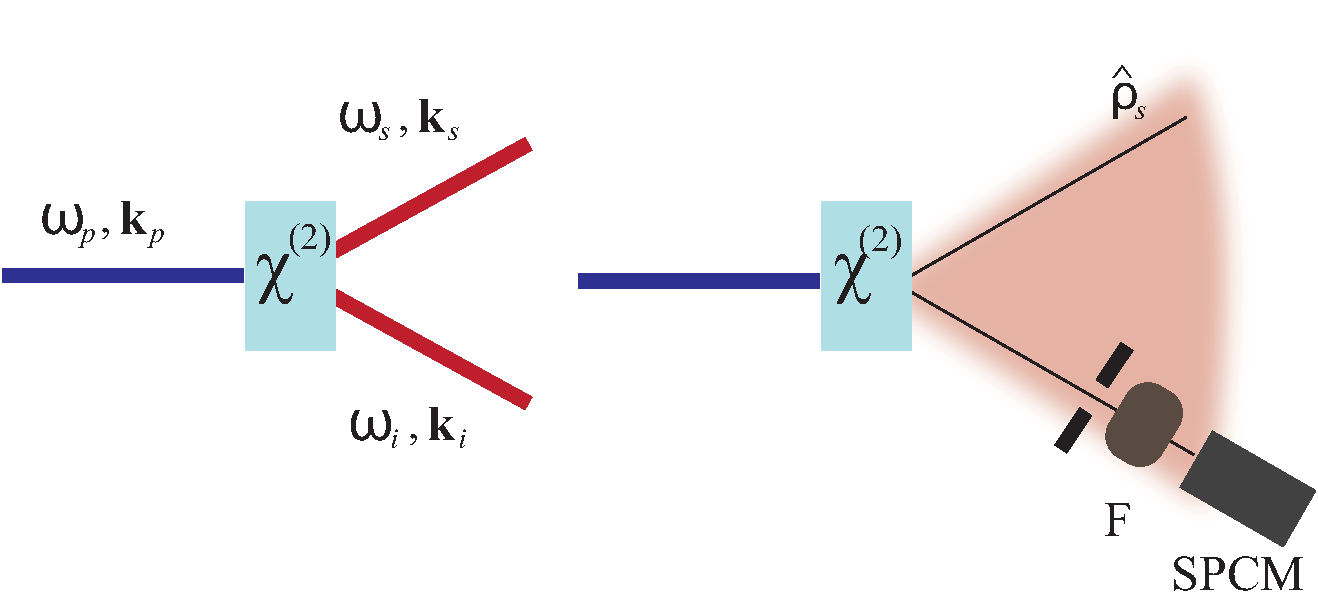}
\end{center}
\caption{(Color online) 
Parametric down conversion. Left: one pump photon with frequency $\omega_{p}$ and wavevector $\mathbf{k}_{p}$ is converted into two lower energy photons $\omega_{s}, \mathbf{k}_{s}$ and $\omega_{i},\mathbf{k}_{s}$. Right: detection via the the single-photon  counting module (SPCM)  after spatial and spectral filtering (F) in the idler channel projects the  signal channel into the state $\hat{\rho}_{s}$ which approaches a pure single-photon state  when the filter bandwidth is much smaller than the pump bandwidth.} \label{SPDCFIG}
\end{figure}

The two photons, called idler and signal,
and the original pump one  satisfy energy and momentum conservation,
which are expressed in the \textit{phase-matching} conditions
$\omega_{p}=\omega_{s}+\omega_{i} $ and $\label{PM2}\mathbf{k}_{p}=\mathbf{k}_{s}+\mathbf{k}_{i}$. The symbols
$\omega_{p,s,i}$ denote the pump, signal and
idler frequencies  and $\mathbf{k}_{p,s,i}$ the
corresponding wave vectors. The  two expressions ensure that
the two down-converted photons are
correlated in momentum and energy, but these attributes are not
defined for either of them until a measurement is performed on the
other one. These features, together with the simultaneous emission
within the pump coherence time, cause  the two photons to exhibit
non-classical behaviour.
The process of parametric down-conversion can be studied in the interaction
picture
\begin{equation}
H_{I}(t)=\chi^{(2)}\int_{V}E^{(-)}_{s}(\mathbf{r},t)E^{(-)}_{i}(\mathbf{r},t)E^{(+)}_{p}(\mathbf{r},t)\dd ^{3}\mathbf{r}+h.c.
\end{equation}
where $\chi^{(2)}$ is the second-order electric susceptibility of the crystal and  $E^{(-/+)}_{p,s,i}(\mathbf{r},t)$ are the negative/positive part of the pump, signal, idler fields.
Using a plane-wave decomposition, the idler/signal field is quantized in each 
spatial-spectral component $ (\mathbf{k}_{s,i},\omega_{s,i})$, and the amplitude of 
the classical pump beam is denoted as $\mathcal{E}^{(+)}_{p}(\mathbf{k}_{p},\omega_{p})$.  
Assuming weak interaction, the solution for the state in the signal and idler channels 
can be expressed in terms of Dyson series \citep{Sakurai1994}, which we write up to the first order:
\begin{align}\label{dyson}
\vert\psi(t)\rangle & =\ket{\mathbf{0,0}}_{si}
-\frac{i}{\hbar}\int_{0}^{t}H_{I}(t')\ket{\mathbf{0,0}}_{si}\dd t'+\ldots\\ \nonumber
 = & \ket{\mathbf{0,0}}_{si}+\int
\phi(k_{s},k_{i}) \,
 \ket{1}_{k_s}\ket1_{k_i}\dd k_s \dd k_i+\ldots
 \end{align}
where we used a single symbol $k=(\mathbf{k},\omega)$ for the spatial and spectral component and $\ket1_{k_{s,i}}$ is the state containing a single photon in the mode $k_s$ of the signal or $k_i$ of the idler channel and vacuum elsewhere. The second term in the above equation describes the simultaneous emission of one photon in the idler and in signal channel
is sometimes called biphoton; its spatial and spectral properties are described by the
function $\phi(k_{s},k_{i})$ which depends both on the pump spatial and spectral distribution and on the phase-matching term related with the crystal properties. Higher-order terms describes the less probable processes with  two and more photons in each channel.

\begin{figure}[tb]
	\begin{center}
		\includegraphics[width=\columnwidth]{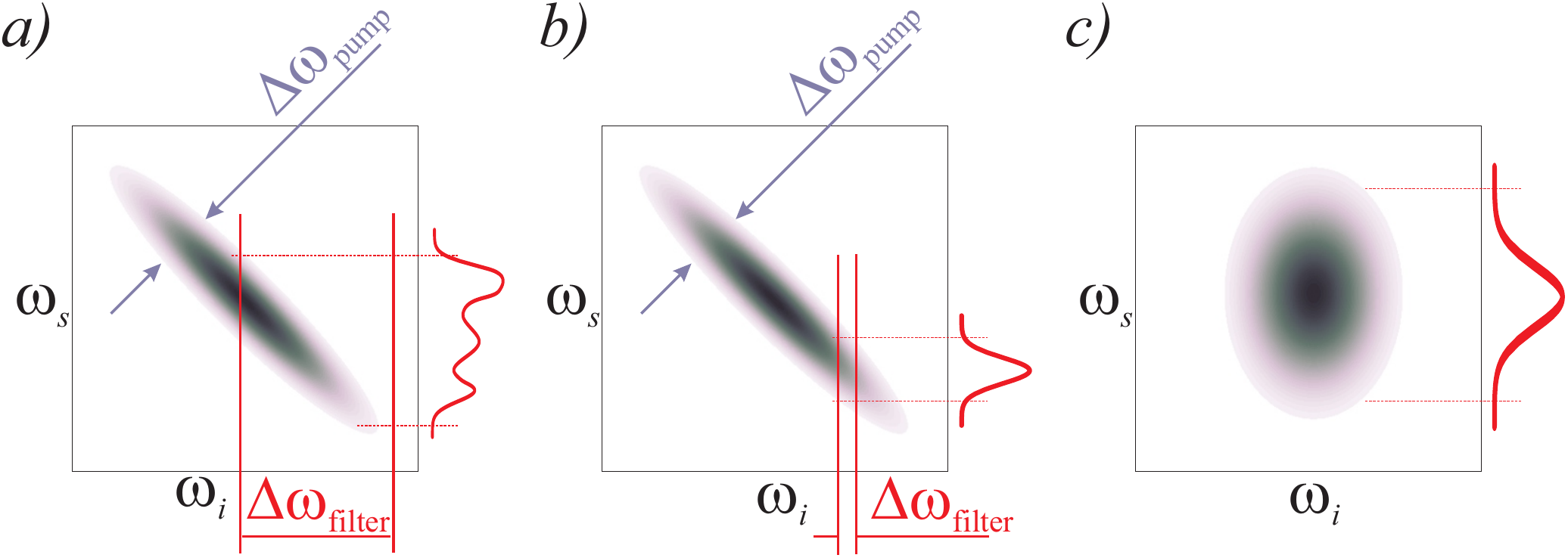}
	\end{center}
	\caption{(Color online) Filtering of the idler channel for heralded preparation of the signal photon. The contour plots show the biphoton spectra $\phi(k_{s},k_{i})$. a) Broadband filtering results in a non-pure state. b) Narrowband filtering produces the heralded photon in the pure state, with the spectrum resembling that of the pump. c) If the biphoton is separable,  the state of the signal photon is independent of the filtering conditions.
} \label{fig:filter}
\end{figure}

Heralded single photon preparation in the signal channel can be described by the projection \cite{Aichele2002}
\begin{equation} \label{SP-spdc1}
\hat{\rho}_{s}=\tr_{i} \lbrace \hat{\rho }_{i}\vert \psi(t)\rangle \langle\psi(t)\vert \rbrace
\end{equation}
where $\tr_{i}$ is the trace taken over the trigger states (idler) and $\hat{\rho}_{i}$ is the state ensemble selected by the filters $T(k_i)$:
\begin{align}
\hat{\rho}_{i}&= \int  \;  T(k_{i} ) \; _{k_i}\hspace{-1mm}\langle 1|\psi(t)\rangle \langle\psi(t)|1\rangle_{k_i}\dd k_i\\ \nonumber
&= \iiint  T(k_{i} )\phi(k_{s},k_{i})\phi^*(k'_{s},k_{i})  |1\rangle_{k_s\,k_s'}\hspace{-1mm}\langle 1|\dd k_i\dd k_s\dd k_s'.
\end{align}
The state purity of the generated heralded photon depends on the ratio between the filter bandwidth and the characteristic width of the function $\phi(k_{s},k_{i})$; generally, narrowband filtering is required [Fig.~\ref{fig:filter}(a,b)], in which case the spectrum of the heralded photon largely replicates that of the pump. However, if the SPDC configuration is such that the biphoton spectrum is separable, i.e.~$\phi(k_{s},k_{i})=\phi_s(k_{s})\phi_i(k_{i})$, the state of the signal photon is independent of the filtering conditions [Fig.~\ref{fig:filter}(c)]. This is a highly desired condition because it permits eliminating the filtering in the idler channel altogether, thereby greatly enhancing the heralded photon production rate. For the first time, this condition has been achieved in \cite{Mosley2008}. 

The  probability of producing one biphoton 
in one pump pulse is $\label{prob1phot}p_{1}=\langle \psi_{1}|\psi_{1}\rangle=
\int |\phi(k_{s},k_{i})|^{2}\dd k_{s}\dd k_{i}$.  The larger the parametric gain, the higher the success rate of generating the heralded photon, provided that the number of photons in the trigger mode can be precisely determined. If a detector that cannot discriminate the exact photon number is used, the gain must be sufficiently low in order to reduce the probability of generating several photons in the signal mode, instead of only one.

%
%
\subsection{Continuous regime}
%
%

Besides the pulsed regime considered in the previous section, 
continuous-wave squeezed light
generated by an optical parametric oscillator (OPO)
offers excellent sources for non-Gaussian operations
with high state purity. 
Compared to pulsed experiments, cw squeezed states produced by OPOs have much narrower spectral bandwidths (of the order of a few MHz rather than a few THz). This requires the use of optical cavities and makes the experimental control more involved but, as a pay-off, all states are prepared in very well-defined spatial and spectral modes. This allows one to reach very high purities for the prepared states, and to detect them with almost unity homodyne efficiency. 

The use of cw squeezed light for quantum information was triggered by seminal papers on quantum teleportation 
\citep{Braunstein1998,Furusawa1998}, and subsequently extended to the non-Gaussian realm. 
In this section we review important non-Gaussian operations in the cw regime.

%
%
%
%
%
%
%
%
%
%
%
%
\subsubsection{CW squeezed state}
%
%

\begin{figure}[tb]
	\centering {\includegraphics[width=0.95\linewidth]{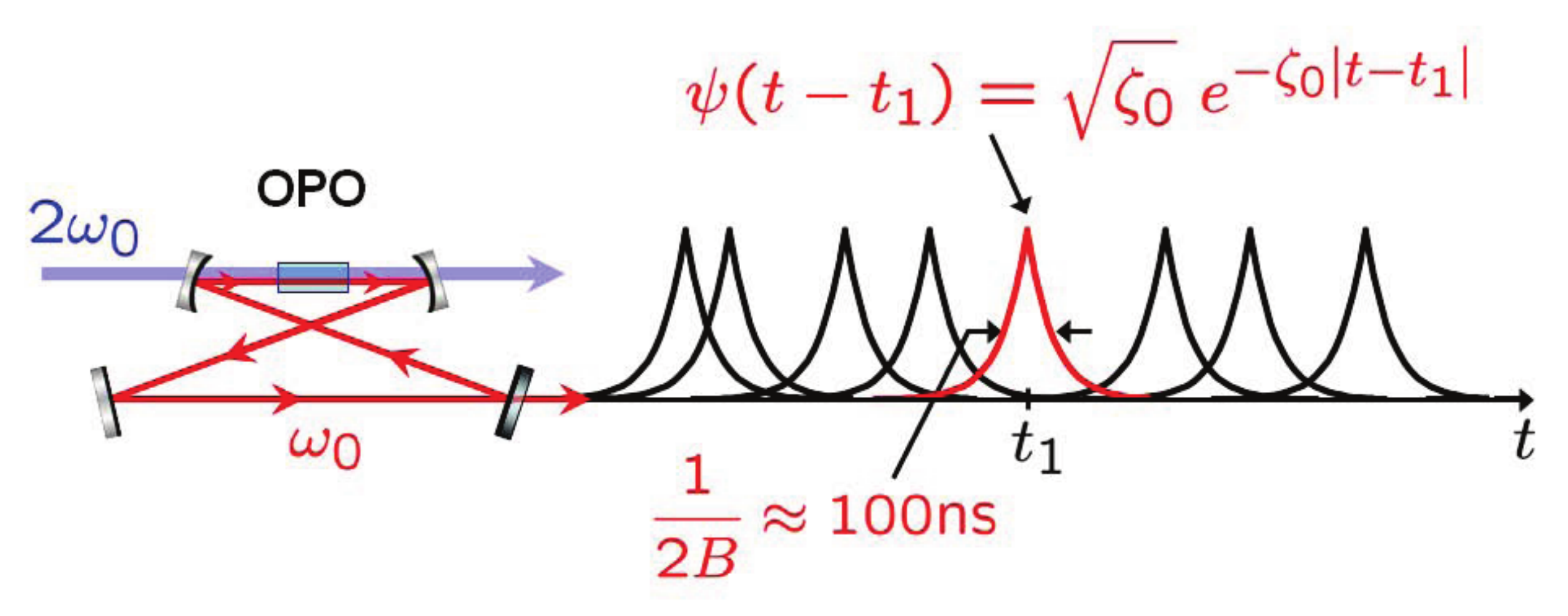}
		\caption{(Color online) 
			Left: OPO cavity preparing a cw squeezed vacuum state. Right: a photon detection event at time $t_1$ conditionally prepares a (squeezed) single photon in the temporal mode
			$\psi(t-t_k)$.
		}
		\label{fig.cw_sq_st_time_domain}}
\end{figure}

Figure \ref{fig.cw_sq_st_time_domain} depicts
a cw squeezed beam generated by
a typical scheme (bow-tie cavity) of OPO. 
It has a Lorentzian spectral shape. 
The OPO is driven
by a cw pump beam at an angular frequency $2\omega_0$.
A cw squeezed beam is generated at the frequency $\omega_0$.

Continuous-variable squeezed states are typically described in the frequency domain, 
because they are generated by a narrow-band-limited OPO. On the other hand, a photon detection event occurs at a specific moment in time, and is hence best treated in the time domain. Therefore, in order to develop the theory of preparing non-Gaussian states from cw squeezing, we need to establish the connection between these approaches \cite{Sasaki2008b,Lvovsky2015}.

The frequency-dependent field operators obey the continuum commutation relation
\be
\label{continuum commutation relation in frequency}
[\hat a(\omega_0+\Omega), \hat a^\dagger(\omega_0+\Omega')]
=2\pi\delta(\Omega-\Omega').
\ee
These operators are related to the time-dependent field operator as 
\be
\hat a(t)=\frac{1}{2\pi}\int_{-\infty}^{\infty} d\Omega \; 
\hat a(\omega_0+\Omega) \; 
e^{-i(\omega_0+\Omega)t}.
\label{eq.01}
\ee
The operator $\hat a(t)$ is defined in the interval $(-\infty,\infty)$ 
and obeys the commutation relation
\be
\label{continuum commutation relation in time}
[\hat a(t), \hat a^\dagger(t')]=\delta(t-t').
\ee
Squeezed light fields can be conveniently described
in the rotating frame about
the center frequency $\omega_0$,
\be
\label{operator in rotating frame}
\hat A(t)=\hat a(t) e^{i\omega_0t}
=
\frac{1}{2\pi}\int_{-\infty}^{\infty}  
\hat A(\Omega) \; e^{-i\Omega t}\dd\Omega
\ee
where
$\hat A(\Omega)=\hat a(\omega_0+\Omega)$.
This angular frequency $\Omega$ is the variable 
for the side-bands around the center frequency $\omega_0$, 
within the OPO bandwidth. 

The single-mode squeezing process can be described by a unitary operator, 
and the pure cw squeezed state is represented as  
$\hat S \ket{\mathbf{0}}$
with
\be\label{Squeezing operator S_{A}}
\hat S
=\mathrm{exp}
\Biggl(
\int_{-\infty}^{\infty} 
r(\Omega)
\biggl[
\hat A(\Omega)\hat A(-\Omega)
-
\hat A^\dagger(\Omega)\hat A^\dagger(-\Omega)
\biggr]\dd\Omega
\Biggr),
\ee
where $r(\Omega)=r(-\Omega)$ is a real squeezing parameter and $\ket{\mathbf{0}}$ is the tensor product of vacuum states in all temporal modes. That is, continuous single-mode squeezing in the time domain corresponds to two-mode squeezing in the frequency domain applied to the continuum of pairs of modes at frequencies $(\Omega,-\Omega)$.

Its spectrum is determined by the cavity characteristics of the OPO, 
and in the idealized lossless limit, 
it is given by \citep{Walls2008}:
\be\label{SqSpectrum}
\mathrm{exp}\left[ r(\Omega) \right]
\equiv
\sqrt{
	\frac{(\zeta_0+\epsilon)^2+\Omega^2}
	{(\zeta_0-\epsilon)^2+\Omega^2}
},
\ee
where 
$\zeta_0$ corresponds to the OPO resonant bandwidth and the value of $\epsilon$ is proportional to
the non-linear $\chi^{(2)}$ coefficient
and the pump field amplitude; the OPO ocillation threshold corresponds to $\epsilon=\zeta_0$.  
On the other hand, for $\epsilon\ll\zeta_0$, Eq.~(\ref{SqSpectrum}) becomes 
$r(\Omega)\approx4\zeta_0\epsilon/(\zeta_0^2+\Omega^2)$. Thus, in the limit of low squeezing, the cw squeezed state has a Lorentzian spectrum with the characteriztic band $[-\zeta_0,\zeta_0]$.

The Bogoliubov transformation associated with this squeezing is
\be
\label{Bogolubov transformation by S_{A}}
\hat S^\dagger \hat A(\Omega) \hat S
=\mu(\Omega) \hat A(\Omega)
+\nu(\Omega) \hat A^\dagger(-\Omega),
\ee
where, as per Eq.~\eqref{EqAAdagOPA},
\beqa
\label{cosh and sinh}
\mu(\Omega) =
\cosh r(\Omega),
\nonumber \;  \;  \;   
\nu(\Omega) =
-\sinh r(\Omega).
\eeqa

For two-mode squeezing, Eqs.~(\ref{Squeezing operator S_{A}}) and (\ref{Bogolubov transformation by S_{A}}) take the form 
\begin{align}\label{Squeezing operator S2_{A}}
&\hat S_{si}\\ \nonumber
&=\mathrm{exp}
\Biggl(
\int_{-\infty}^{\infty} 
\frac{r(\Omega)}2
\biggl[
\hat A_s(\Omega)\hat A_i(-\Omega)
-
\hat A_s^\dagger(\Omega)\hat A_i^\dagger(-\Omega)
\biggr]\dd\Omega
\Biggr)
\end{align}
and 
\be
\label{Bogolubov transformation by S2_{A}}
\hat S_{si}^\dagger \hat A_{s,i}(\Omega) \hat S_{si}
=\mu(\Omega) \hat A_{s,i}(\Omega)
+\nu(\Omega) \hat A_{i,s}^\dagger(-\Omega),
\ee
where the signal $s$ and idler $i$ are two modes that are distinguistable either spatially or in polarization.

In a practical setting, 
the above pure-state treatment needs to be modified
by including imperfections of losses and noises, 
which results in the output squeezed state becoming mixed and losing its minimum-uncertainty nature.

%
%
\subsubsection{Heralded photon preparation} \label{cwHeraldedSec}
%
%
Next, we consider heralded preparation of the photon in the low-gain regime, i.e.~the same problem as in Sec.~\ref{PDC-PLG}, but for the cw regime. In the case of low gain, we start with the two-mode squeezed vacuum $\hat S_{si}\ket{\mathbf{0,0}}_{si}$ and decompose the exponent in Eq.~\ref{Squeezing operator S2_{A}} to the first order. We find
\begin{align}\label{cwSq}
\hat S_{si}\ket{\mathbf{0,0}}_{si}&\\ \nonumber 
&\hspace{-1cm}=\left(\mathbb{I}
-\int_{-\infty}^{\infty} \frac{r(\Omega)}2\biggl[\hat A_s^\dagger(\Omega)\hat A_i^\dagger(-\Omega)\biggr]\dd\Omega
\right)\ket{\mathbf{0,0}}_{si}
\end{align}
Now suppose a photon is detected in the idler mode at the moment $t_1$. The state of this photon can be written as $\hat A_i^\dag(t_1)\ket{\mathbf{0}}_i$, so the projection onto that state will prepare the following state of the signal:
\begin{align}\label{} _i\hspace{-1mm}\bra{\mathbf{0}}\hat A_i(t_1)\hat S_{si}\ket{\mathbf{0,0}}_{si}&\\ \nonumber
&\hspace{-2cm}=\frac1{2\pi}\int_{-\infty}^{\infty} \frac{r(\Omega)}2 \hat A_s^\dagger(\Omega) \; e^{i\Omega t_1}\dd\Omega\ket{\mathbf{0}}_s=A_1^\dagger\ket{\mathbf{0}}_s,
\end{align}
where
\begin{align}
	\label{A_1a}
	\hat A_1^\dagger&=\frac{1}{4\pi}\int_{-\infty}^{\infty} e^{i\Omega t_1} \frac{r(\Omega)}2 \hat A_s^\dag(\Omega) d\Omega
	\\&=\frac12\int_{-\infty}^{\infty} 	\hat A^\dagger(t) \psi (t-t_1) \dd t,\label{A_1a2}
\end{align}
with 
\be
\label{psi(t)a}
\psi (t-t_1)=\frac{1}{2\pi} \int_{-\infty}^{\infty}
e^{-i\Omega (t-t_1)}
r(\Omega) \dd\Omega
.
\ee
We see that the state of the signal mode is a single photon in the temporal mode given by Eq.~\ref{psi(t)a}  \citep{Lund2004,Suzuki2006}. This mode is a  two-sided exponential centered at $t_1$: 
\begin{equation}\label{cwTempMode}\psi (t-t_1)\propto e^{-\zeta_0 \vert t-t_1\vert} 
\end{equation}because the function $r(\Omega)$ approaches Lorentzian in the limit of low squeezing (Fig.~\ref{fig.cw_sq_st_time_domain}).

In deriving the above results, we assumed that the time moment $t_1$ of the heralding event is known precisely. In fact, the heralded wavepacket is incoherently smeared over the time interval $T$ that is equal to the uncertainty of that event. However, as long as this uncertainty is significantly below the width of that wavepacket, i.e.,~$\zeta_0T\ll1$, it can be neglected \citep{Sasaki2006}. 

This is indeed the case  in a typical cw experiment. Unless the heralding photon is deliberately subjected to spectral filtering, typically $\zeta_0\sim30\times 10^6$\,s$^{-1}$
whereas $T\lesssim$1\,ns for photon counting modules based 
on silicon avalanche photodiodes), thus $\zeta_0T\sim0.03$, resulting in a high-purity heralded mode. In contrast, the short pulse scheme typically has $\zeta_0\sim10^{12}$\,s$^{-1}$, so $\zeta_0T\sim10^3$, therefore narrow filtering in the trigger channel is required as discussed in the previous section.

%
%
\subsubsection{Photon subtraction from squeezed vacuum}
%
%

Let us now analyze a more complex case of subtracting a single photon from a cw single-mode squeezed vacuum by means of a low-reflectivity beam splitter and a SPCM, as described in Sec.~\ref{ScCatTheoSec}. Detecting a trigger signal at $t=t_1$ is equivalent to applying the annihilation operator $\hat A(t_1)$ to the state $\hat S_A \ket{\mathbf{0}_{A}}$. Using the unitarity of the squeezing operator, we write 
\begin{align}\label{}
\hat A(t_1)\hat S \ket{\mathbf{0}}&=\frac{1}{2\pi}\int_{-\infty}^{\infty} \; e^{-i\Omega t_1}
\hat A(\Omega)  \hat S \ket{\mathbf{0}} d\Omega\\
&=\frac{1}{2\pi}\int_{-\infty}^{\infty} \; e^{-i\Omega t_1}\hat S\hat S^\dag
\hat A(\Omega)  \hat S \ket{\mathbf{0}} d\Omega\nonumber
\end{align}
Now substituting the Bogoliubov transformation \ref{Bogolubov transformation by S_{A}}, we find
\begin{equation}\label{ASA0A}
\hat A(\Omega)  \hat S \ket{\mathbf{0}} =\hat S\hat S^\dag\hat A(\Omega)  \hat S \ket{\mathbf{0}} =\hat S \hat A_1^\dag \ket{\mathbf{0}},
\end{equation}
where the mode operator $\hat A_1^\dag$ is given by Eq.~(\ref{A_1a2})
with the mode function
\be
\label{psi(t)}
\psi (t-t_1)=\frac{1}{2\pi} \int_{-\infty}^{\infty}
e^{i\Omega (t-t_1)}
\nu(\Omega) d\Omega
.
\ee
In other words, in the case of finite squeezing, the conditionally prepared state is the squeezed single photon in the temporal mode (\ref{psi(t)}) and squeezed vacuum in other temporal modes. Although we are no longer working in the limit of infinitesimal squeezing, the temporal mode is typically well approximated by Eq.~(\ref{A_1a2}).

This interpretation is only approximate because the operator $\hat S_A$, given by Eq.~(\ref{Squeezing operator S_{A}}), is frequency dependent, so different spectral modes of the single photon $\hat A_1^\dag\ket{\mathbf{0}_{A}}$ in Eq.~(\ref{ASA0A}) are squeezed differently  \citep{Yoshikawa2007}. This leads to effective detection efficiency losses that are insignificant for small squeezing.  This issue can be addressed, for example, by spectrally filtering the heralding photon with a linewidth that is much narrower than that of the squeezing cavity \citep{Asavanant2017}.

The typical feature of the cw scheme, $\zeta_0T\ll1$,
leads not only to the ideal mode preparation condition,
but also to a regime
where
multiple photons can be subtracted within
the coherence time of the squeezed light,
because
the trigger photon counting 
has a faster temporal resolution than
the OPO time scale.
This provides rich physics in the time domain, which we discuss in Sec.~\ref{PhotonSubExpSec}.

%
%
%
%
%
%
%
%
%
%
%
%
%
%

\vskip 1 cm

\part{Results \& analysis}
\label{Results}

\section{Production and tomography of Fock states} 
The first tomographic reconstruction of a single-photon Wigner function was realized in the 1.6-picosecond pulsed regime \citep{Lvovsky2001}, by heralding on a biphoton prepared via low gain SPDC (Sec.~\ref{PDC-PLG}).
All experimental imperfections in the generation stage (e.g. limited state-purity given by the finite filter bandwidth in the idler channel, and dark counts of the single photon counter) and in the detection process (e.g. optical losses, detector electronic noise and mode-mismatch between the LO and the signal field) led to a global efficiency $\eta=0.55$. 
In this first experiment the mode-locked pulse repetition rate was lowered from the $\sim$80MHz to $\sim$800kHz by means of a pulse picker because the  bandwidth of the homodyne detector was too low to discriminate consecutive pulses. 
Afterwards high-frequency bandwidth homodyne detectors for fast states analysis were developed (Sec.~\ref{Homodynedet}) allowing the generation and analysis of quantum states at the full laser repetition rate. A 0.3 nm bandwidth interference filter in the trigger channel was used to make the signal photons sufficiently pure (Fig.~\ref{fig:filter}(b)) and to make their mode match that of the maser laser which cold then be used as the LO for homodyne detection \cite{Aichele2002}.

Single-photon states can also be conditionally prepared using 
cw OPAs, where the spontaneous parametric down conversion occurs in a cavity \citep{Nielsen2007c}.  The output state of an OPA operating far below threshold is a two-mode squeezed state described by Eq. (\ref{Squeezing operator S2_{A}}) which, in the ideal case, corresponds to a spontaneous parametric down conversion emission in a well-defined spatial and temporal mode. 
One possible configuration is the non-degenerate  Type-I OPA, where the signal and idler photon are emitted in two distinct frequency modes $\omega_+$ and $\omega_-$,   separated by one cavity free spectral range \citep{Neergaard-Nielsen2007}.  In this case the idler photon is split  from 
the signal by a cavity which is resonant only with the idler field, and then  transmitted and strongly filtered by subsequent cavities in order to suppress the contributions from neighbouring uncorrelated cavity modes.
Otherwise, in the degenerate Type-II scheme the two channels have the same frequency but different polarizations \citep{Morin2012}, and  the signal and idler channel are separated by a polarizing beam splitter before going into the detection stages. The detection of the idler photon on an avalanche photodiode triggers the homodyne acquisition in the signal channel, which is done by integrating the photocurrent acquired with a continuous local oscillator over time with the weight corresponding to the detection mode \cite{Neergaard-Nielsen2006,Lvovsky2009,Wakui2007}:
\be \label{IntCWmode}
\hat x_1(t_1)\equiv \int_{t_1-T'/2}^{t_1+T'/2} \psi(t-t_1) \hat x(t) dt.
\ee
Here $t_1$ is the idler detection event time, $ \psi(t-t_1)$ is the double-sided exponentially-decaying temporal mode in which the photon is prepared (see Sec.~\ref{cwHeraldedSec}) and $T'$ is the acquisition period of homodyne signals around the trigger, which must be  larger than the duration of that mode. In a typical experiment, the quadrature data $x(t)$ is first acquired with a transient digitizer and then the convolution (\ref{IntCWmode}) is calculated off-line.  An elegant variant of this calculation has been implemented in \cite{Ogawa2016,Asavanant2017}: it was performed in an analog, rather than digital, fashion, by means of a custom designed electronic circuit with the response function that is equal to $\psi(t)$. Then one can observe the quadrature $x_1(t_1)$ in real time on the oscilloscope screen, so the momentary quadrature statistics at the trigger event moment corresponds to the state of interest. 

\begin{figure}[!h]
\begin{center}
\includegraphics[width = 7cm]{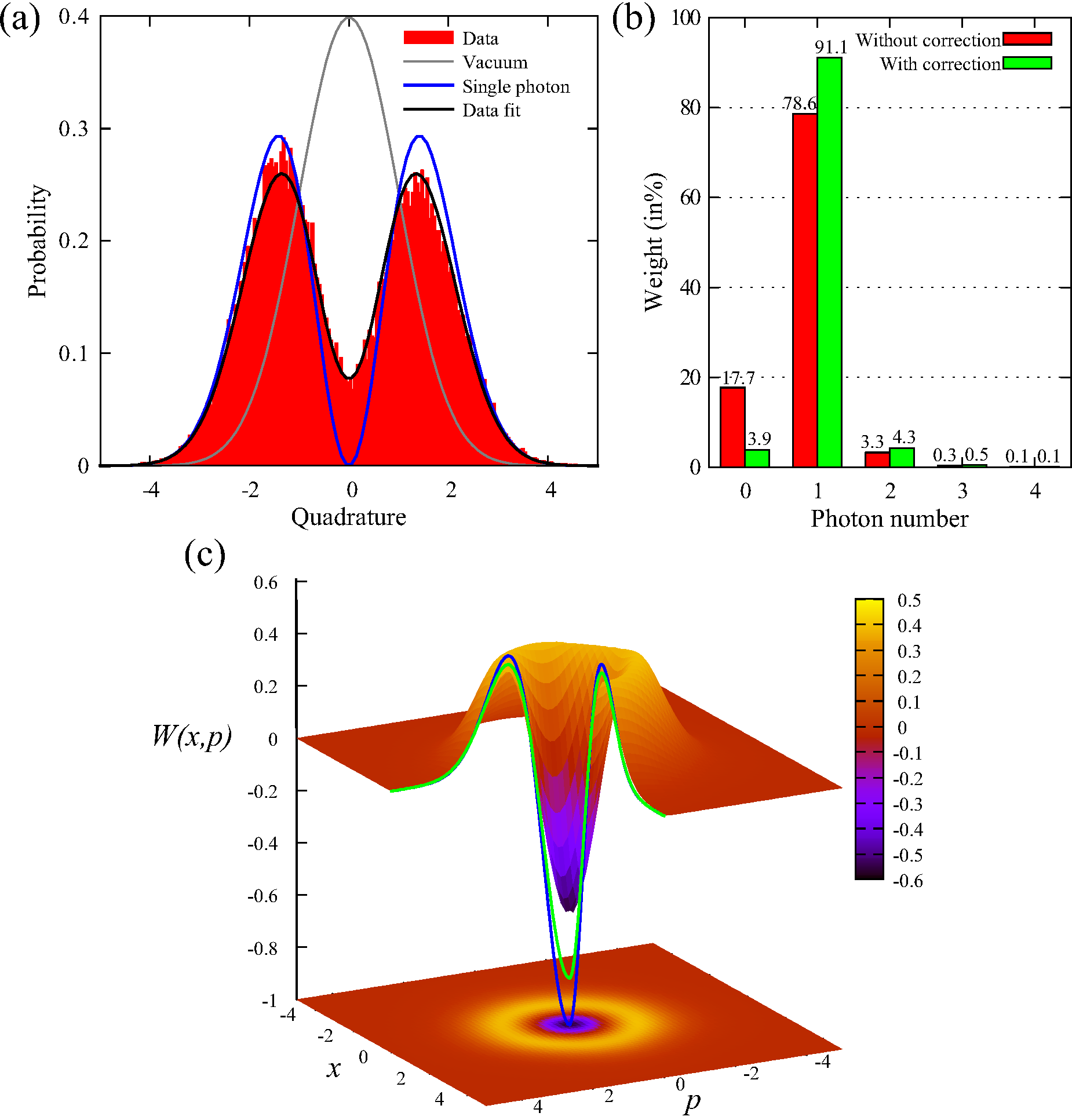}
\vskip 2.5mm
\includegraphics[width = 7cm]{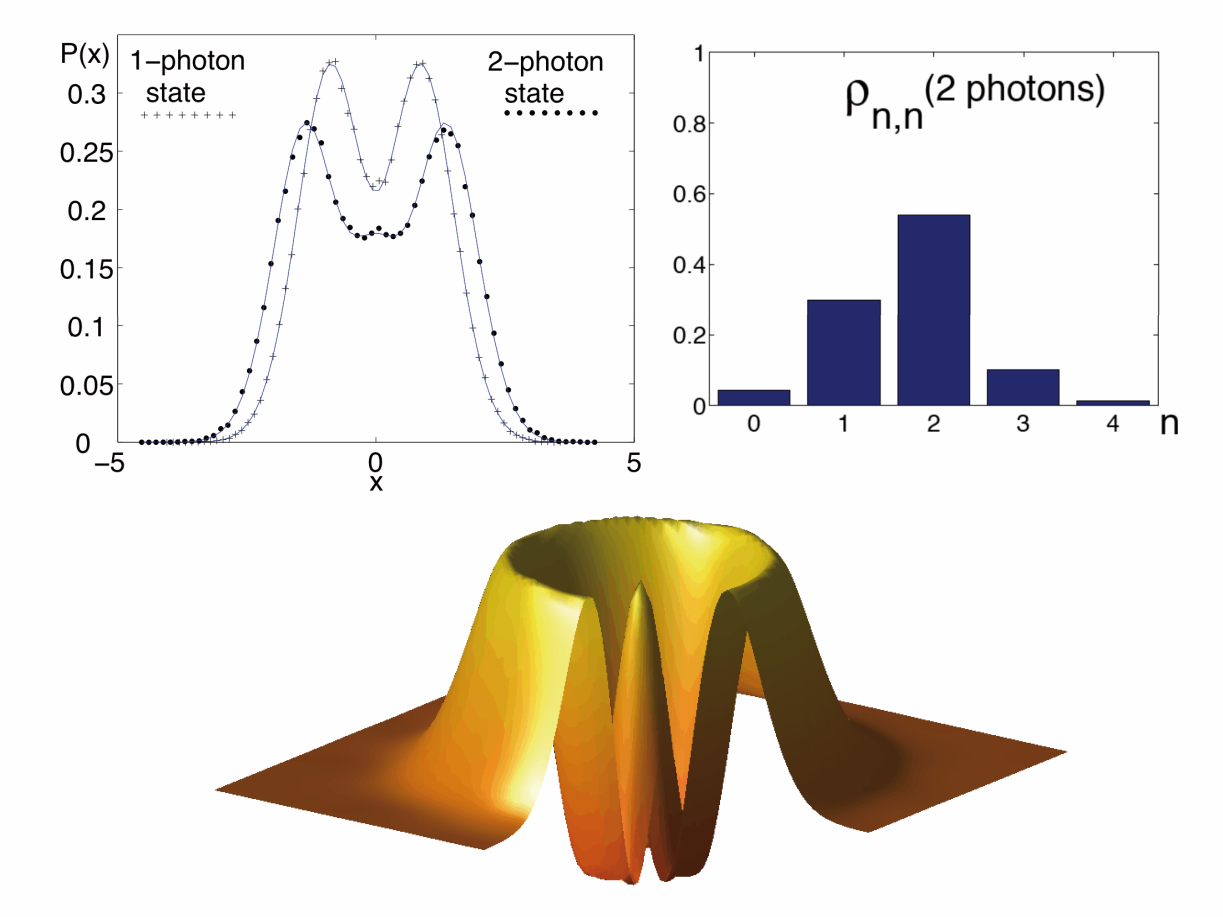}
\vskip 2.5mm
\includegraphics[width = 7cm]{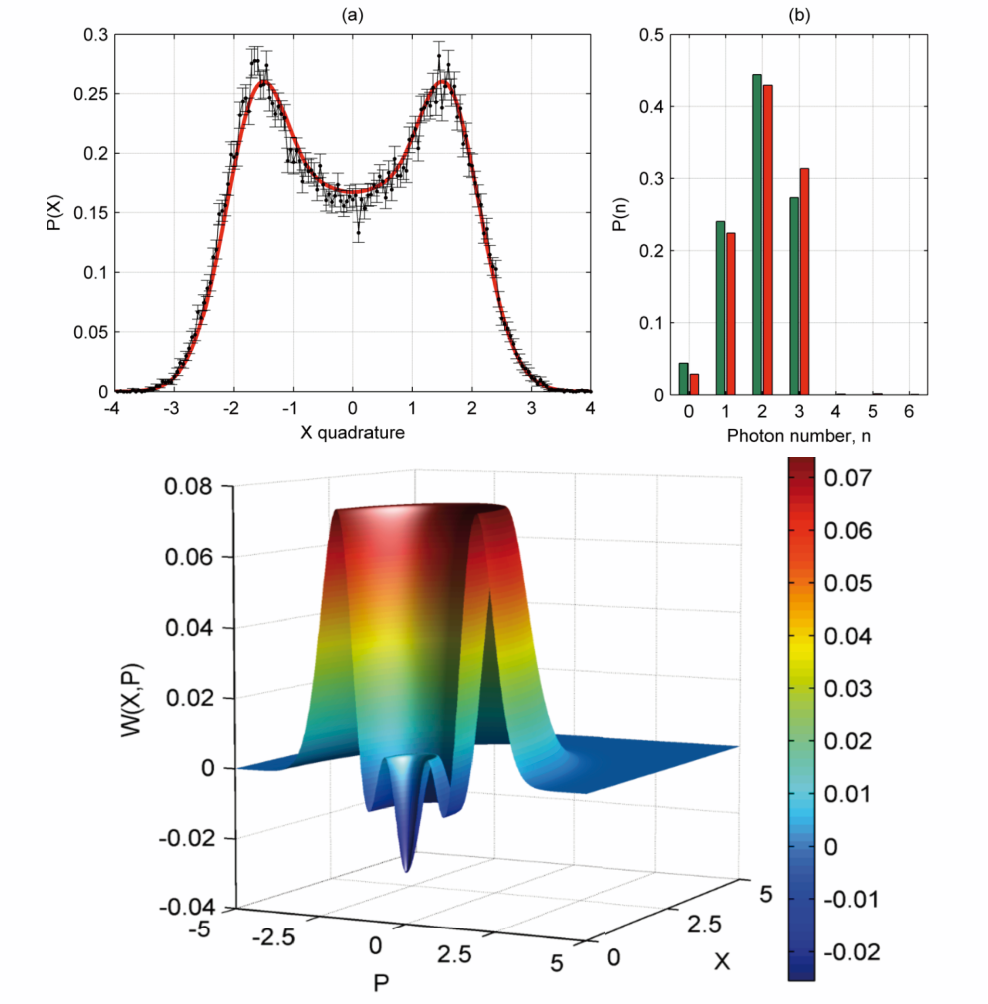}
\end{center}
\caption{ (Color online) 
Experimental quadrature distributions, reconstructed density matrix elements and Wigner function of Fock states. From top to bottom: single-photon \citep{Morin2012}, two-photon \citep{Ourjoumtsev2006b}, and three-photon states  \citep{Cooper2013b}} \label{Focktom}
\end{figure}

The conditional scheme in the parametric process can produce higher number states, even though the success rate decreases exponentially with the photon number.
Until now, up to  three-photon Fock  states have been generated and analyzed by tomographic techniques, as
shown on Fig.~\ref{Focktom}. 
The two-photon state has been generated by conditioning on coincident two-photon detections in the idler channel, both with the pulsed non-degenerate parametric amplifier scheme  \citep{Ourjoumtsev2006b} and with the cw approach \citep{Zavatta2008}.
Three-photon Fock states have been created similarly by conditioning on triple coincidences \citep{Cooper2013b}. The latter experiment took advantage of the pulsed SPDC configuration with a separable biphoton [Fig.~\ref{fig:filter}(c)], eliminating the need for spectral filtering in the trigger channel and dramatically increasing the heralding rate.

Besides the conditional preparation on spontaneous parametric down conversion process, which produces the purest number-states in a well-defined spatio-temporal mode (fidelity larger than $ 90 \% $), 
techniques based on atomic ensembles have been thoroughly investigated in order to produce single photons which are compatible with atomic memories in both their bandwidths and central frequencies \citep{Laurat2006,Simon2007}. Experimental efforts made these sources efficient enough to analyze them by homodyne tomography and obtain a complete characterization of the state in a well-defined spatio-temporal mode \citep{MacRae2012,Bimbard2014,Brannan2014}.

\section{Photon addition and subtraction in experiments}\label{AssSubSec}

In Sec.~\ref{NonGauss:PhotonCount}, we discussed the heralded implementation of the photon subtraction ($\hat a$) and addition ($\hat a^\dag$) operators. The experimental realization of these schemes opened many interesting scenarios for non-Gaussian states and photonic operation engineering, as well as for testing fundamental quantum mechanical effects, which we discuss next. 

While the photon subtraction operator  can be realized by means of a linear optical ``tapping" scheme [Fig.~\ref{FigAlexei:PrincipePhotSubtractAdd}(a)], the realization of the photon addition operator is more complicated and involves SPDC [Fig.~\ref{FigAlexei:PrincipePhotSubtractAdd}(b)]. The first  experiment on photon addition to a non-vacuum state (specifically, the coherent state) was made by \textcite{Zavatta2004b}. In this experiment, the ability of this operation to convert a classical state into a non-classical one  is evidenced by  the  transformation of the Wigner function, which loses its Gaussian shape and acquires negative values.

Later, single-photon added thermal states (SPATs) were generated and tomographically reconstructed. SPATs possess non-classical features such as negative Wigner functions, but these features become harder to detect for input states with higher mean photon numbers. SPATs have been therefore used as a benchmark for testing 
different  nonclassicality criteria which can be applied to experimental data in order to assess the quantum character of the generated states \citep{Thadd2007}. 
SPATs are also interesting as non-classical states whose P function may be negative but regular, which permits direct experimental reconstruction of that function \citep{Kiesel2008}.


Combined in the  same experimental setup, sequences of the photon addition and subtraction operators permit implementing complex quantum state engineering \citep{Jeong2014,Morin2014} as well as useful protocols \citep{Zavatta2011,Coelho2014}. By conditioning on coincident photon detections in various heralding channels, it is possible to implement an arbitrary function of the two operators \citep{Fiurasek2009}. This can be useful, in particular, to realize noiseless amplification (Sec.~\ref{section:NLA}) or simulate Kerr nonlinearities (Sec.~\ref{QCompSec}).  However, the increasing number of implemented operations has to be paid for by a decreased success rate of the whole process.

\begin{figure}[tb]
	\begin{center}
		\includegraphics[width=\columnwidth]{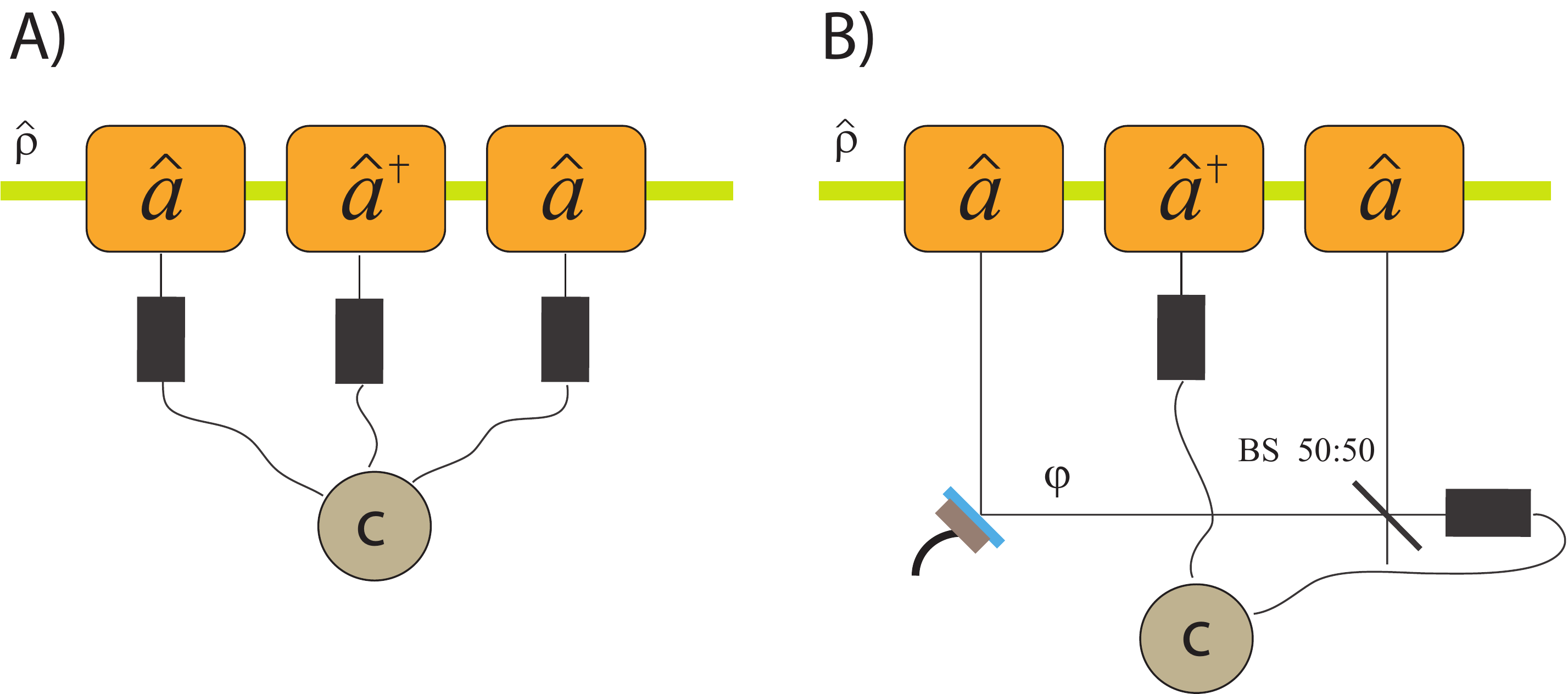}
	\end{center}
	\caption{(Color online)  a) Scheme for implementing sequences of single photon subtraction and addition operators. By observing a  coincidence between the first two stages the $\hat{a}^{\dag}\hat{a}$ operation is realized, while a coincidence of the last two stages implements the $\hat{a}\hat{a}^{\dag}$ operator.  b) Superposition of the two sequences can be achieved by indistinguishably observing an event from one of the two photon-subtraction operations.} \label{addsubsc}
\end{figure}

\begin{figure}[tb]
\begin{center}
\includegraphics[width=0.8\columnwidth]{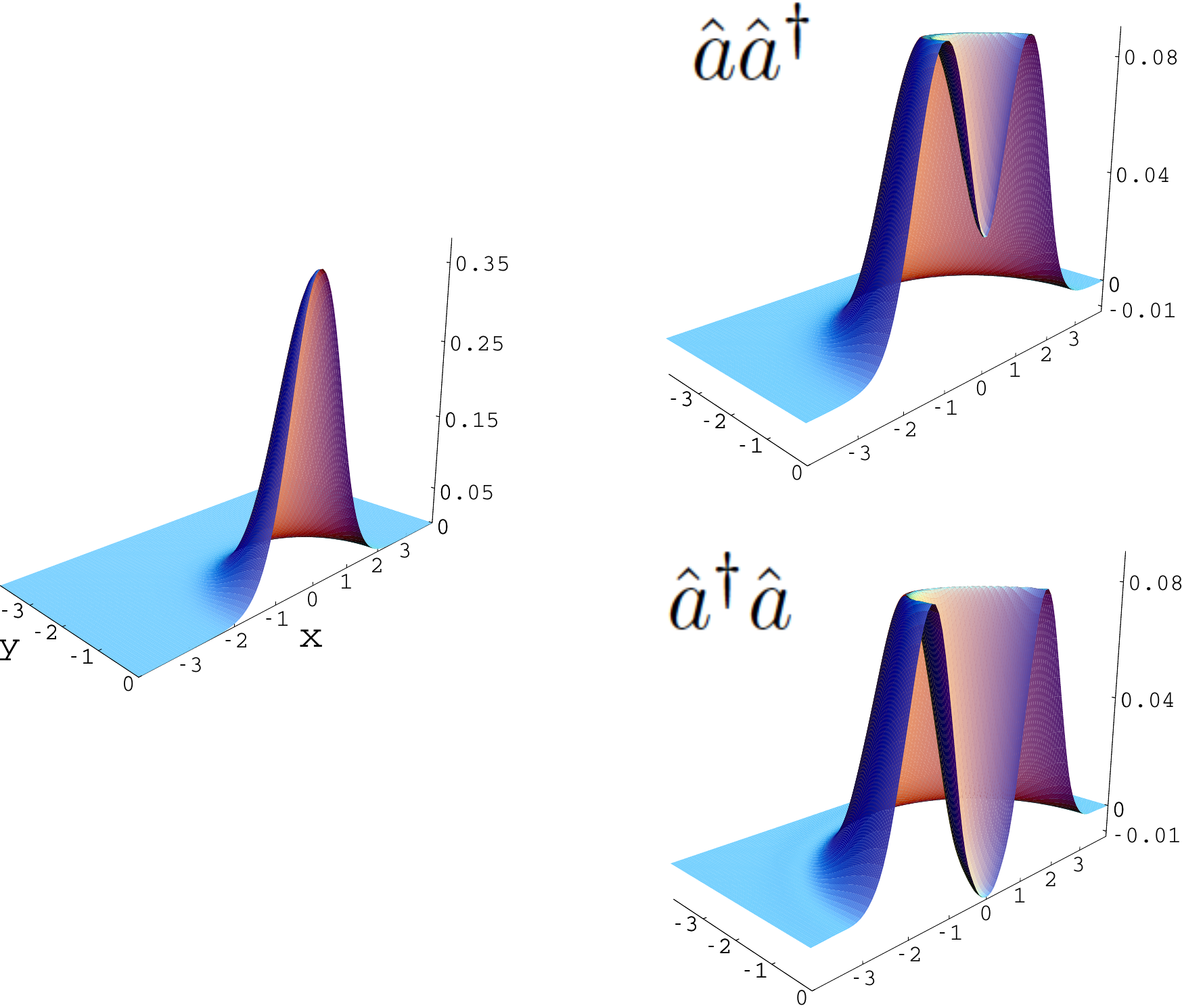}
\end{center}
\caption{(Color online) Left: Wigner function of a thermal state. Right: the two Wigner functions after the $\hat{a}\hat{a}^{\dag}$ and the $\hat{a}^{\dag}\hat{a}$ sequences have been applied to that state, showing that the two operations do not commute \citep{Parigi2007}.} \label{comm}
\end{figure}

In  \textcite{Parigi2007}, the two sequences $\hat{a}\hat{a}^{\dag}$ and $\hat{a}^{\dag}\hat{a}$ were applied to thermal states in order to test the bosonic commutation rules. To that end, three different stages have been concatenated, by placing two single-photon subtraction setups  (each composed of a half-wave plate and a polarizing beam splitter, with an APD in the reflected channel) before and after photon-addition arrangements (composed of the parametric crystal with an APD in the trigger channel), as shown on Fig.~\ref{addsubsc}(a). By conditioning on a coincidence click from the photon addition state and one of the photon subtraction stages, it is possible to implement the two operators in either sequence. The double-click conditioned state is then analyzed by homodyne detection. The evident difference (see Fig.~\ref{comm}) between the reconstructed Wigner functions of the states resulting from the two sequences applied on a thermal state shows that $\hat{a}\hat{a}^{\dag} \neq \hat{a}^{\dag}\hat{a} $ \citep{Parigi2007}.

A complete quantitative test of the commutation relation   $[\hat{a},\hat{a}^{\dag}]=\hat{\mathbb{I}}$, was performed using the scheme in Fig.~\ref{addsubsc}(b),  where the heralding fields of the two photon-subtraction stages interfere on a  $50:50$ BS \citep{Zavatta2009}. When the photon detector after the BS clicks, it is not possible to know
which photon-subtraction has been performed. The double event from this detector and that in the addition stage  heralds the  superposition of operators $\hat{a}\hat{a}^{\dag}- \textit{e}^{i \varphi} \hat{a}^{\dag}\hat{a}$, where $\varphi$ is the relative phase of the two interfering subtracted photons that can be controlled with a piezo-actuated mirror.

\begin{figure}[tb]
\begin{center}
\includegraphics[width=\columnwidth]{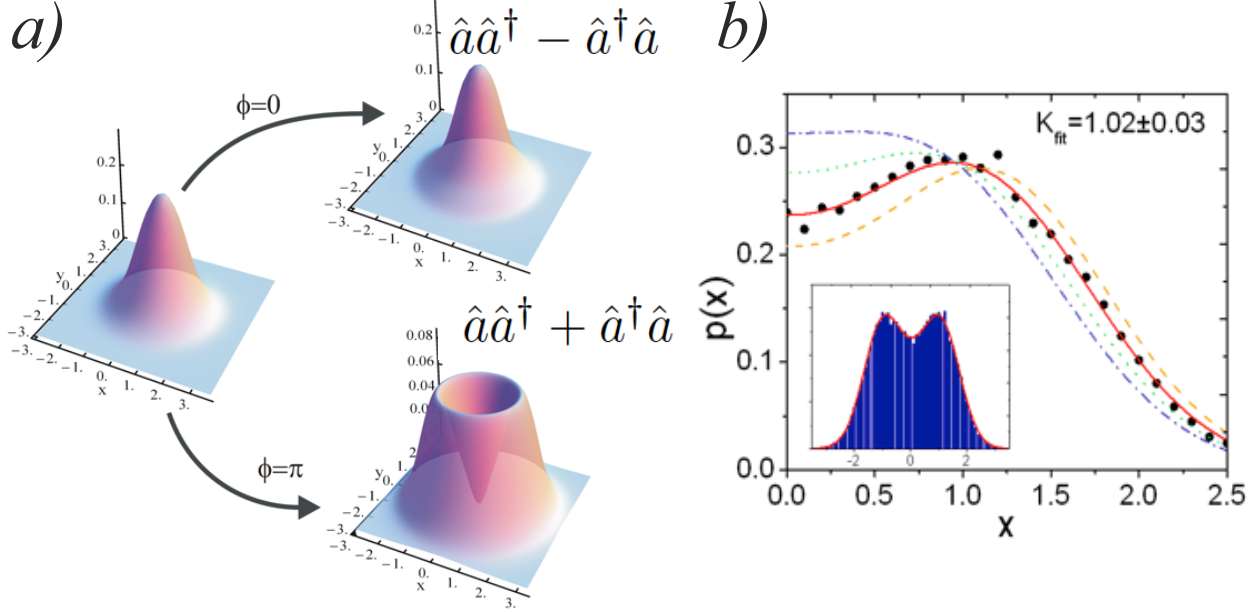}
\end{center}
\caption{(Color online) a) Wigner function of an initial thermal state and the resulting states after applying the commutator and the anti-commutator sequences of creation and annihilation operators. b) histogram of the raw quadrature data (solid dots) for the anticommutator operation output state with theoretical curves for  different values of the commutator ($K= 0$: dashed orange; $K=2$: dotted green; $K=
3$: dash-dotted blue). The solid red curve is the result of the best
fit to the experimental data \citep{Zavatta2009}.} \label{comm-quant}
\end{figure}
 
In particular, by setting $\varphi=0$ or $\varphi=\pi$, one can implement the commutator or the anticommutator of the creation and annihilation operators, respectively. The application of the commutator operator to a thermal state yielded a state that is largely identical to the input [Fig.~\ref{comm-quant}(a), top right], from which \textcite{Zavatta2009} concluded that this commutator is proportional to the identity: $[\hat{a},\hat{a}^{\dag}]= K\hat{\mathbb{I}}$. Importantly, because of the probabilistic nature of the implementation of the non-unitary operators $\hat a$ and $\hat a^\dag$, the above measurement does not establish the value of the proportionality constant $K$. This value can be determined by observing that the anticommutator of the creation and annihilation operators must be equal to $2\hat{a}^{\dag}\hat{a}+  K$ --- so that, when applied to the thermal state, the resulting state will depend on $K$. \textcite{Zavatta2009} subjected this state to homodyne tomography and found the measurement statistics to be consistent with  $K=1$ [Fig.~\ref{comm-quant}(a, bottom right) and (b)],  thus quantitatively demonstrating the bosonic commutation relation.

\section{Optical ``Schr\"odinger's cat" experiments} 
\label{Cats}

\subsection{Photon subtraction from squeezed vacuum}\label{PhotonSubExpSec}
\label{Cats:1PhotSubtract}
In Sec.~\ref{ScCatTheoSec}, we discusssed how the ``Schr\"odinger's cat'' states (coherent superpositions of coherent states) can be produced by means of photon subtraction from squeezed vacuum. The first attempt to implement this protocol was made in the pulsed regime by \textcite{Wenger2004a}. The  Wigner function of the prepared state was clearly non-Gaussian but remained positive due to experimental imperfections. Improving the experimental apparatus allowed the same team to create ``Schr\"odinger's kittens" with $|\alpha|^2\approx 0.8$ and to observe negative values on reconstructed Wigner functions  without correcting for the homodyne detection losses \citep{Ourjoumtsev2006a}. In the pulsed domain, this protocol has been implemented by several other groups \citep{Gerrits2010,Namekata2010,Tipsmark2011}.

In parallel with pulsed experiments, ``kitten'' states have been generated in the cw regime, where squeezing can be generated with high purity, 
and detected with high efficiency. 
After the first demonstration by \textcite{Neergaard-Nielsen2006}, significant experimental improvements allowed \textcite{Wakui2007} to produce extremely pure photon subtracted squeezed states with strongly negative Wigner functions. This result was further improved by \citep{Asavanant2017}, where nearly perfect squeezed single photons have been demonstrated.

While being relatively simple to generate, ``kitten'' states are very non-classical \citep{Kim2005,Jezek2012}: like single photons they are antibunched and present negative Wigner functions, but at the same time they are phase-dependent and may even feature quadrature squeezing. Therefore, they are an excellent benchmark for testing quantum devices such as quantum memories or quantum teleportation systems, which must operate with high coherence and low loss levels in order to preserve all of these non-classical features. 

An interesting twist to the technique of cat state preparation in the cw regime was realized by \textcite{Serikawa2018b}.  Here the two-mode squeezing was first produced at two sidebands 500 MHz above and below the carrier (LO) frequency. Both these modes were then subjected to electro-optical modulation at the frequency of 500 MHz, resulting in a part of the optical energy from both sidebands transfered to the carrier mode. This mode was then separated from the sidebands by a set of cavities and subjected to photon detection. A click heralded the preparation of the cat state in the mode that is the sum of the two sideband modes. This cat was reconstructed using a homodyne detector that was custom designed to have a gain peak at the 500 MHz sideband frequency.

%
%
%
%

As discussed in Sec.~\ref{ScCatTheoSec}, subtracting photons from the cat state increases its amplitude $|\alpha|$. Hence the task of producing larger-amplitude cat states can be solved by subtracting multiple photons from squeezed vacuum
\citep{Dakna1997}. In the pulsed scheme, this was done  by
\textcite{Gerrits2010}, where
two and three photons were simultaneously subtracted
from 140-fs pulsed squeezed vacuum pulses
at 861.8-nm wavelength.
A superconducting transition edge sensor (TES)
was used as a photon number resolving counter.
Three-photon subtraction could generate a non-Gaussian state
which had a fidelity $F\sim0.59$ with an ideal cat state of
$|\alpha|\sim1.76$.
Similar experiments with TES were done at the telecom wavelength \citep{Namekata2010}.

\begin{figure}[tb]
	\centering {\includegraphics[width=0.8\linewidth]{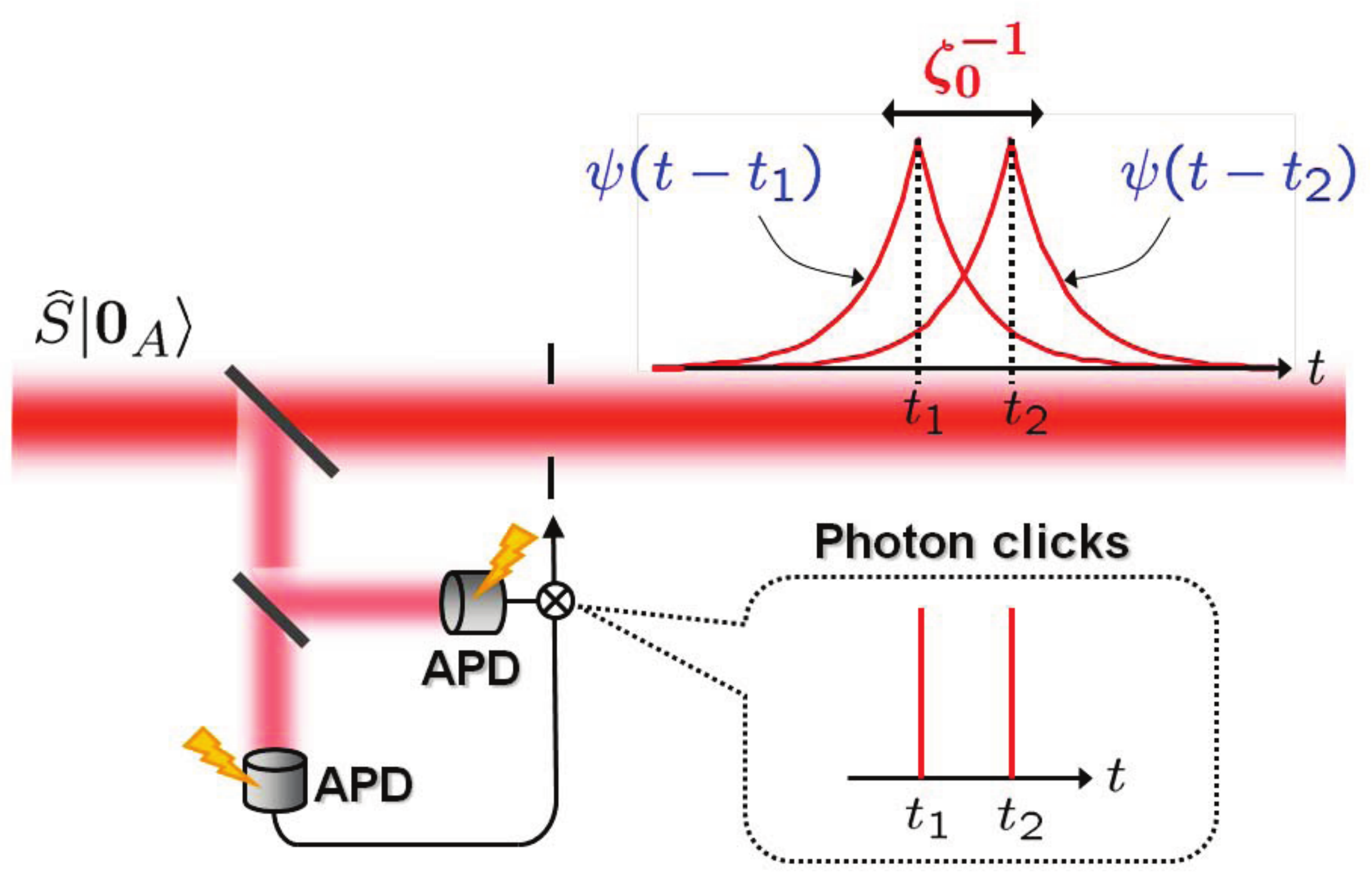}
		\caption{(Color online)
			Two-photon subtraction from the cw squeezed state.
			Small fraction of squeezed beam is tapped,
			and directed to two APDs via a 50:50 BS.}
		\label{fig.2-ph_subtraction_from_cw_sq}}
\end{figure}

In the cw scheme,
multiple photons can be subtracted at different times
within the coherence time of the squeezed light,
which provides rich physics in the time domain
\citep{Sasaki2008b,Takeoka2008,Takahashi2008}.
Suppose that one photon is subtracted at $t=t_1$ and another one
at $t=t_2$, as in
Fig.\ref{fig.2-ph_subtraction_from_cw_sq}.
Depending on the relation between
the time-separation $\Delta\equiv|t_2-t_1|$
and the coherence time scale $\zeta_0^{-1}$,
three regions can be identified.

(i) $\Delta \ll \zeta_0^{-1}$
where
the two-photon subtraction takes place in a single mode,
producing a small even-number cat state
as originally proposed by
\cite{Dakna1997}.

(ii) $\Delta\sim\zeta_0^{-1}$ where the two packets
overlap with each other, and nontrivial interference occurs.

(iii) $\Delta\gg\zeta_0^{-1}$ where
two small odd-number cat states are generated in separated packets
$\psi(t-t_1)$ and $\psi(t-t_2)$.
Each cat state is the squeezed single-photon state
as described in the previous subsection.

\begin{figure}[tb]
	\centering {\includegraphics[width=0.7\linewidth]{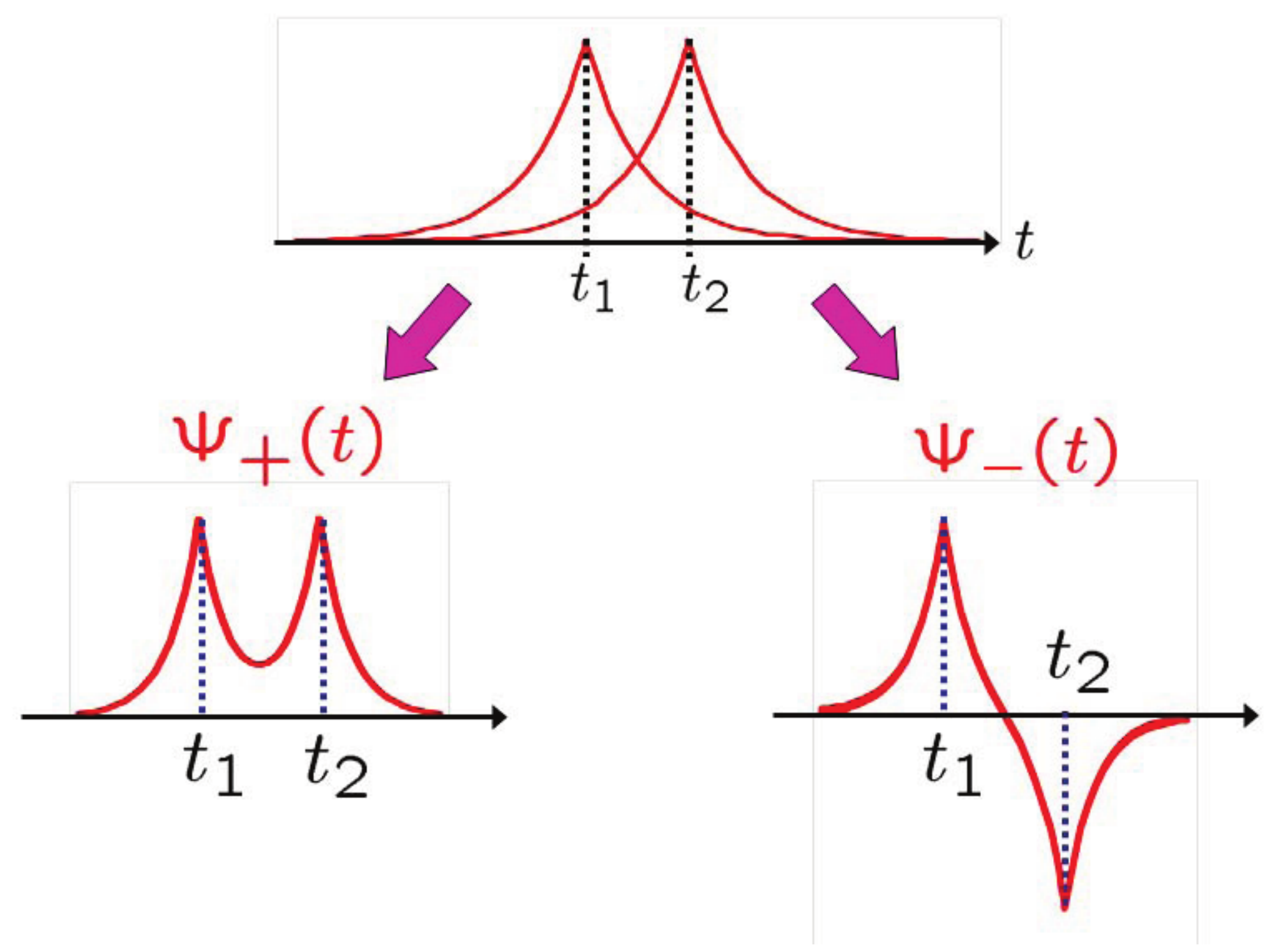}
		\caption{(Color online) Temporal shape of the biased modes.
			$\Psi_+(t)$ is symmetric,
			while  $\Psi_-(t)$ is antisymmetric.}
		\label{fig.PsiP_PsiN_psi1_psi2}}
\end{figure}

In the intermediate region (ii), there is a crossover
between (i) and (iii),
and both even- and odd-number cat states exist,
being entangled over the two modes.
To study this entanglement, \cite{Sasaki2008b}
 introduced the ``biased" orthonormal set consisting of the symmetric and antisymmetric modes, whose temporal shape is given by (Fig.~\ref{fig.PsiP_PsiN_psi1_psi2})
\beq\label{biased}
\Psi_\pm(t)=\frac{\psi(t-t_2)\pm\psi(t-t_1)}{\sqrt{ 2\left(		1\pm  I_\Delta \right) } },
\eeq
where
\beq
I_\Delta\equiv\int_{-\infty}^{\infty} \psi(t-t_1)\psi(t-t_2)\dd t\approx ( 1+\zeta_0\Delta )e^{-\zeta_0\Delta}.
\eeq

\begin{figure*}
	\centering {\includegraphics[width=0.8\linewidth]{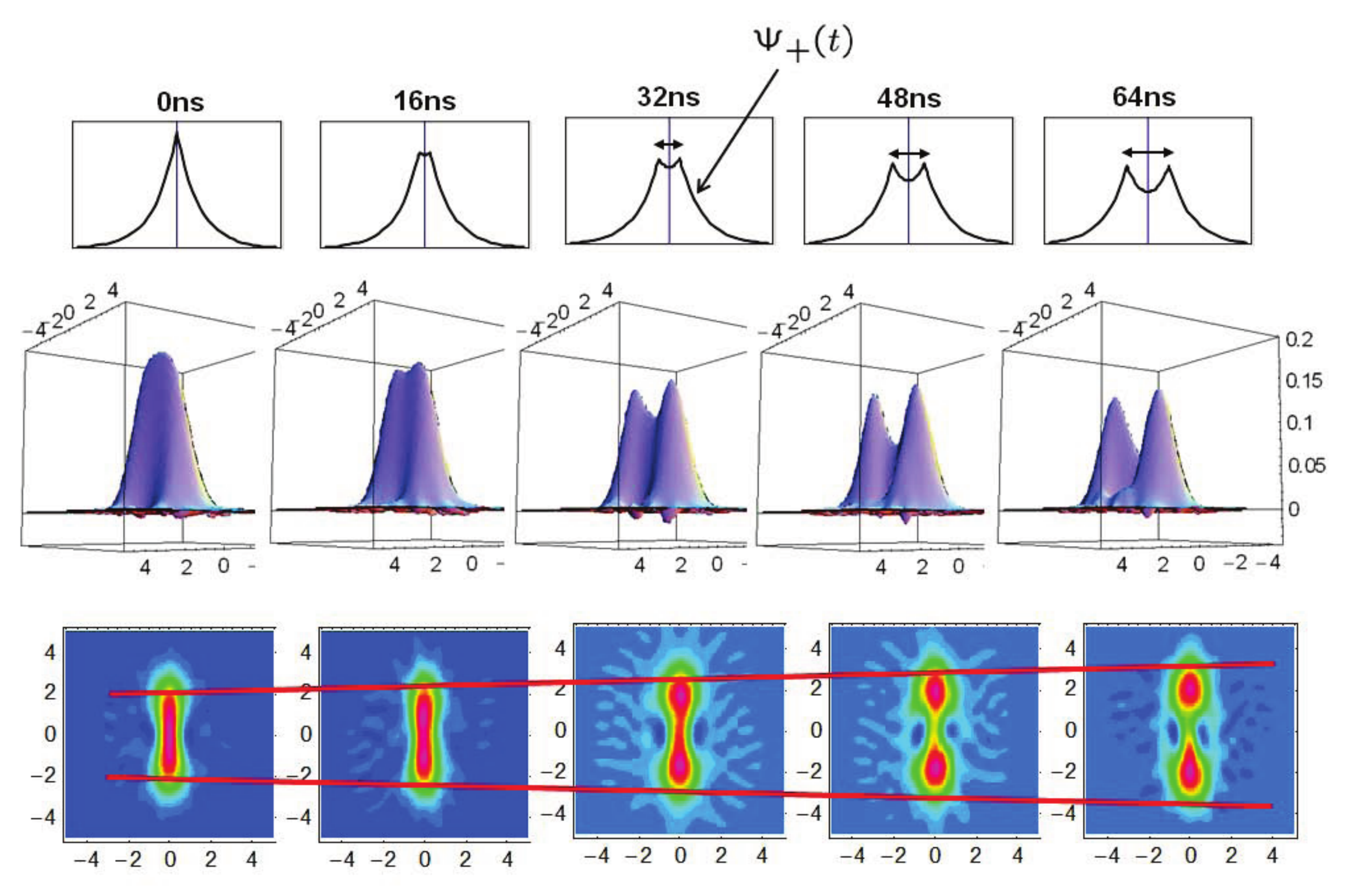}
		\caption{(color online) 
			Amplitude enhancement of a cat state
			in the symmetric mode $\hat A_+$ by controlling the time separation $\Delta$.
			The top panels show the temporal shape of $\Psi_+(t)$.
			The middle panels are the experimental Wigner functions.
			The bottom panels are their contour plots.
			From left to right the time separation is increased, and 
			the amplitude of the generated state gradually increases accordingly.
			For $\Delta\,=\,$32\,ns,
			the negative dips are observed at the sides of the origin 
			\citep{Takahashi2008}. The OPA linewidth is $\zeta_0=2\pi\times$ 4.5 MHz.
		}
		\label{fig.Cat_breeding}}
\end{figure*}
 $\Psi_+(t)$ is the mode that extracts the features emerging
due to the overlap of the packets
$\psi(t-t_1)$ and $\psi(t-t_2)$,
while  $\Psi_-(t)$ is the mode that extracts
the differential features of these localized packets.
The field operators in the biased modes are defined by
\beq
\label{signal mode field}
\hat A_\pm
\equiv
\int_{-\infty}^{\infty}  \hat a(t) \Psi_\pm(t) \dd t.
\eeq

By replacing 
the order of the squeezing and the annihilation operators  
akin to 
Eq. (\ref{ASA0A}), 
the two-photon subtracted state can be represented as \citep{Sasaki2008b}
\beqa
\label{Two photon subtracted state in biased modes}
&&
\hat A(t_2) \hat A(t_1)\hat S\ket{\mathbf{0}}
\sim
\hat S
\Biggl[
\frac{1+ I_\Delta}{\sqrt2}\ket{2_+,0_-,\mathbf{0}_{\tilde A}}
\\\nonumber
&&\quad-
\frac{1- I_\Delta}{\sqrt2}\ket{0_+,2_-,\mathbf{0}_{\tilde A}}+
\frac{\zeta_0}{\epsilon} e^{-\zeta_0 \Delta}
\ket{0_+,0_-,\mathbf{0}_{\tilde A}}
\Biggr],
\eeqa
where $\ket{\mathbf{0}_{\tilde A}}$, hereafter omitted for brevity, is the vacuum state of all modes that are orthogonal to both $\hat A_+$ and $\hat A_-$, and the squeezing operator $\hat S$ applies to all three states inside the ket. Thus the output state is generally a squeezed entangled state
of the 2-photon and vacuum states over modes $\Psi_\pm(t)$ \citep{Sasaki2008b}. 

In the region $\zeta_0\Delta\ll1$, the wavepacket overlap $I_\Delta$ approaches unity, so the second term in Eq.~(\ref{Two photon subtracted state in biased modes}) becomes negligible and the state becomes separable:
\beq\label{Two photon subtracted state in biased modes 3}
\hat A(t_2) \hat A(t_1)\hat S \ket{\mathbf{0}_{A}}\rightarrow
\hat S\bigg[\biggl( \sqrt{2}\ket{2_+}+ \frac{\zeta_0}{\epsilon}e^{-\zeta_0 \Delta}\ket{0_+} \biggr)\otimes \; \ket{0_-}\bigg]
\eeq
The mode $\hat A_+$ is in a squeezed superposition of the zero and two-photon states, which approximates the even cat state (Sec.~\ref{ScCatTheoSec}) while mode $\hat A_-$ is approximately in the squeezed vacuum state. 

The amplitude of the cat in the symmetric mode can be controlled by the OPA pumping rate, which determines the parameter $\epsilon$ and the degree of squeezing in the OPA output. However, one can also increase the quantity $\zeta_0\Delta$, which will reduce the vacuum term in the cat state. This effect has been  experimentally demonstrated by
\citep{Takahashi2008} (Fig.~\ref{fig.Cat_breeding}) and in \cite{Huang2016}.

As the separation $\Delta$ increases, the second term in Eq.~(\ref{Two photon subtracted state in biased modes}) becomes significant and the state of the two biased modes is no longer separable. In the regime $\zeta_0\Delta\gg1$, we have $I_\Delta\ll 1$, and state takes the form 
\beq
\label{Two photon subtracted state in biased modes 4}
\hat A(t_2) \hat A(t_1)\hat S\ket{\mathbf{0}}\rightarrow
\frac{1}{\sqrt{2}}\hat S\biggl(\ket{2_+,0_-}-\ket{0_+,2_-}\biggr).
\eeq
This means that, when the symmetric mode is analyzed separately, it becomes a mixture of the form $\hat S(\ketbra{2_+}{2_+}+\ketbra{0_+}{0_+})\hat S^\dag$. 

Equation (\ref{Two photon subtracted state in biased modes 4}) is easy to interpret. When the photon detection events are well separated, two independent squeezed single photons are produced in the wavepackets $\psi(t-t_1)$ and  $\psi(t-t_2)$, so we can write 
\beq\label{Two photon subtracted state in unbiased modes}
\hat A(t_2) \hat A(t_1)  \hat S \ket{\mathbf{0}}\sim\hat S\ket{1_1,1_2}
\eeq
The biased modes are related to these wavepackets by the  beam-splitter-like transformation (\ref{biased}), resulting in the Hong-Ou-Mandel effect \cite{Hong1987} visible in Eq.~(\ref{Two photon subtracted state in biased modes 4}).

\subsection{Cats as even or odd photon superpositions}\label{CatsEOSec}
The Hong-Ou-Mandel effect, in combination with homodyne projective measurements (Sec.~\ref{NonGauss:HomodyneProject}), has been used for the preparation of cat states in a direct fashion by \textcite{Etesse2015} as follows. Two heralded photons have been overlapped on a 50:50 beam splitter, generating the state $\ket\Psi=\frac{1}{\sqrt{2}}(\ket{2,0}-\ket{0,2})$. Subsequently a homodyne measurement of the $x$ quadrature has been applied in one of the beam splitter outputs, preparing the superposition 
\begin{equation}\label{Etesse}\braket x\Psi=\frac{1}{\sqrt{2}}(\braket x2 \ket 0-\braket x0 \ket 2) 
\end{equation}of the vacuum and two-photon states in the other channel. The inner products $\braket x0$ and $\braket x2$ are simply wavefunctions of Fock states, given by $H_n[x]\exp[-x^2/2]$ where $H_n$ is the $n$-th Hermite polynomial \citep{Walls2008}. Depending on the value $x$ of the quadrature detected in the first channel, these wavefunctions take different values, thereby determining the coefficients of the superposition (\ref{Etesse}). In particular, postselecting on the observation of $x=0$ prepares the superposition $(\ket 0+\sqrt2\ket2)/\sqrt3$ which has a 99\% fidelity with an even cat state of amplitude $\alpha=1.63$ squeezed by $s=1.52$ along the quadrature $x$ \citep{Etesse2015}.

This idea was extended by \textcite{Ulanov2016} who prepared the state (\ref{Etesse}) by means of entanglement swapping. They initially used two OPAs to generate two weakly squeezed vacuum states in pairs of modes $(\hat A,\hat B)$ and $(\hat C, \hat D)$. They then overlapped modes $\hat B$ and $\hat C$ on a symmetric BS and selected for further analysis those events in which a single photon has been observed in both outputs of that BS. Because of the reversible character of the beam splitter transformation, these events projected modes $\hat B$ and $\hat C$ onto  the state (\ref{Etesse}). This, in turn, prepared  modes $\hat A$ and $\hat D$ in the same state thanks to the entangled nature of the two-mode squeezed state and the fact that this state carries the same number of photons in both of its channels as per Eq.~(\ref{EqStateEPR}). 

This approach is operational --- i.e. the state  $\frac{1}{\sqrt{2}}(\ket{2,0}-\ket{0,2})$ is produced in modes $\hat A$ and $\hat D$ with high efficiency --- even if the heralding photons in modes  $\hat B$ and $\hat C$ experience attenuation. Moreover, the technique can be extended to prepare a general ``N00N" state $\frac{1}{\sqrt{2}}(\ket{N,0}-\ket{0,N})$ with an arbitrary $N$. These states are useful for  quantum-enhanced phase metrology \citep{Dowling2008}, and the technique of  \textcite{Ulanov2016} makes it possible to prepare and use them in spite of the modes $\hat A$ and $\hat D$ being separated by a lossy medium. This method has been generalized to non-weak SPDC sources in a theoretical proposal for long-range distribution of multiphoton entanglement \cite{Mycroft2019}.

The above procedures approximate even low-amplitude cats as superpositions of $\ket 0$ and $\ket 2$. But, as mentioned previously, ``usual'' cat states consist of either only even or only odd Fock components. When the amplitude of the cat is not too high, one can identify two terms that dominate in the state's Fock decomposition, while other terms are significantly lower. Accordingly, any Fock state superposition of the form  $\beta\ket{n}+\gamma\ket{n-2}$, with $n$ not exceeding 3--5,  approximates the cat state reasonably well. In \citep{Huang2015}, it was proposed to generate this superposition by mixing two single-mode orthogonally squeezed states on an imbalanced BS and detecting $n$ photons in one of the output modes. The other mode is then projected on a state $\beta\ket{n}+\gamma\ket{n-2}$ where $\beta$ and $\gamma$ depend on the BS's reflectivity and the squeezing. In the experiment by \textcite{Huang2015}, the above protocol was implemented for $n=2$. Two-photon detection with highly efficient superconducting single photon detectors were applied to  the two-mode squeezed vacuum in a cw beam at 1064 nm, and the generated superposition states had the fidelity of 67\% to a coherent cat state with a size $|\alpha|^2=3$.

\subsection{Homodyne projective measurements on Fock states }

\begin{figure}[tb]
\begin{center}
\includegraphics[width=7cm]{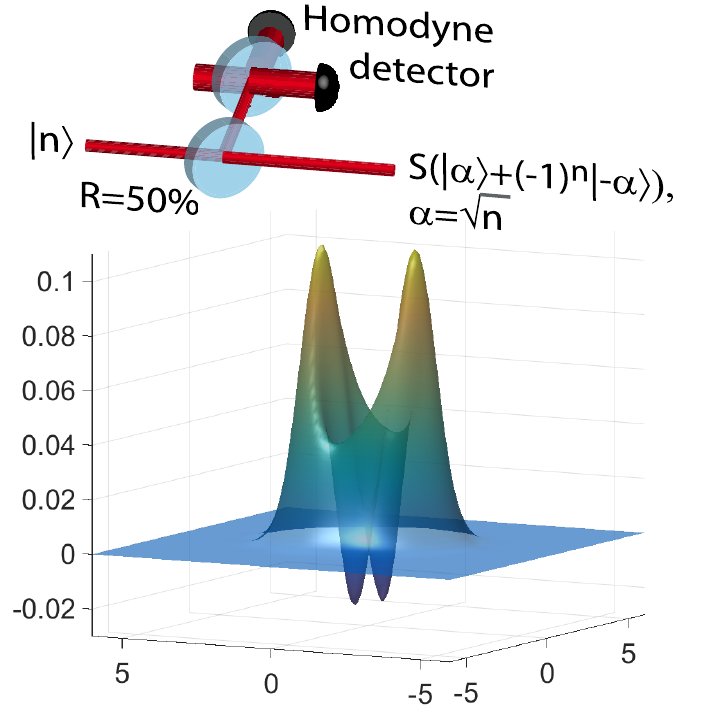}
\end{center}
\caption{(Color online) Generation of arbitrarily large ``Schr\"odinger cat'' states. (Top) A Fock state $\ket{n}$ containing $n$ photons is divided on a $50:50$ BS, and a homodyne measurement is performed in the reflected port. When the measurement's outcome $p$ satisfies $|p|<\epsilon\ll 1$ the transmitted beam is projected into a ``cat'' state squeezed by $3$ dB, with the same parity as $n$ and with a coherent state amplitude $\alpha=\sqrt{n}$. (Bottom) Experimentally reconstructed Wigner function for a state generated with $\ket{n=2}$ \citep{Ourjoumtsev2007a}. See text for details.} \label{FigAlexei:SqzCat}
\end{figure}

Although multiple photon subtractions from squeezed vacuum allow one to produce cat states with increasingly large sizes, the success rate decreases and the role of the imperfections increases exponentially with the number of subtracted photons. An alternative rather simple approach \citep{Ourjoumtsev2007a} consists in preparing Fock states $\ket{n}$, sending them through a 50:50 BS, and making a homodyne measurement of the momentum quadrature $p$ in the reflected mode (Fig.~\ref{FigAlexei:SqzCat}). When the result of the measurement is $0$, the remaining transmitted beam is projected on a superposition of only even or only odd photon numbers ranging from $0$ to $n$. The resulting state closely resembles a ``Schr\"odinger's cat'' $c(\ket{\alpha}+(-1)^n\ket{-\alpha})$ with $\alpha=\sqrt{n}$, squeezed by a factor of 2 along the $x$ quadrature. Since the initial Fock state is phase-invariant, the $x$ and $p$ quadratures are actually determined by the projective homodyne measurement: by definition, the $x$ quadrature of the cat state is the one orthogonal to the quadrature measured by the heralding homodyne detector.

\begin{figure}[tb]
	\centerline{\includegraphics[width=\columnwidth]{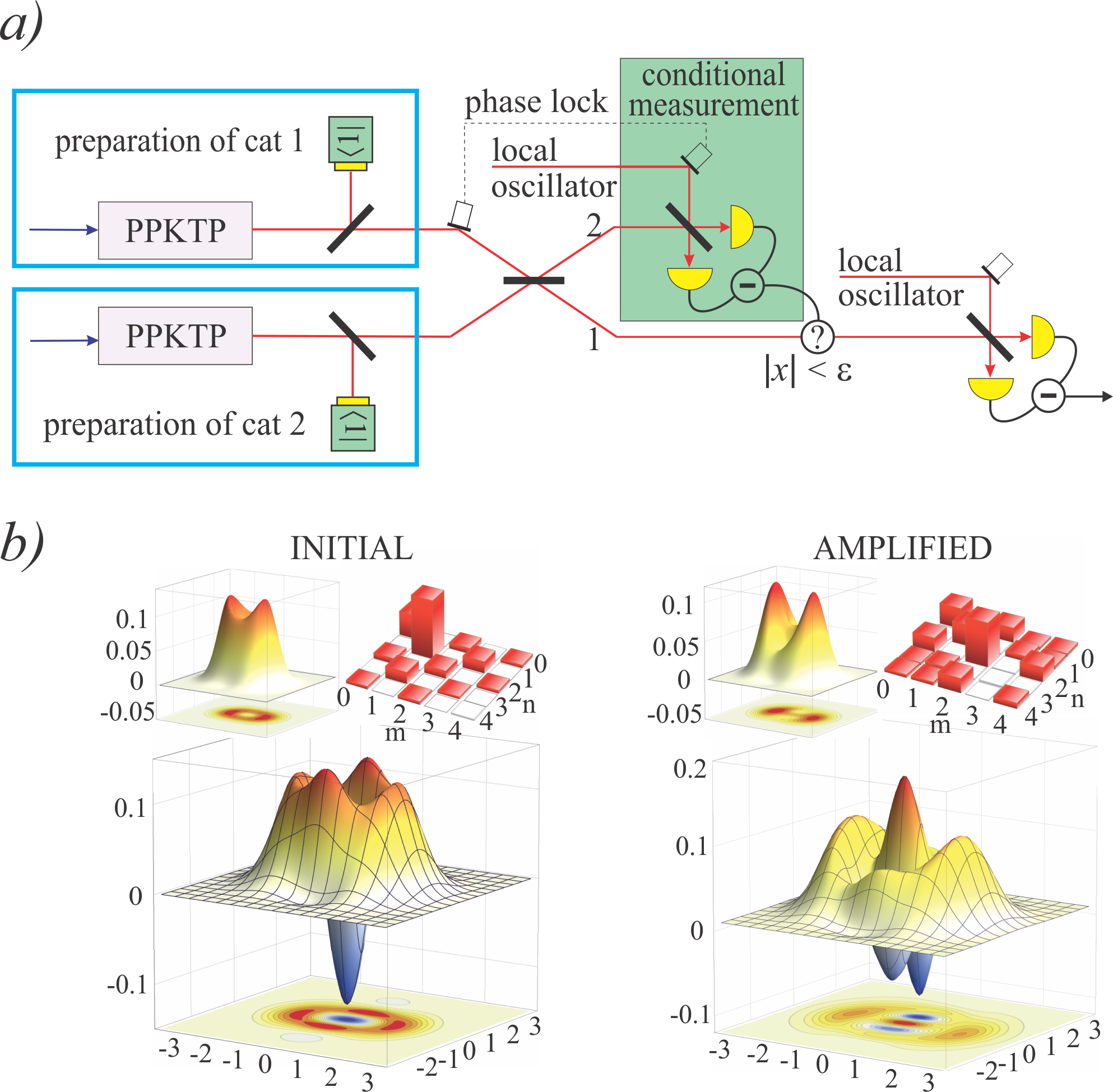}
	}
	\caption{The experiment on ``breeeding" Schr\"odinger cats \cite{Sychev2017a}. a) Scheme. Detecting the quadrature $x\approx0$ in mode 2 heralds an enlarged cat in mode 1. b) Experimentally reconstructed density matrices and Wigner functions of the initial (left) and amplified (right) cat states, corrected for the detection quantum efficiency of 62\%. The best-fit state is $\ket{\psi_{{\rm cat},-}[1.15]}$ for the initial cat and  $\ket{\psi_{{\rm cat},+}[1.18]}$ for the amplified one, squeezed by 1.74 dB (3.04 dB).}
	\label{BreedingFig}
\end{figure}

The intuition behind this technique is best seen in the quadrature basis. The unnormalized wavefunction of the initial Fock state, expressed using the momentum quadrature $p$, is $H_n[p]\exp[-p^2/2]$ where $H_n$ is the $n$-th Hermite polynomial. Mixed with a vacuum mode with the wavefunction $\exp[-p_0^2/2]$, where $p_0$ is the momentum quadrature of that mode on a 50:50 BS, the two-mode wavefunction becomes $H_n[(p-p_0)/\sqrt{2}]\exp[-(p^2+p_0^2)/2]$ according to Eq.~(\ref{EqAAdagBS}). A homodyne measurement's result $p_0=0$ leaves the transmitted beam in the state $H_n[p/\sqrt{2}]\exp[-p^2/2]$. Its expression as a function of the $x$ quadrature, obtained by a Fourier transform, is proportional to $x^n\exp[-x^2/2]$ and has double-peaked structure characteristic of a cat state. 

By construction, the state prepared in this manner cannot contain more than $n$ photons whereas the squeezed cat state has an infinite expansion in the Fock basis. However, neglecting experimental imperfections, the fidelity between the two states is $\mathcal{F}\approx 1-0.028/n$, already at the $99\%$ level for $n=2$ photons and increasing towards unity for larger $n$ \citep{Ourjoumtsev2007a}. In practice, a finite preparation success rate requires one to accept homodyne measurement outcomes within a narrow but finite window  $|p|\leq\epsilon\ll 1$. However, the fidelity decreases only quadratically with $\epsilon$ whereas the success rate increases linearly, and, even when the probabilistic generation of resource Fock states is taken into account, this protocol is very competitive compared to others in terms of success rates and state purities.

\subsection{``Breeding" larger cats from smaller ones}\label{BreedingSec}
Most of the methods listed above enable the preparation of Schr\"odinger cat states of relatively low amplitudes. While this may suffice for  many applications --- for example, coherent-state quantum computing, see Sec.~\ref{QCompSec}, it is also important to develop methods to prepare cat states of unlimited size. This would, in particular, help answering a fundamental question: at what degree of macroscopicity, if any, does the world stop being quantum? 

A practically viable technique in this context was proposed by \textcite{Lund2004}. The idea is to ``breed" larger size cat states from smaller ones my means of linear optics and conditional measurements.

Suppose a pair of identical cat states $\ket{\psi_{{\rm cat},\pm}[\alpha]}=\ket\alpha\pm\ket{-\alpha}$ is made to interfere on a symmetric BS [Fig.~\ref{BreedingFig}(a)].  The BS output state is then
\begin{equation}
\label{eq2}
\ket{\psi_{{\rm cat},+}[\sqrt{2}\alpha]}_1\ket{0}_2 \pm \ket{0}_1\ket{\psi_{{\rm cat},+}[\sqrt{2}\alpha]}_2.
\end{equation}
This result is readily obtained under the assumption that two identical coherent state incident on the BS  experience constructive interference in the first output channel and destructive in the second: $\ket\alpha\ket\alpha\to\ket{\sqrt2\alpha}\ket 0$, whereas two coherent states of opposite amplitudes exhibit opposite behavior: $\ket\alpha\ket{-\alpha}\to\ket 0\ket{\sqrt2\alpha}$. 

If we now perform a measurement on mode 2 of the state \eqref{eq2} to distinguish the states $\ket 0$ and $\ket{\psi_{{\rm cat},+}[\sqrt{2}\alpha]}$, mode 1 will collapse onto either $\ket{\psi_{{\rm cat},+}[\sqrt{2}\alpha]}$ or $\ket 0$, respectively. The positive cat state of  amplitude $\sqrt{2}\alpha$ can thus be conditionally prepared in mode 1. 

Experimentally, conditioning on the detection of the vacuum in mode 2 is challenging because a photon detector with imperfect efficiency may fail to click even if photons are present at the input. A more practical scheme was proposed by \textcite{Laghaout2013}, in which the required conditioning is realized by homodyne measurement of the position quadrature in mode 2 and looking for the result that is close to zero: $|x|<\epsilon\ll1$. For $\alpha\gg 1$, the probability of observing $x=0$ is much higher in the vacuum state than in the state  $\ket{\psi_{{\rm cat},+}[\sqrt{2}\alpha]}$, as one can see by comparing the wavefunctions of these two states.

The experiment by \textcite{Etesse2015}, discussed above, can be viewed as a simple realization of this proposal, with the role of input cats played by heralded single photons. In the experiment by \textcite{Sychev2017a}, two squeezed cat states of amplitude $\alpha\approx 1.15$ have been prepared by photon subtraction from squeezed vacuum. After the ``breeding" operation, the amplitude has increased to $\approx 1.85$ (Fig.~\ref{BreedingFig}). Remarkably, this operation did not significantly affect the cat's fidelity. However, the non-ideality of the heralding homodyne detection results in additional squeezing of that state.

An interesting feature of the ``breeding" protocol is that its performance \emph{improves} with higher amplitudes. This is because higher amplitude cats contain lower amounts of the vacuum state and hence are more distinguishable from that state. If the homodyne heralding is used, the width of the acceptance window $\epsilon$ can be increased for higher $\alpha$ without sacrificing the state preparation fidelity. In the limit of high amplitudes, the probability of a successful heralding event tends to $\sim\frac12$, which allows the application of this protocol in an iterative fashion with a moderate exponential overhead.  

%
%
\subsection{Cat state qubit}\label{CatQubitSec}
%
%

The opposite-phase coherent states $\ket\alpha$ and $\ket{-\alpha}$ of sufficiently high amplitude can be used to encode a qubit:
\be\label{Logical encoding}
\ket{0}_L=\ket{\alpha},\quad \ket{1}_L=\ket{-\alpha}.
\ee
This encoding has many helpful properties for both quantum communication and quantum computation, as discussed below. \textcite{Neergaard-Nielsen2010} demonstrated a technique to generate such a qubit, based on the idea of 
\cite{Takeoka2007}. Starting with the cw squeezed vacuum state $\hat S(r) \ket{0}$ with the squeezing parameter $r = 0.38$, which approximates the state $\ket{\psi_{{\rm cat},+}[\sqrt r]}$, they applied a superposition of the photon subtraction and identity operators using the scheme shown in  Fig.~\ref{fig:AddSubGen}(a). Because the photon subtraction converts a positive cat into a negative one, the superposition of positive and negative cats (more precisely, of the cw squeezed vacuum and squeezed single photon) is generated in the output, which can be represented on a Bloch sphere as:
\beqa\label{ket_rho(theta,varphi)}
\ket{\rho(\theta,\varphi)} =
\cos\frac{\theta}{2} \hat S(r) \ket{0} + e^{i\varphi}
\sin\frac{\theta}{2} \hat S(r) \ket{1}.
\eeqa
The coefficients of this superposition can be controlled by the amplitude and phase of the auxiliary coherent state.


\begin{figure}[tb]
	\centering {\includegraphics[width=\columnwidth]{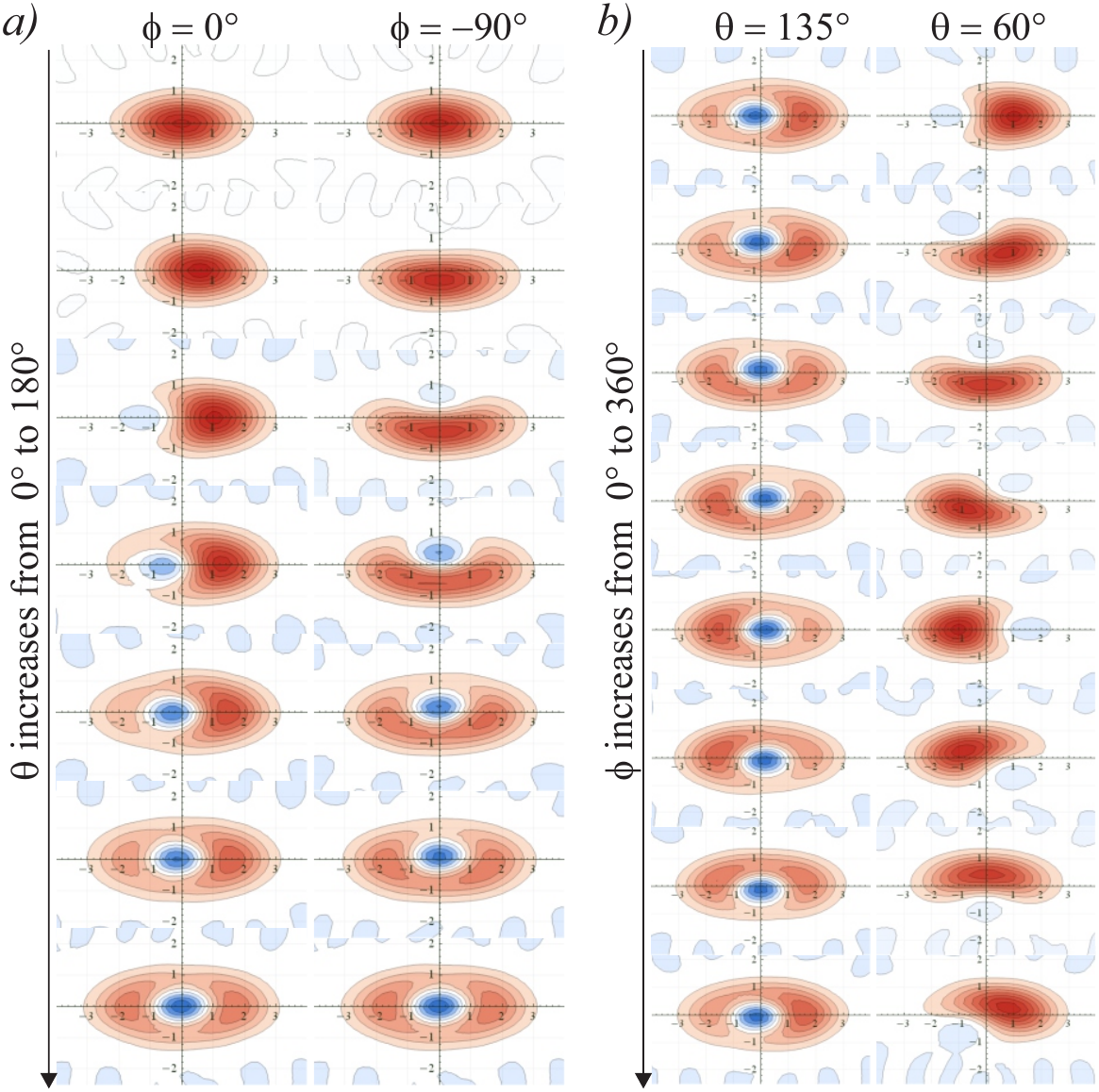}
		\caption{(Color online) 
			Wigner functions of measured cat qubit states  \citep{Neergaard-Nielsen2010}. Red colors correspond to positive values, blue to negative.  
		 a) The Bloch sphere polar angle $\theta$ that determines the relative weights of the terms in the superposition (\ref{ket_rho(theta,varphi)}) is varied with the asimuthal angle $\phi$ fixed at
			0$^{\circ}$ (left) and $-90^{\circ}$ (right). b) The azimuthal angle $\phi$ is varied while $\theta$ is fixed at 60$^{\circ}$ (left) and $135^{\circ}$ (right). }
		\label{fig:CatQubit}}
\end{figure}

The  Wigner functions of the generated states are presented in Fig.~\ref{fig:CatQubit}.
Fig.~\ref{fig:CatQubit}(a)
shows the control of the superposition weight $\theta$,
while keeping the phase constant. Conversely, Fig.~\ref{fig:CatQubit} (b) shows the control of the complex phase
$\varphi$ for fixed weights of the superposition.


%
%
%
%
%
%
%
%
%
%
%
%
%
%

\section{Generating arbitrary quantum states} \label{ArbStateGenSec}
Mastering the quantum technology of any physical system implies an ability to prepare arbitrary quantum states of that system. There has been significant effort to develop this ability in application to light, particularly to single-mode states thereof. Existing approaches to optical quantum state engineering involve linear optics, conditional measurements, coherent states, as well as sources of nonclassical light. The role of the latter can be played by single- or two-mode squeezers, on-demand Fock state sources, or nonclassical operators such as photon addition. 

An arbitrary linear combination of the first $n$ Fock states can be produced by applying a sequence of $n+1$ operators of the form $x_i\hat a^\dag+y_i$ (where $i$ runs from $0$ to $n$) to the vacuum state. This is the basis of some of the first quantum state engineering proposals \citep{Dakna1999a}. A subsequent proposal \cite{Clausen2001} uses an optical feedback loop based on a ring resonator, in which a parametric amplifier is seeded with a weak coherent state and single photon detection is performed in the idler channel [Fig.~\ref{fig:AddSubGen}(b)]. Later, \textcite{Fiurasek2005} demonstrated that  linear combinations $x\hat a+y$ of the photon subtraction and identity operators  [Fig.~\ref{fig:AddSubGen}(a)] can be used in a similar fashion, if a high-efficiency squeezer is available as an additional resource. 
These schemes can be extended, with multiple photon additions and subtractions, to the engineering of quantum \emph{operations} on travelling light beams 
\citep{Fiurasek2009},  such as noiseless linear amplification of light and the emulation of Kerr nonlinearity.


Sequential application of photon addition or squeezing operations involves transmitting the state through multiple OPAs. Such schemes would involve significant losses and are therefore not advisable in experimental practice. Instead, one can use a single non-degenerate OPA to generate a two-mode squeezed vacuum, and subsequently implement a sophisticated measurement on the idler channel to \emph{remotely prepare} the desired state in the other channel. This approach has been used to produce arbitrary  coherent superpositions of Fock states up to two- \citep{Bimbard2010} and three-photon states \citep{Yukawa2013}.

The measurement on the idler channel is implemented as follows. The idler channel is split into two (in \textcite{Bimbard2010}) or three (in \textcite{Yukawa2013}) modes, each of which is  overlapped on a BS with an auxiliary coherent state. A single-photon detector is placed into one of the outputs of each BS, \emph{\`a la}  Fig.~\ref{fig:AddSubGen}(a). A click in each detector projects the corresponding mode onto a superposition of the single-photon and vacuum states. Therefore an event in which all the detectors have clicked projects the idler channel onto a superposition of Fock states, where the coefficients of that superposition can be controlled by the amplitudes and phases of the ancillary coherent states. Due to the entanglement of the two-mode squeezed vacuum, such a projection remotely prepares the same state in the signal channel.

\begin{figure}[tb]
	\centerline{\includegraphics[width=\columnwidth]{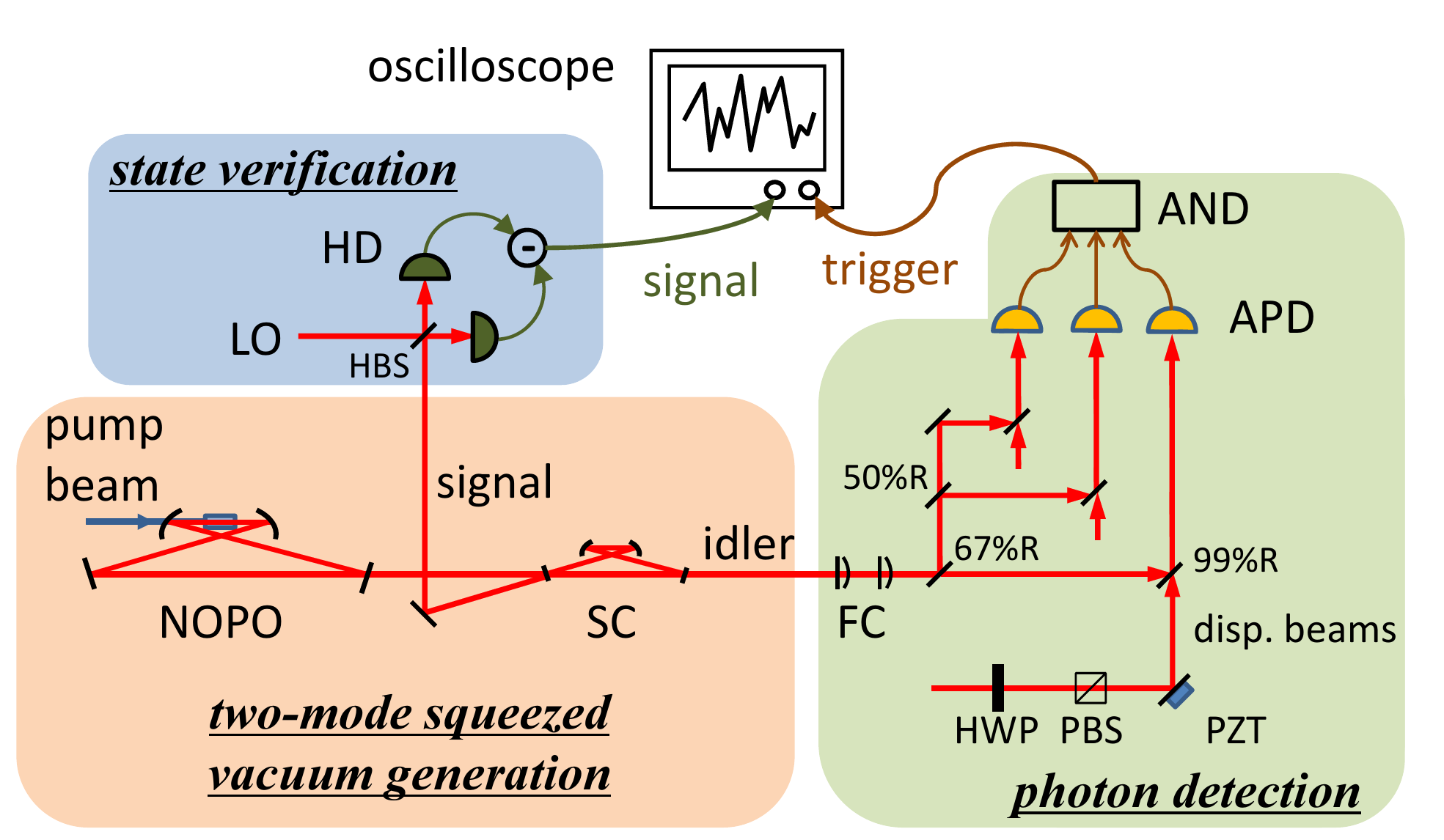}
	}
	\caption{A scheme for generating arbitrary Fock state superpositions up to three photons. The desired state preparation event is heralded by a triple coincidence click of the APDs in the idler channel. NOPO, non-degenerate optical parametric oscillator;
		SC, split cavity; FC, filter cavity; HD, homodyne detector; APD, avalanche photo diode;
		HBS, half beamsplitter; HWP, half-wave plate; PBS, polarization beamsplitter; PZT, piezo
		electric transducer.
From \cite{Yukawa2013}. }
	\label{YukawaFig}
\end{figure}

Another avenue for quantum-state engineering is to employ conditional measurements of quadratures instead of photon counting. 
\textcite{Lance2006} proposed and experimentally demonstrated the generation of squeezed single photons and superpositions of coherent states from input single- and two-photon Fock states, respectively, via the interaction with an ancilla squeezed vacuum state on a BS, followed by postselecting on a continuous-observable measurement outcome of the ancilla state. A ``state synthesizer" based on the iterative mixing of simple resource states of the form $x_i\ket 0+y_i\ket 1$ on a BS, followed by homodyne conditional measurements, has been proposed by \textcite{Etesse2014b}. This protocol builds the state in a piece-by-piece fashion, and potentially takes advantage of quantum memories in order to improve the success probabilities. Experimentally, \textcite{Babichev2004a} have shown that superpositions of the single-photon and vacuum states can be prepared by splitting a single photon between two modes and realizing a homodyne measurement in one of them to prepare the desired state in the other.

%
%
\section{Applications to quantum communications}
%
%

\subsection{From cat state ebits to quantum repeaters} 
\label{EntCatSec}
Although the states $\ket{\alpha}$ and $\ket{-\alpha}$ are never perfectly orthogonal, their overlap $|\braket{\alpha}{-\alpha}|^2=\e^{-4|\alpha|^2}$ decreases very rapidly, and for $\alpha \gtrsim 1$ they can be used as a basis for encoding qubits and using them for various quantum information processing tasks \citep{Jeong2007}, in particular for long-distance quantum communications. To this end, one must be able to prepare Bell-like ``entangled cat states'' $\ket{\Phi_\pm}=c(\ket{\alpha}\ket{\alpha}\pm\ket{-\alpha}\ket{-\alpha})$ and  $\ket{\Psi_\pm}=c(\ket{\alpha}\ket{-\alpha}\pm\ket{-\alpha}\ket{\alpha})$, allowing two distant partners, Alice and Bob, to share one ebit of entanglement as a resource for quantum teleportation. In principle, one can prepare the states $\ket{\Phi_\pm}$ simply by splitting a single-mode cat state $c(\ket{\sqrt{2}\alpha}\pm\ket{-\sqrt{2}\alpha})$ on a 50/50 BS, and phase-shifting one of the output modes by $\pi$ to transform $\ket{\Phi_\pm}$ into $\ket{\Psi_\pm}$. In practice, however, this approach is rather unrealistic because cat states are notoriously fragile, all the more as $\alpha$ is large. Preparing them locally and distributing them to distant sites through lossy channels will very rapidly destroy their entanglement, and though the resulting errors can in principle be corrected \citep{Glancy2008}, doing it in practice is technically challenging.

\begin{figure}[tb]
\begin{center}
\includegraphics[width=7cm]{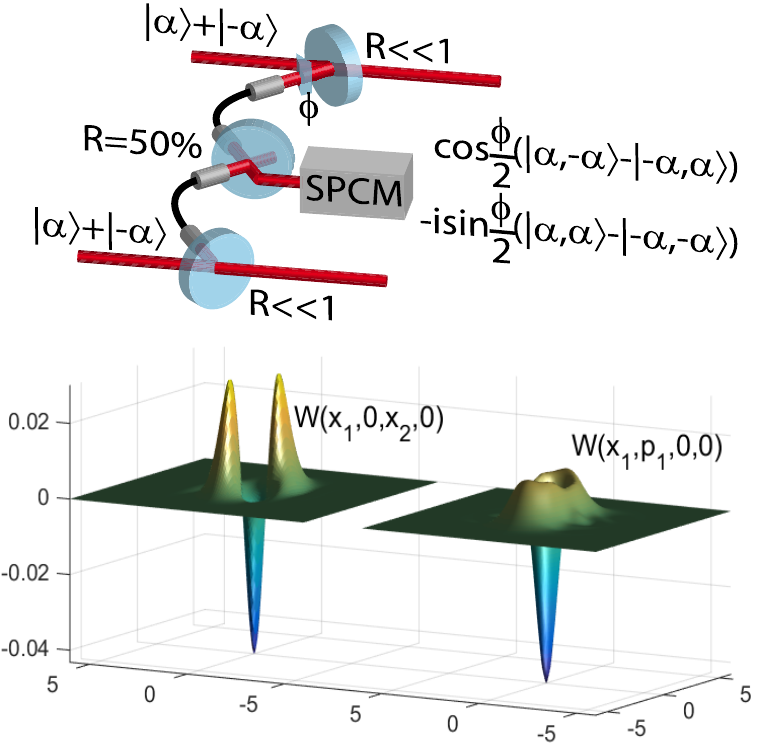}
\end{center}
\caption{(Color online) Sharing entangled ``cat'' states between distant sites. (Top) Small fractions of two independent cat states propagate through lossy optical fibers and interfere on a $50/50$ BS with a phase $\phi$. Detecting a photon in one of the output ports creates an entangled superposition of coherent states in a way relatively immune to fiber losses. Bottom: ``cuts'' of experimentally measured two-mode Wigner functions for $\alpha=0.8$ and $\phi=\pi/2$, where initial even ``kitten'' states were approximated by squeezed vacuum \citep{Ourjoumtsev2009}. } 
\label{FigAlexei:NonLocalCat}
\end{figure}

To circumvent this problem, one can transfer the loss from the cat state to the single photon used for its heralding. In the photon subtraction process, losing a photon before it reaches the detector mainly decreases the success rate, without affecting the quality of the prepared state to the lowest order.
Based on this idea, a method to share entangled cat states between distant sites has been proposed by \citet{Ourjoumtsev2009}, and demonstrated experimentally using small ``kittens'' (Fig.~\ref{FigAlexei:NonLocalCat}). First, Alice and Bob independently prepare two cat states $\ket{\psi_{{\rm cat},+[\alpha]}}_{A,B}$: since these states are prepared locally, they do not suffer from propagation losses. Then, the two modes are subjected to the scheme in Fig.~\ref{fig:AddSub2mode}(a): a small fraction of each cat state is reflected and sent through the lossy quantum channel to an intermediate node (Charlie), where they interfere on a 50/50 BS. Detecting a photon in one of the output modes of this BS applies the operator $\a_A-\e^{i\phi}\a_B$ to the initial state $\ket{\psi_{{\rm cat}}}_{A}\ket{\psi_{{\rm cat}}}_{B}$, where $\phi$ is the relative phase of the interfering beams. Since $\a\ket{\psi_{{\rm cat},+}[\alpha]}\propto\ket{\psi_{{\rm cat},-}[\alpha]}$, this delocalized photon subtraction entangles the two distant modes and projects them into the state
\begin{align*}
\ket{\Psi_\phi}&=\ket{\psi_{{\rm cat},-}[\alpha]}_A\ket{\psi_{{\rm cat},+}[\alpha]}_B\\
&-\e^{i\phi}\ket{\psi_{{\rm cat},+}[\alpha]}_A\ket{\psi_{{\rm cat},-}[\alpha]}_B
\end{align*}
which reduces to $\ket{\Psi_-}$ for $\phi=0$ and  to $\ket{\Phi_-}$ for $\phi=\pi$. 
The other pair of states ($\ket{\Psi_+}$ and $\ket{\Phi_+}$) can be generated by using initial cat states with different parities (one odd, the other even). Moreover, for $\phi\neq 0$, using this state as a resource for quantum teleportation allows one to perform phase-space rotations on the teleported qubit which are difficult to realize by other means. 

The above scheme has been proposed as a basis for a quantum repeater \citep{Sangouard2010,Brask2010,Brask2012b}, and has the following advantage. A necessary component of the quantum repeater is the Bell measurement, which enables the distribution of entanglement to end user parties via entanglement swapping \citep{Sangouard2011}. A Bell measurement performed on a pair of dual-rail qubits, such as the commonly used photon polarization qubit, cannot have a success probability above 50\% \citep{Vaidman1999,Lutkenhaus1999}. Therefore any quantum repeater based on such qubits will inevitably show exponential decay of the success probability with distance. With  coherent-state qubits, on the other hand, one can design a Bell measurement scheme with a unit success probability \citep{Brask2010}. However, taking advantage of this feature will require nearly perfect quantum-optical memory and on-demand cat state sources, which are beyond the current technology.

\subsection{Teleportation of cats}

Quantum teleportation is an important instrument
both for quantum communications and quantum computing. The CV teleportation scheme initially proposed by \textcite{Vaidman1994} and realized experimentally 
in  the seminal work  by 
\textcite{Furusawa1998} enables, in principle, the teleportation of any single mode state. However, perfect teleportation requires, in addition to the elimination of all the losses, an infinitely squeezed EPR resource state. For this reason, many variations of the CV teleportation protocol  \citep{Yoshikawa2007, 	Zhang2003,Bowen2003, 	Li2002, Mizuno2005} that emerged after the original experimental demonstration have been limited to Gaussian states. 

\begin{figure}[tb]
	\centering {\includegraphics[width=0.9\linewidth]{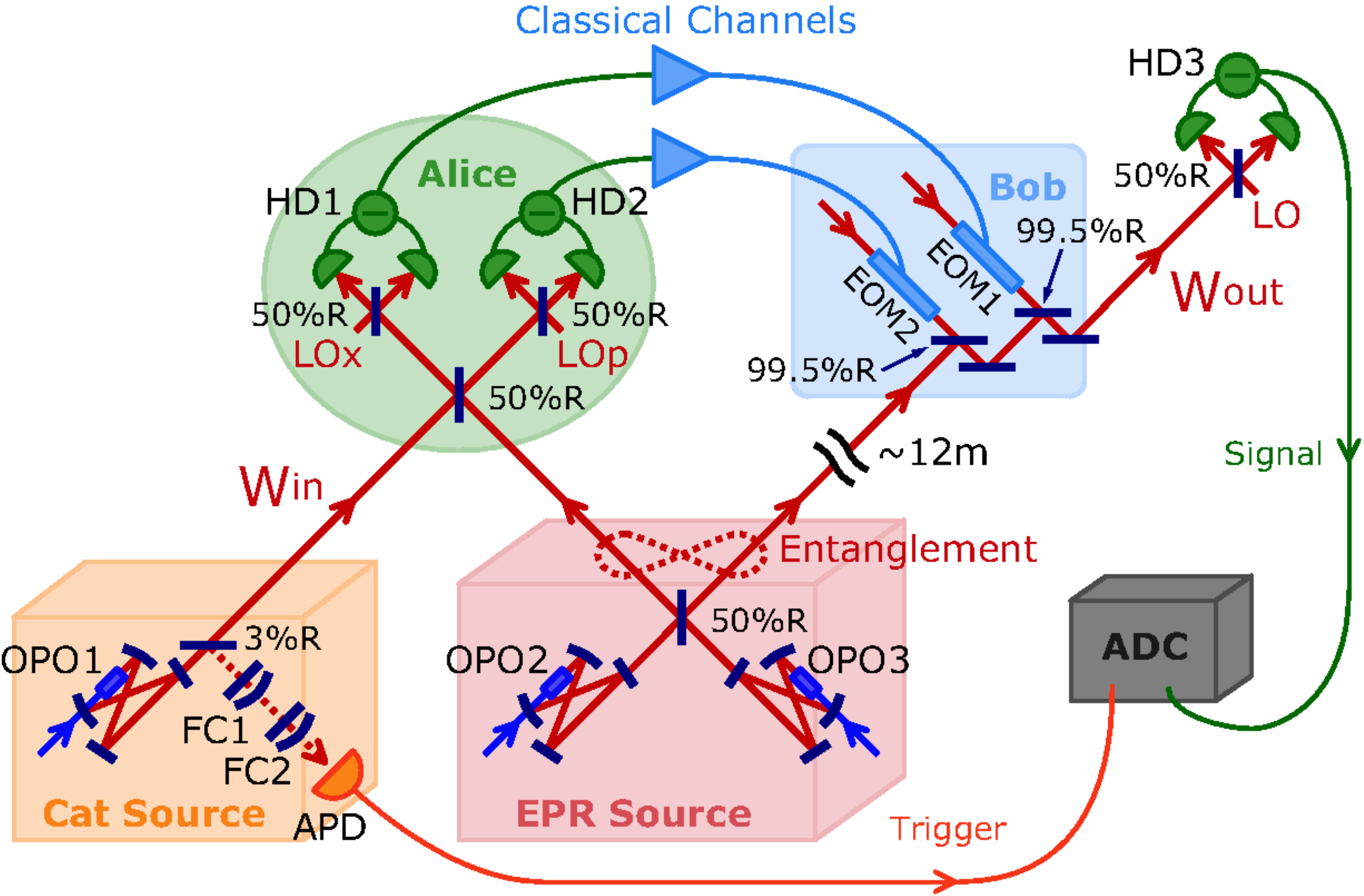}
		\caption{(Color online) 
			Experimental setup for the teleportation of cats,
			from \textcite{Lee2011}.
			OPO, optical parametric oscillator;
			APD, avalanche photodiode;
			HD, homodyne detector;
			LO, local oscillator;
			EOM, electro-optical modulator;
			ADC, analog-to-digital converter;
			FC, filtering cavity. } 
		\label{fig:Cat_teleportation_scheme}
	}
\end{figure}

Teleportation of non-Gaussian and non-classical states,
i.e. cat states whose Wigner functions had negative distributions, 
was demonstrated by \textcite{Lee2011} 
using the scheme shown in
Fig.~\ref{fig:Cat_teleportation_scheme}. 
Three OPOs were used to generate the necessary squeezed vacua. 
One produced a 2.4 dB single-mode squeezed vacuum state for the cat-state preparation, 
and the other two were for preparing the two-mode squeezed state with the measured squeezing factor of 6.9 dB. 
Alice received both the input cat state $\hat\rho_{\rm in}$ 
and one of the entangled beams, 
and then performed the continuous-variable Bell measurement 
consisting of two homodyne detectors denoted as HD1 and HD2. 
Bob received Alice's measurement results through
the classical channels and applied the displacement operation 
on the other component of the entangled beam. 
The output state $\hat\rho_{\rm out}$ was finally evaluated 
by the homodyne tomography using the detector HD3.

Although the three  OPOs produced cw squeezed vacua, 
the APD click at a time $t_1$ heralded the preparations of a cat state in a wave packet of the form similar to $\psi(t-t_1)$ in 
Fig.~\ref{fig.cw_sq_st_time_domain}. 
To teleport this cat state faithfully, 
all operations had to be realized with less noise and 
with a much broader bandwidth (corresponding to the wave packet of the cat state) than in the previous CV Gaussian teleportation experiments. 

\begin{figure}[tb]
	\centering {\includegraphics[width=0.9\columnwidth]{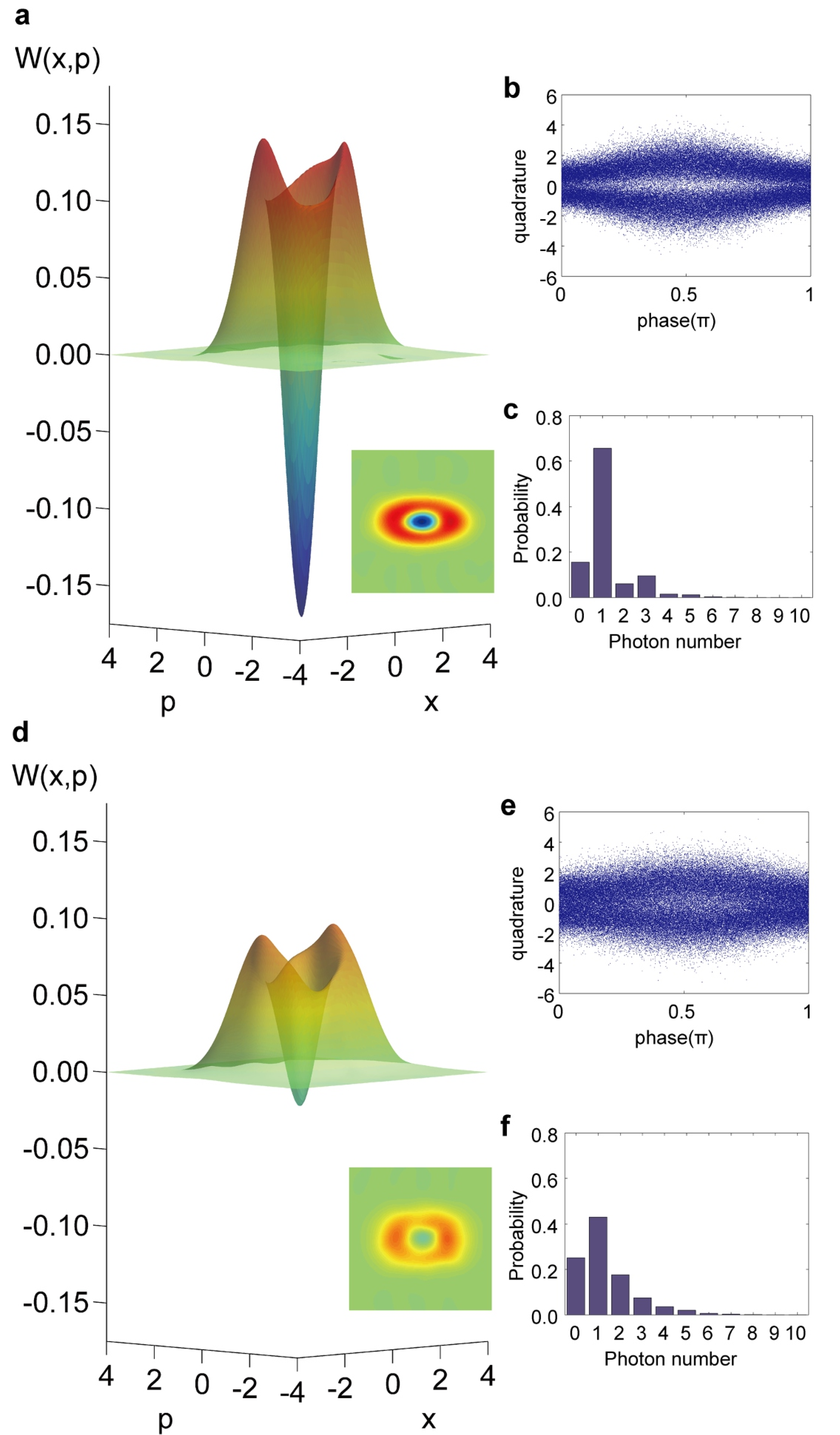}
		
		\caption{(color online)
			Experimental results of teleportation of cat states  by  \cite{Lee2011}.
			Experimentally measured input state's Wigner function (a),
			marginal distribution (b),
			and photon number distribution (c.
			Experimentally measured output state's Wigner function (d),
			marginal distribution (e),
			and photon number distribution (f).
		}
		\label{fig:Cat_teleportation_result}}
\end{figure}

Experimental results are presented 
in Fig.~\ref{fig:Cat_teleportation_result}.
As seen in the right panels, 
the Wigner function of the output state show a negative
$W_{\rm out}(0,0) = - 0.022 \pm 0.003$. 
This could demonstrate successful teleportation of 
a highly non-classical input state
($W_{\rm in}(0,0) = - 0.171 \pm 0.003$), 
shown in the left panel of Fig.~\ref{fig:Cat_teleportation_result}. 
The teleportation fidelity was greater than the no cloning limit of 2/3 
\citep{Grangier2011}.

In the limit of low squeezing, the negative cat state resembles the single-photon state Fig.~\ref{fig:Cat_teleportation_result}(c,f). In this experiment, this state --- the particle of light --- was "assembled"  in the output mode by Gaussian displacement operations on the electric field in one of the modes of the Gaussian EPR state. This result constitutes a remarkable illustration of wave-particle duality of quantum physics, as it cannot be explained by either the particle or wave pictures of light individually.

This work has been developed further in \cite{Takeda2015}. Here, the input of the teleportation device of Fig.~\ref{fig:Cat_teleportation_result} was fed with  one mode of the delocalized single-photon state $\sqrt{1-R}\ket{1,0}+\sqrt R\ket{0,1}$, obtained by splitting a photon on a BS with the reflectivity $R$. The teleported state was then verified to be entangled with the other mode of the delocalized  photon. In this way, entanglement swapping between discrete and continuous variables has been implemented for the first time (\emph{cf.}~Sec.~\ref{CVDVIntSec}).

\subsection{Non-deterministic amplifiers }  \label{section:NLA} 

The performance of deterministic linear amplifiers is limited by quantum physics: quantum states cannot be amplified without introducing additional noise \citep{Caves1982,Caves2012}. This is strongly connected with the no cloning theorem  \citep{Wootters1982}.
Nevertheless, non-deterministic  cloning \citep{Duan1998} and  non-deterministic noiseless linear amplifiers are possible \citep{Ralph2009}. 

If we consider a single mode field $\hat E=(\hat{a}\emph{e}^{-i \omega t}+\hat{a}^{\dag} \emph{e}^{i \omega t})/\sqrt{2}$, the quantum noise 
can be described in term of the quadrature variance as $\Delta E^2  = (\langle \Delta x ^2 \rangle+\langle \Delta p ^2 \rangle) /2 $.
In the case of a (non existing) ideal phase-independent amplifier, one would have $\hat{a}_{out}= g \, \hat{a}_{in}$ , and  the signal to noise ratio would be conserved, because the signal power and the variance would be multiplied by the same factor $G=g^2$. However the requirement to conserve  the bosonic commutation rules through a unitary transformation  imposes including an extra operator $\hat{a}_{out}= g\hat{a}_{in}+ \hat{L}^{\dag}$ which  adds to the signal variance an excess noise factor of at least $(1/2)|g^2-1|$. This bound can be reached by a NDOPA, or by a simple BS in the case $g<1$ \citep{Caves1982}. 

\begin{figure}[tb]
 	\includegraphics[width=\columnwidth]{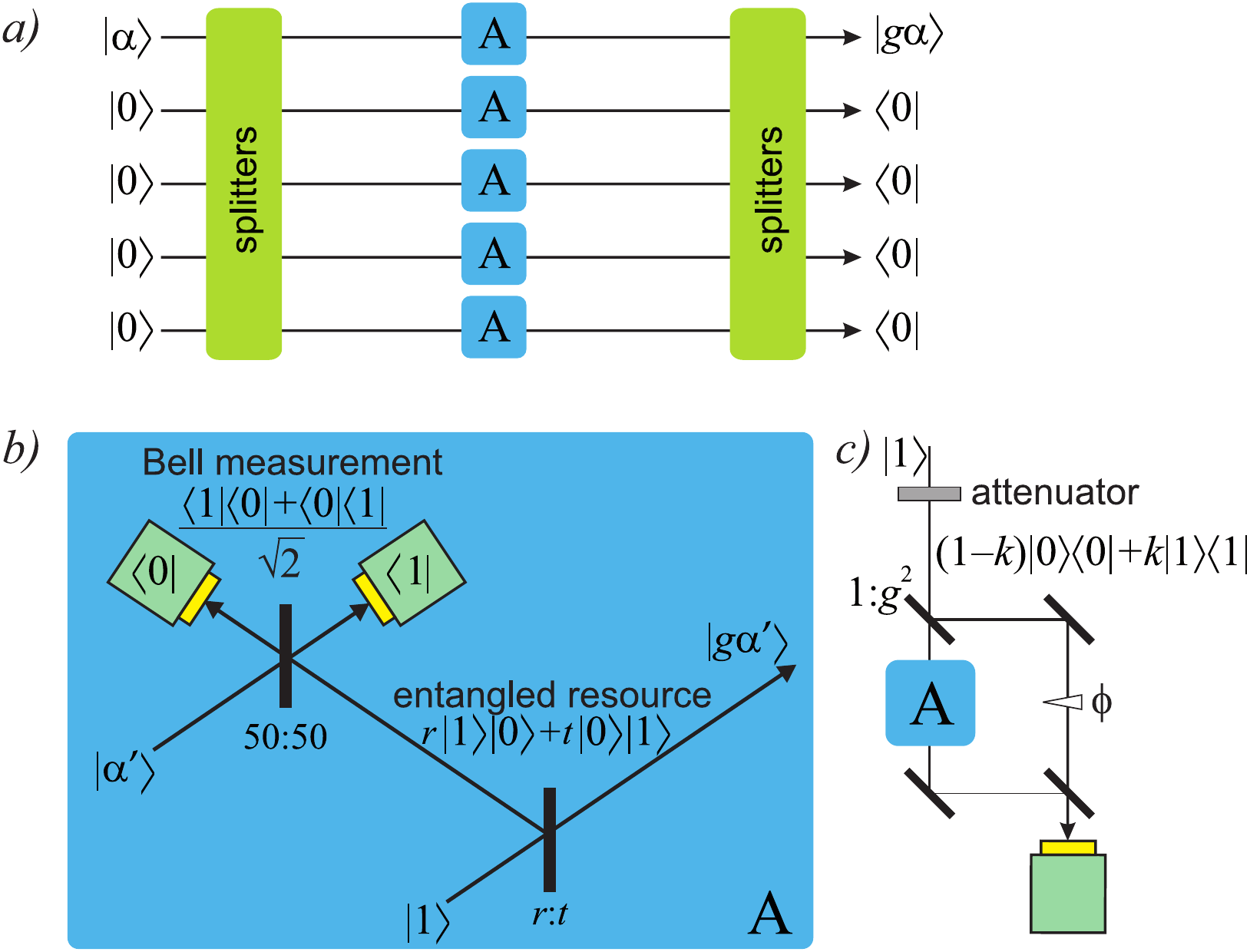}
 	\caption{(Color online) Noiseless amplification by means of ``asymmetric quantum scissors" \citep{Ralph2009}. a) overall scheme. b) a single amplification stage [marked ``A" in (a)] operational for low-amplitude inputs. c) Interferometric scheme of \cite{Xiang2010} to prove the noiseless character of the amplification.}
 	\label{NLA}
 \end{figure}

Nevertheless, 
it is  possible to circumvent this limitation by implementing {\it non-deterministic} linear amplification \citep{Barbieri2011}. 
T.C. Ralph and A.P. Lund made the first proposal for the realization of such a non-deterministic noiseless linear amplifier (NLA) in 2008 \citep{Ralph2009} which has been followed by two experimental implementations in  2010 \citep{Xiang2010,Ferreyrol2010}.
The scheme is displayed in Fig.~\ref{NLA}(a): the optical state to be amplified, which we assume to be a coherent state $\ket\alpha$, is divided, by using an arrangement of beam splitters, into $N$ modes each containing a coherent state $\ket{\alpha'}$ with $\alpha'\ll 1$. Each of these is sent in an amplification stage indicated by ``A''. 

Each amplification stage [Fig.~\ref{NLA}(b)] constitutes a modified teleportation scheme in which the delocalized single photon state is used as the entangled resource \citep{Pegg1998,Babichev2003}. This resource, given by  $\ket{\Psi}=r\ket{1}\ket{0}+t\ket{0}\ket{1}$, is generated when a single photon $\ket{1}$,  incident upon a beam splitter of amplitude reflectivity $r$ and transmissivity $t$, entangles itself with the vacuum $\ket{0}$. The input coherent state $\ket{\alpha'}\approx\ket 0+\alpha'\ket1$, along with one of the modes of $\ket\Psi$, is subjected to a Bell measurement in the Fock basis. Specifically, the modes are mixed at a 50:50 BS and its outputs are measured with single-photon counters. A detection event in one of the counters and the absence thereof in the other projects the two modes onto the Bell state $\frac{1}{\sqrt{2}}(\ket{1}\ket{0}+\ket{0}\ket{1})$. The state of the remaining mode of the entangled resource will then contain the teleported state.

If the first BS is symmetric ($r=t=\frac1{\sqrt2}$), the teleported state is identical to the input state in terms of the vacuum and single-photon terms in its Fock decomposition; higher Fock terms of the input are ``cut off" from the output because they are absent in the entangled resource (hence this protocol is termed ``quantum scissors"). If the input coherent state is weak ($\alpha'\ll1$), its higher-order Fock terms are negligible and the teleported state is similar to the original input. \textcite{Ralph2009} observed that, if $r\ne t$, the teleported output is given by 
\begin{equation}\label{truncalpha}
\vert 0 \rangle \pm g \alpha ' \vert 1 \rangle \approx  \vert g \alpha ' \rangle,
\end{equation}
where $g=t/r$ and the approximation is valid if $g\alpha'\ll 1$. For $g>1$, the output is a coherent state of higher amplitude than the input, i.e. the NLA takes place. 
 
The outputs of individual amplification stages can be recombined interferometrically with the inverse arrangement of beam splitters. In this case, the desired state $\ket{g\alpha}$ will come out from one of the ports given that no photon is detected in the remaining $N-1$ ports. In this way, a coherent state of any arbitrary amplitude $\ket\alpha$ can in theory be amplified.

 To date, only a single stage of the NLA has been implemented experimentally. 
 The experiment realized  by \textcite{Xiang2010} used one attenuated channel of a SPDC source to generate the state  $  (1-k) \vert 0 \rangle \langle 0\vert + k \vert 1 \rangle \langle 1\vert $ which is very close to a phase mixed coherent state with the amplitude $|\alpha|^2=k$ and it is used as the input state of the amplifier. The other arm of the same source produces the single photon ancilla. To verify that amplification has occurred, \textcite{Xiang2010} used photon counting to compare the measured average photon number at the input and output of the amplifier stage. Linear gain up to $\vert g\vert^{2}=4$  has been realized for state with   $\vert\alpha\vert$ up to $\simeq 0.04$.  To verify that the gain process is coherent and does not add noise, they embedded the amplifier stage in an inbalanced Mach-Zehnder interferometer. The splitting ratio of the interferometer's initial beam splitter was set to $g^2$, such that the arms became balanced after the amplification. High-visibility fringes ovserved in this experiment showed that the noise introduced by the NLA was significantly below what would be expected from an linear amplifier with the same gain. 
 
\begin{figure}[tb]
	\includegraphics[width=0.82\linewidth]{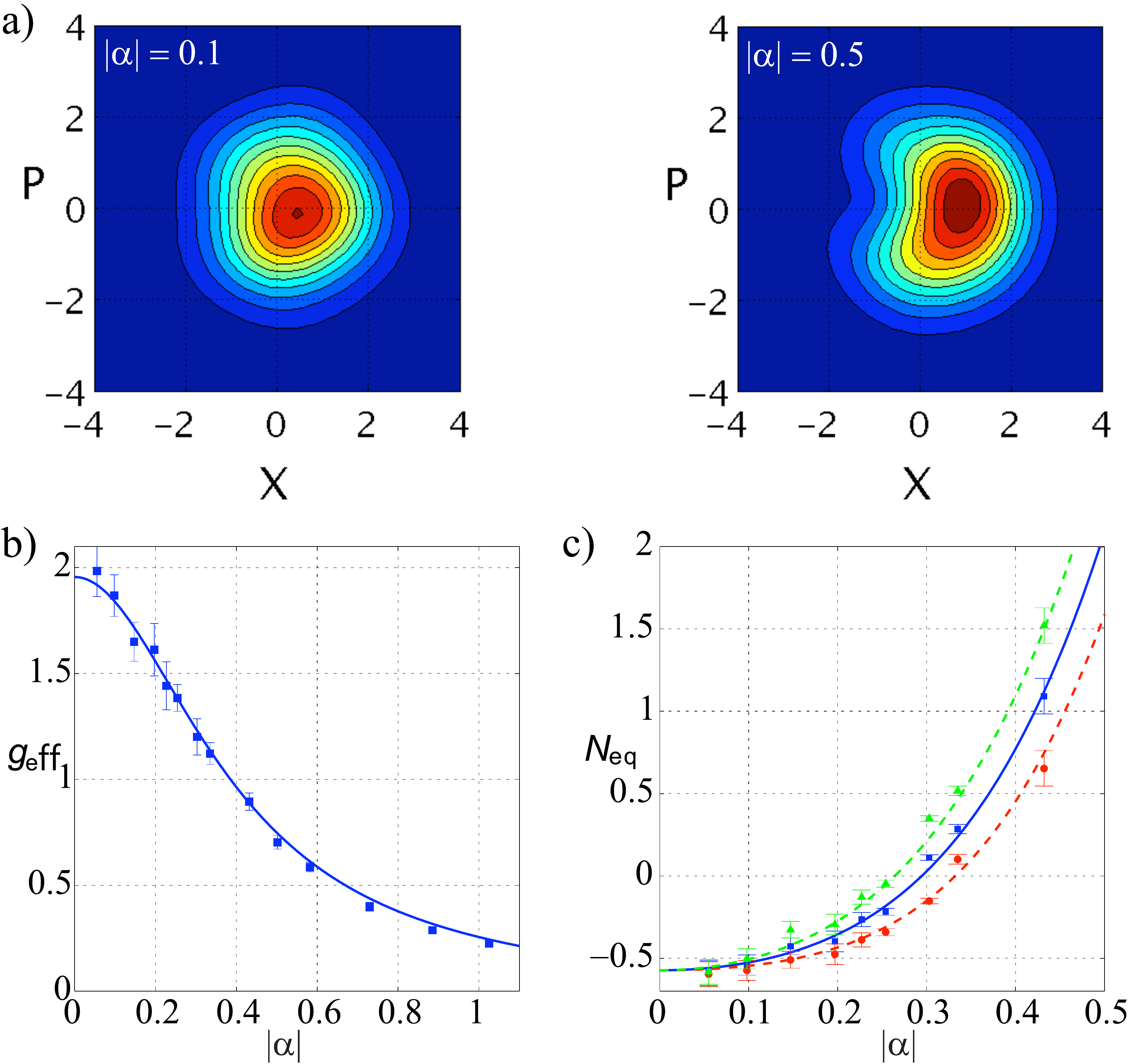}
	\caption{(Color online) Results of the experiment by \textcite{Ferreyrol2010}. a) Reconstruction of the Wigner quasi-distributions of the amplifier outputs, for two values of $\vert \alpha\vert$. The circular shape of the coherent state is approximatively preserved for low $|\alpha|$ (left), but not for high $|\alpha|$ (right). b) Effective phase-independent gain as a function of the input state amplitude compared with the model (solid line). c) Equivalent input noise (\ref{Neq}) for three different quadratures: the average, maximal and minimal value are potted as a function of the state amplitude.
	}
	\label{NLA-exp-F}
\end{figure}

\textcite{Ferreyrol2010} used homodyne tomography to completely characterize the output state. 
The effective amplification can then be quantified by $g_{\rm eff}=\langle x_{\rm out} \rangle / \langle x_{\rm in} \rangle$ and the noise behavior of the amplifier is analyzed in terms of its  ``equivalent input noise":
 \begin{equation}\label{Neq}
 N_{\rm eq}=\dfrac{\langle\delta x_{\rm out}^{2}\rangle}{g_{\rm eff}}- \langle\delta x_{\rm in}^{2}\rangle
 \end{equation}
 This figure is the quantum optical analogue to the one adopted in electronics \citep{Roch1993}, it tells how much noise must be added to the input noise level in order to mimic the observed output noise for the given gain. The value of $N_{\rm eq}$ cannot be negative for a classical amplifier, but is observed in the experimental data for $\vert \alpha \vert \lesssim 0.3 $ [Fig.~\ref{NLA-exp-F}(c)] thereby making the quantum character of noiseless amplification manifest.
 
The use of heralded noiseless amplifiers has been proposed \citep{Gisin2010}  to overcome the problem of channel losses in discrete-variables device-independent quantum key distribution and to improve the maximum transmission distance of continuous-variable quantum key distribution \citep{Blandino2012c}.   Heralded noiseless photon amplification at telecom wavelengths has been tested
obtaining a gain $ >100 $ associated with a heralding probability greater than $80\%$ up to a
distance (in fiber) of 20 km \citep{Bruno2013}. Noiseless linear amplification of a qubit encoded in the polarization state of a single photon has also been demonstrated \citep{Kocsis2013}.

Asymmetric quantum scissors is not the only possible implementation of NLA. An alternative scheme is shown in Fig.~\ref{fig:AddSubGen}(a) for $\ket\psi=\ket 1$. Indeed, the output state in this case is given by 
\begin{equation}\label{catalysis}(\sqrt R\hat a+\alpha)\ket 1=\sqrt R(\ket 0+\frac\alpha {\sqrt R}\ket 1)\approx\ket{\frac\alpha {\sqrt R}},
\end{equation} 
(where $R$ is the reflectivity of the BS and the approximation is valid for $\frac\alpha {\sqrt R}\ll 1$). In other words, the input coherent state $\ket\alpha$ is amplified by the factor $g=\frac1{\sqrt R}$. Experimentally, this scheme has been developed and experimentally realized by \textcite{Lvovsky2002}, however this paper did not interpret its result as NLA. Subsequently, this scheme has been utilized in a theoretical proposal to amplify collective states of atomic spins for quantum metrology applications \citep{Brunner2011}. Furthermore, it was used for CV entanglement distillation as discussed in the next section. 

A further alternative strategy to realize non-deterministic noiseless amplifiers has been proposed by  \cite{Fiurasek2009} and realized by \cite{Zavatta2011}.
It employs the observation that one can associate the NLA of gain $g$ with the operator 
\begin{equation}\label{Gg0}\hat G=g^{\adag\a}. 
\end{equation}One can see this intuitively by applying this operator to a low-amplitude coherent state: $\hat G\ket\alpha\approx\hat G(\ket 0+\alpha \ket 1)=\ket 0+g\alpha \ket 1\approx\ket{g\alpha}$. To prove the relationship (\ref{Gg0}) rigorously, let us define a fictitious Hamiltonian $\hat H=i\hbar\adag\a\ln g/\tau$, where $\tau$ is an arbitrary positive value in units of time. Then $\hat G$ is the evolution operator under this Hamiltonian for time $\tau$ because $\exp(-\frac i\hbar \tau)=\exp(\adag\a\ln g)=g^{\adag\a}$. On the other hand, the annihilation operator evolves in the Heisenberg picture according to $\dot{\a}=\frac i\hbar[\hat H,\a]=\ln g/\tau \hat a$, so the evolution for the time $\tau$ transforms it as follows: $\a(\tau)=\a(0)e^{\ln g}=g\a(0)$, consistent with the definition of the ideal NLA given in the beginning of this section. 

Decomposing $\hat G$ into the Taylor series with respect to $\adag\a$ up to the first order, we find
\begin{equation}\label{Gg}
\hat{G}\approx(g-1){\adag\a}+1=(g-2)\hat{a}^{\dag}\hat{a}+\hat{a}\hat{a}^{\dag}. 
\end{equation} Such combinations of the photon addition and photon subtraction operations can be realized using the methods described above in the context of Fig.~\ref{addsubsc}.  
The advantage of this scheme is that it is possible to realize higher amplification without reducing the fidelity with the ideal target state $\vert g \alpha \rangle$ because it is able to amplify also the higher-photon terms of the state.

\begin{figure}[tb]
\includegraphics[width=0.95\linewidth]{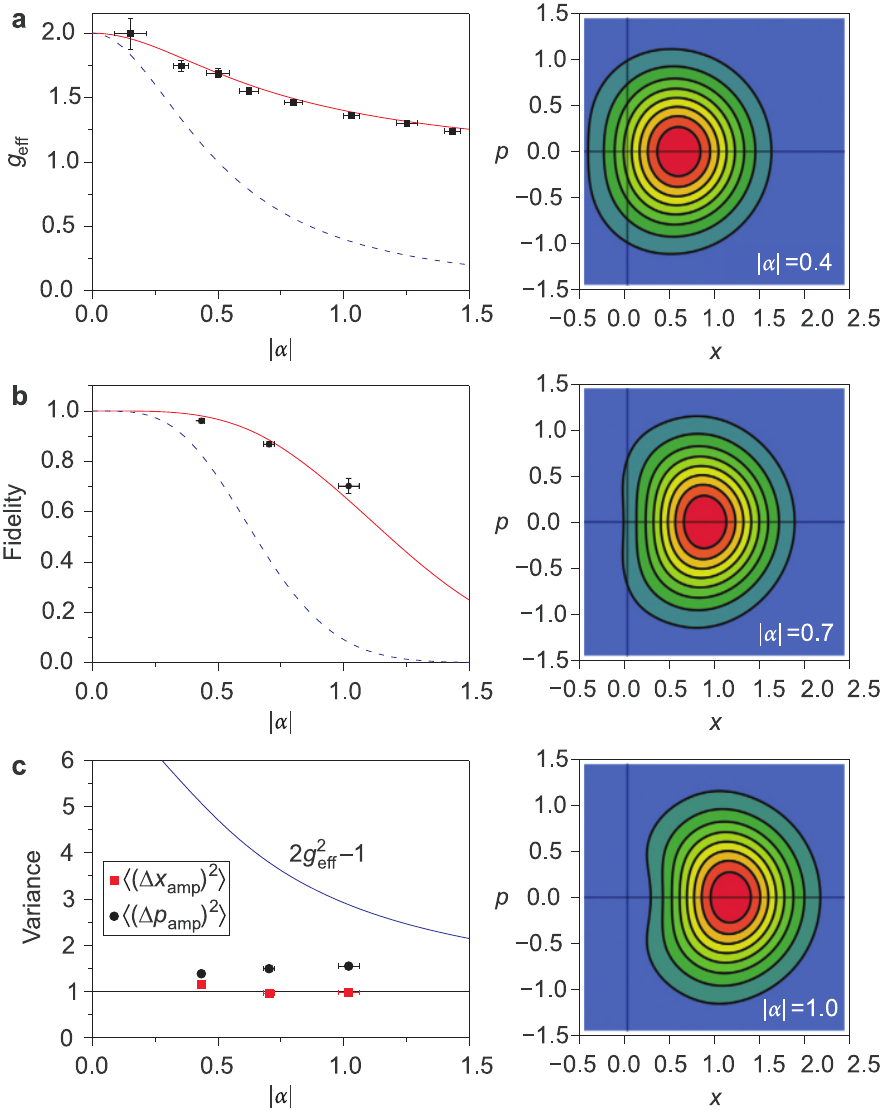}
\caption{(Color online)  Results of the experiment performed by \textcite{Zavatta2011}. Right column: reconstructed Wigner functions for 	three amplified coherent states of different amplitudes. Left column: a) Effective gain; b) Final state fidelity vs. the input state amplitude $\vert \alpha\vert$. Red solid curves: model of the addition/subtraction scheme \cite{Zavatta2011}; blue dashed curves: model based on the asymmetric scissors \cite{Ralph2009}. c) Measured variances (corrected for the detection efficiency) of the position and momentum quadratures of the amplified coherent state and the corresponding (blue solid) curve for the best deterministic amplifier. 
}
\label{NLA-exp-Z}
\end{figure}

The results in  Fig.~\ref{NLA-exp-Z} correspond to the case of $g=2$, for which the experimental setup is particularly simplified because Eq.~(\ref{Gg}) then reduces to $\hat{G}=\hat{a}\hat{a}^{\dag}$. In the experiment by \textcite{Zavatta2011}, coherent states of different amplitudes have been subjected to the sequential action of the creation and annihilation operators, and then characterized by homodyne tomography.
A fidelity above $90\%$ has been reached for $\vert \alpha \vert \lesssim 0.65$ corresponding to the effective gain $g_{\rm eff}\gtrsim 1.6$. The equivalent input noise is $N_{eq}<-0.48$ for $\vert \alpha \vert \leq 1.4$. The advantage in fidelity with respect to the asymmetric scissors is visible by visually comparing the Wigner functions in  Figs.~\ref{NLA-exp-F}(a) vs. \ref{NLA-exp-Z}(right column)  and is quantified in  Fig.~\ref{NLA-exp-F}(b). In a follow-up theoretical paper, \textcite{Park2016} proved that even better preservation of fidelity for higher-amplitude coherent states can be achieved by constructing the NLA using linear combinations of higher order products of the creation and annihilation operators.
 

\begin{figure}[tb]
\includegraphics[width=0.95 \linewidth]{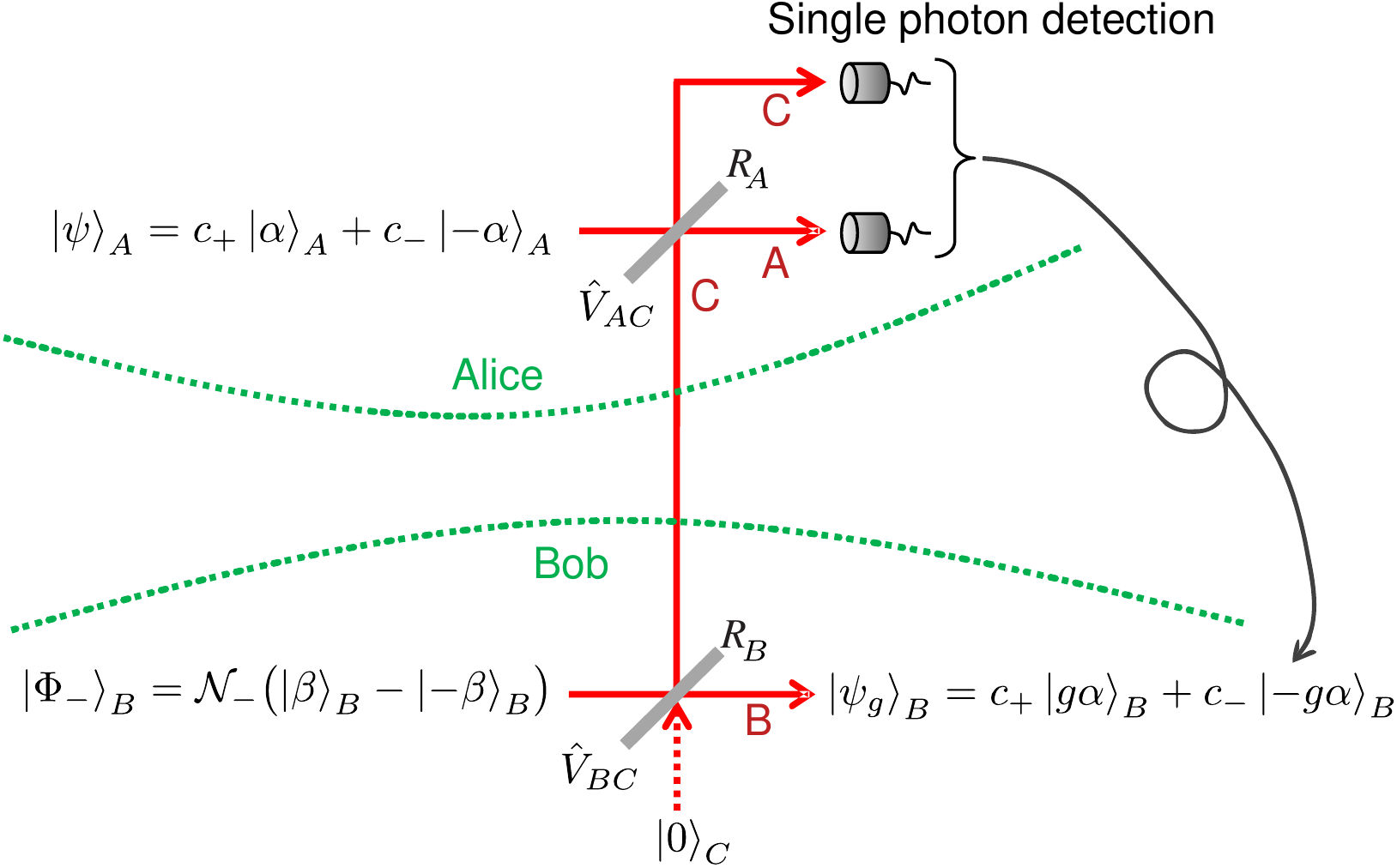}
\caption{(Color online)  Quantum tele-amplification of a binary cat-state in an ideal lossless channel. From \cite{Neergaard-Nielsen2013}.
}
\label{teleamp}
\end{figure}

Yet another strategy for coherent state amplification \citep{Ralph2003,Neergaard-Nielsen2013} is based on teleportation with an asymmetric entangled resource [akin to Ref.~\cite{Ralph2009}], but this teleportation is implemented in the coherent-state basis rather than the Fock basis. The scheme is presented in Fig.~\ref{teleamp}: Bob prepares an odd-cat state $\emph{N}_{-}(\vert \beta \rangle - \vert -\beta \rangle)$ and splits it using a BS with reflectivity $r_{B}$, denoted in Fig.~\ref{teleamp} by the symbol $\hat V_{BC}$, thereby generating an delocalized cat state $\ket{(1-r_B)\beta}_B\ket{r_B\beta}_C-\ket{-(1-r_B)\beta}_B\ket{-r_B\beta}_C$  in paths $B$ and $C$  (\emph{cf.}~Sec.~\ref{EntCatSec}), which is used as the entangled resource for the teleportation. The input state in channel $A$ has the form $c_+\ket\alpha_A+c_-\ket{-\alpha}_A$, where 
\begin{equation}\label{alphabeta}\beta\sqrt{r_{A}r_{B}}=\alpha \sqrt{1- r_{A}},\end{equation} 
and $r_A$ is the reflectivity of  the beam splitter $\hat V_{AC}$. To perform the Bell measurement,  Alice combines modes $A$ and $C$ at the beam splitter $\hat V_{AC}$ with reflectivity $r_{A}$ and looks for those events in which a single photon at port $A$ and no photon at port $C$ is detected. In this case, Bob obtains the state $c_+\ket{(1-r_B)\beta}-c_-\ket{-(1-r_B)\beta}$, which is the amplified input state with the gain $g=\sqrt{(1- r_{A})(1- r_{B})/ r_{A} r_{B}}$.

This Bell measurement is described in detail in the Supplementary to \citep{Neergaard-Nielsen2013}; here we offer a brief intuition behind it. Notice that the relation (\ref{alphabeta}) is such that the amplitudes of the coherent states that are reflected from mode $C$ by $\hat V_{AC}$ into the detector in mode $A$ and the one that is transmitted from mode $A$ into the same detector are of the same absolute values. Suppose now that the phases of modes $A$ and $C$ are set such that equal-amplitude coherent states at the entrance to  $\hat V_{AC}$ experience constructive interference at that detector, while opposite-amplitude states interfere destructively. This means that the detector in mode $A$ can click only if the states entering  $\hat V_{AC}$ are either $\ket\alpha_A\ket{R_B\beta}_C$ or $\ket{-\alpha}_A\ket{-R_B\beta}_C$. A more precise calculation shows that this click, combined with an absent click in mode $C$, projects the input of $\hat V_{AC}$ onto the coherent Bell-like superposition $\ket\alpha_A\ket{R_B\beta}_C+\ket{-\alpha}_A\ket{-R_B\beta}_C$, therefore enabling the teleportation protocol.


In the experiment by \textcite{Neergaard-Nielsen2013}  the resource odd-cat state is generated   by photon-subtraction from a squeezed vacuum, and $\beta$ is experimentally tuned by the squeezing level. 
Compared with the previous strategies,  this scheme gives a higher fidelity with the target state but requires  the knowledge of $\alpha$ in order to choose the right amplitude of the odd-cat state. Furthermore, the protocol is significantly affected by the losses in channel $C$ between Alice and Bob.

%
%
\subsection{CV entanglement distillation } 
%
%

The two-mode squeezed vacuum state, mentioned extensively in this review, is a workhorse of CV quantum information processing and communication \cite{Weedbrook2012,Braunstein2005}. Applications thereof include  complete quantum teleportation \cite{Furusawa1998},   key distribution \cite{Madsen2012}, 
  repeaters \cite{Campbell2013}, metrology \cite{Anisimov2010} and many others. An obvious advantage of the CV entangled resource, compared to its discrete-variable counterpart, is that it is readily
  available on demand from parametric amplifiers. On the other hand, the CV EPR state, when distributed over large distances, is deteriorated by losses, and therefore requires distillation to recover its entanglement.

\begin{figure}[tb]
	\includegraphics[width=0.85\columnwidth]{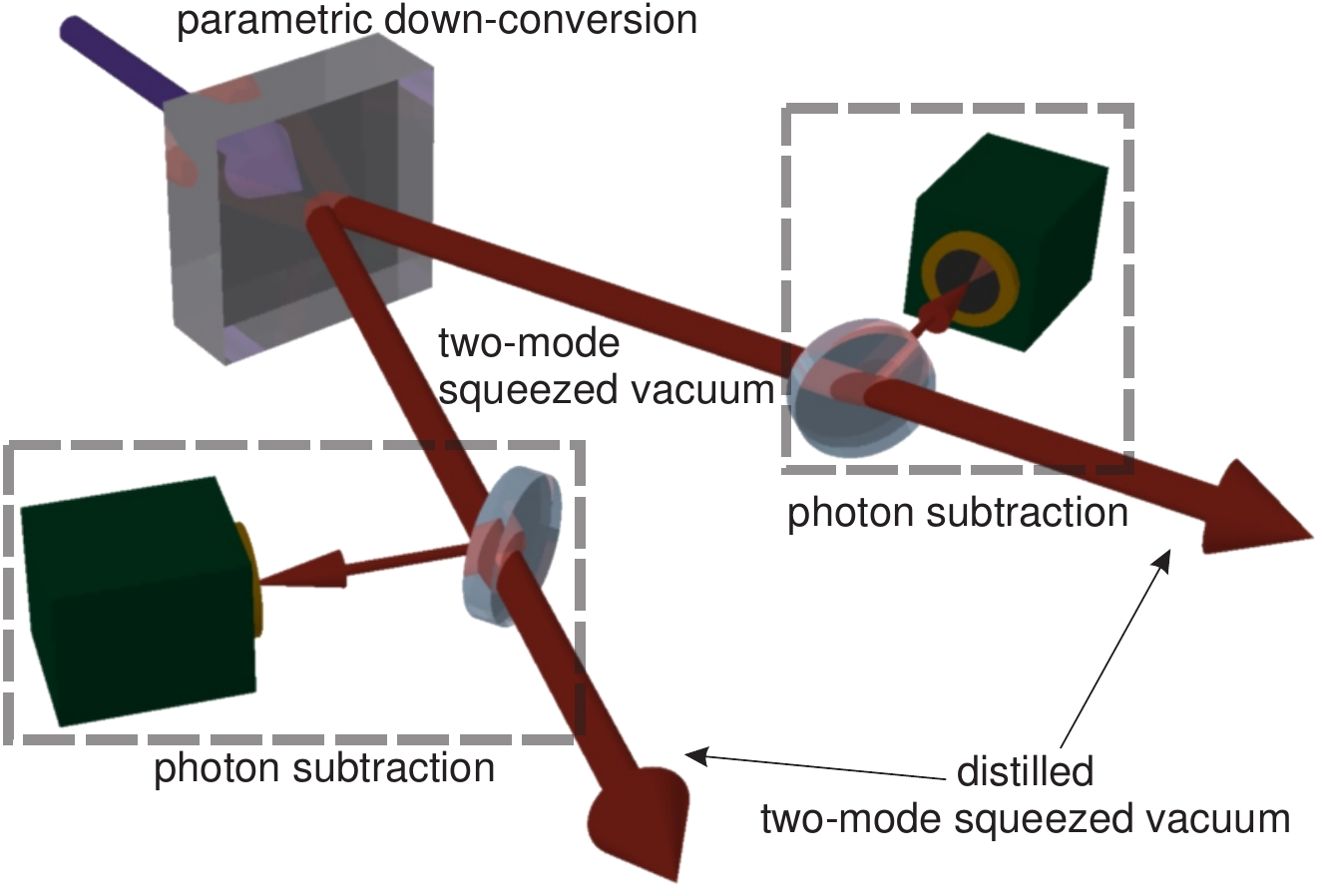}
	\caption{(Color online) 
CV entanglement distillation by photon subtraction.
	}
	\label{fig:Distillation_schematic}
\end{figure}

\begin{figure}[tb]
	\includegraphics[width=\linewidth]{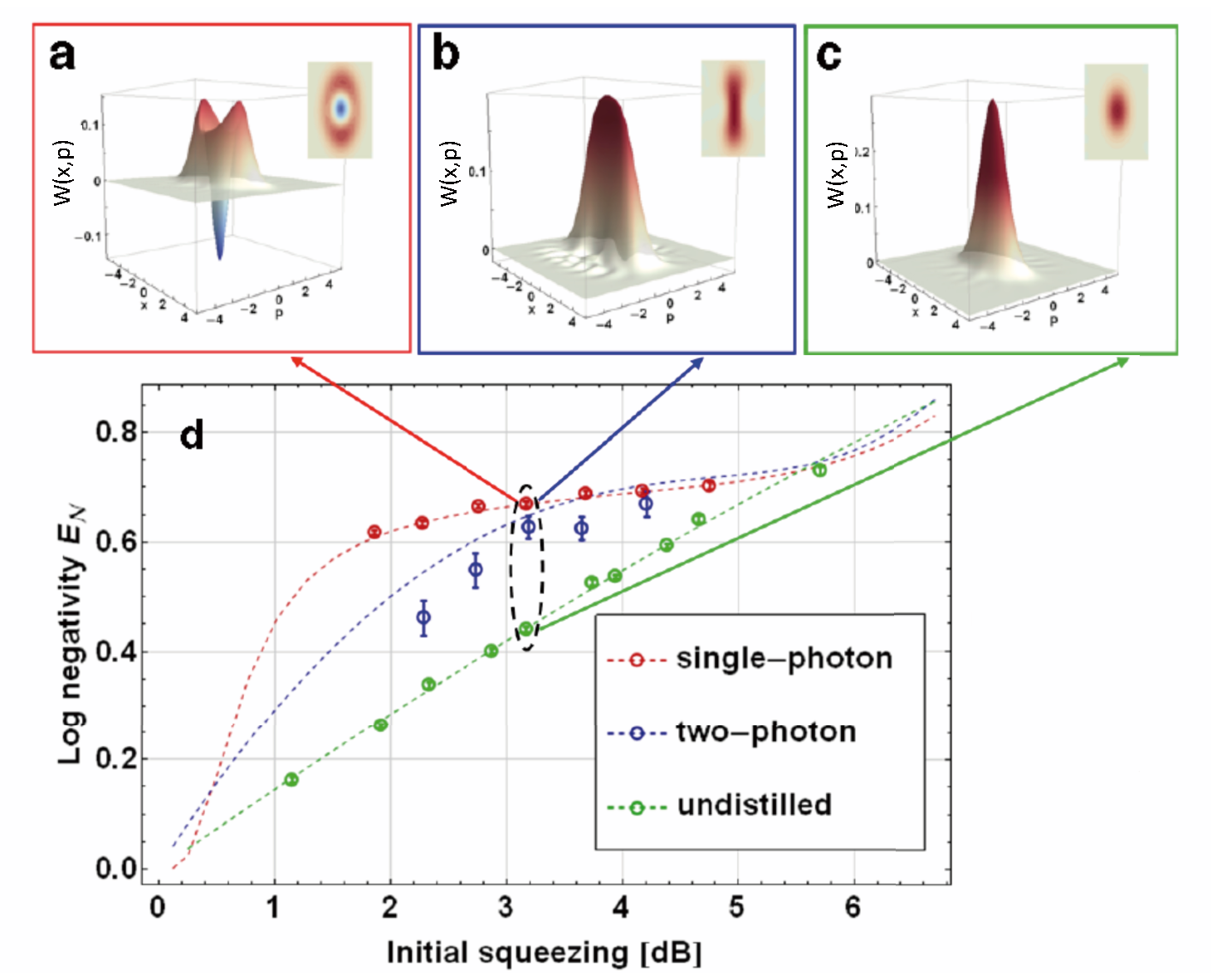}
	\caption{(Color online) 
		Results of the experiment by  \citep{Takahashi2010b} on the distillation of the split single-mode squeezed vacuum. (a-c) 	Experimentally reconstructed Wigner functions
		and their contour plots of the  mode defined by the operator $(\hat a_A-\hat a_B)/\sqrt2$:
		a)  distillation  via single-photon
		subtraction with $R$=5\%; b) distillation via
		two-photon subtraction with $R$=10\%; c)
		undistilled initial state (squeezed vacuum with $R$=0\%),
		all with the initial squeezing of $-3.2$ dB.
		(d) Experimental logarithmic negativities as functions of the initial input
		squeezing. The dashed curves are theoretical predictions
		based on independently  measured experimental parameters.
	}
	\label{fig:negwig} 
\end{figure}

It was theoretically proven as a ``no-go theorem'' that
Gaussian operations can never distill
entanglement from Gaussian state inputs
\citep{Eisert2002,Fiurasek2002,Giedke2002}. Under Gaussian operations, we understand operations that preserve the Gaussian shape of a state's Wigner function. These include interference, conditional homodyne detection, as well as single- and two-mode squeezing. 
For the entanglement distillation from CV Gaussian states,
one should rely on non-Gaussian operations, and
photon subtraction is actually a straightforward way
to circumvent the above no-go restriction. 

The idea of CV entanglement distillation by photon subtraction was first proposed by \textcite{Opatrny2000} 
and further theoretically investigated for photon-number
resolving \cite{Cochrane2002} and threshold \cite{Olivares2003} detectors. To understand it, let us apply  annihilation operators to both modes of the EPR state (Eq. \ref{EqStateEPR}), as shown in Fig.~\ref{fig:Distillation_schematic}. We find
\begin{equation}\label{Psiout}
\hat{a}_A \hat{a}_B \ket{\psi_{\rm EPR}} =\frac1{\cosh(r)} \sum^{\infty}_{k=1} k\tanh^{k}r |k-1,k-1\rangle,
	\end{equation}
where the subscripts $A$ and $B$ correspond to the modes of the EPR state.	In the limit of small squeezing ($r\ll 1$), both states (\ref{EqStateEPR}) and (\ref{Psiout}) can be approximated to their first order:
	\begin{subequations}\label{eq:psiin}
	\begin{eqnarray}
	\ket{\psi_{\rm EPR}}&\propto&|0,0\rangle+r |1,1\rangle; \\
		\hat{a}_1 \hat{a}_2 \ket{\psi_{\rm EPR}}&\propto&|0,0\rangle+2r |1,1\rangle.
			\end{eqnarray}
	\end{subequations}
A higher contribution of the double-photon term enhances both the entanglement and the two-mode squeezing \cite{Bartley2013}. This enhancement comes at the price of losing the state's Gaussian character: this effect is weak for small initial squeezing, but becomes significant otherwise. The above analysis can be readily extended to account for losses in one or both channels of the EPR state. 

In fact, the entanglement can increase even more dramatically if, instead of subtracting one photon in each mode via the operator $\hat{a}_A \hat{a}_B$, one subtracts a single photon delocalized in the two modes of the EPR state via the operator $\hat{a}_A +\hat{a}_B$ implemented as shown on Fig. \ref{fig:AddSub2mode}a. In this case, for $r\ll 1$, one obtains the state $(\ket{0,1}+\ket{1,0})/\sqrt{2}$ with one ebit of entanglement: the initial two-mode squeezing only affects the success rate, and even if the initial entanglement is infinitely small, the final entanglement is finite. This protocol was successfully implemented by \cite{Ourjoumtsev2007b}. However, even if the transmission of the quantum channel used for the non-local photon subtraction can be small, it is still a non-local operation, so this process does not qualify as entanglement distillation.

The first experiment on entanglement distillation of a Gaussian state by local photon subtraction was done by \textcite{Takahashi2010b}. In this experiment, the input state was not the EPR state but a single-mode squeezed vacuum state split into two modes on a symmetric beam splitter, resulting in entanglement of these two modes. Photons were then subtracted from one or both modes at the same time, and the distilled entangled state was measured by two independent homodyne detectors. \textcite{Takahashi2010b} observed that the two-mode state, both before and after the photon subtraction, becomes separable under the beam splitter transformation $\hat a_\pm=(\hat{a}_A\pm\hat{a}_B)/\sqrt2$. Indeed, the entangled state before subtraction is then transformed back into the initial product of vacuum and squeezed vacuum states $\ket{0}_+\ket{s}_-$, and the photon subtraction operators $\hat a_A=(\hat{a}_+ +\hat{a}_-)/\sqrt2$ and  $\hat a_B=(\hat{a}_+ -\hat{a}_-)/\sqrt2$ both reduce to $\hat{a}_-$ as there are no photons to subtract from $\ket{0}_+$. For one- and two-photon subtraction, the mode $\hat a_+$  was indeed observed to remain in the vacuum state, while the state of $\hat a_-$ are the  odd [Fig.~\ref{fig:negwig}(a)] and even [Fig.~\ref{fig:negwig}(b)] cat states, respectively. The entanglement increase induced by photon subtraction was quantitatively measured via the logarithmic negativity [Fig.~\ref{fig:negwig}(d)], defined as the sum of the negative eigenvalues of 
the partially transposed density matrix of a 2-mode entangled state, 
$\hat\rho_{AB}^{T_A}$ 
and known to be a monotone measure of entanglement 
\cite{Vidal2002}.

\textcite{Kurochkin2014} demonstrated entanglement distillation of the proper two-mode squeezed state, which was generated via non-degenerate parametric down-conversion [Fig.~\ref{fig:Distillation_schematic}]. The experiment was conducted in the pulsed regime and the initial squeezing was relatively low: $0.63$ dB. After the photon subtraction, which utilized two tapping beam splitters with the reflectivities $R=0.11$, the squeezing increased to $0.83$ dB. The entanglement negativity has also increased from $0.24$ to $0.3$.

\begin{figure}[tb]
	\includegraphics[width=\columnwidth]{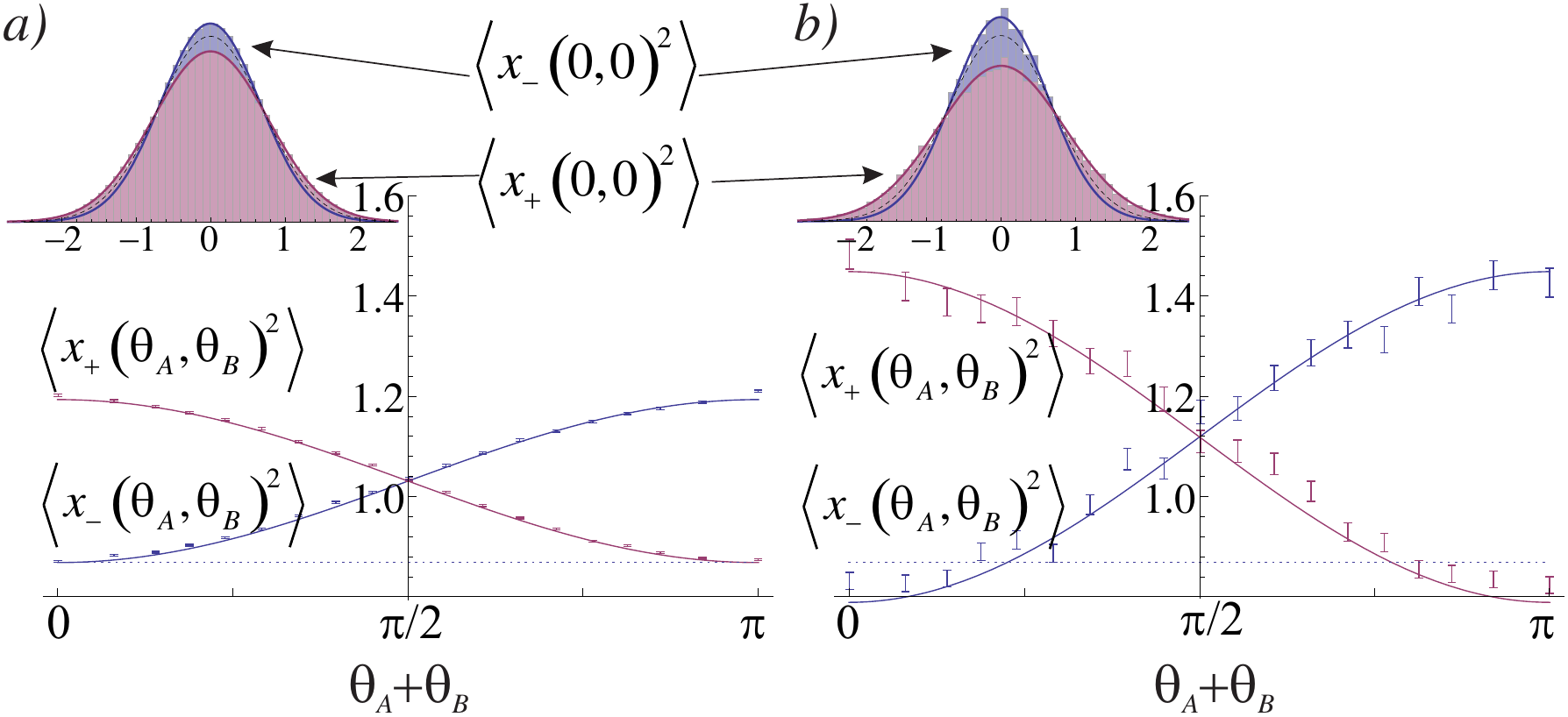}
	\caption{(Color online) CV entanglement distillation by photon subtraction from an EPR state.
		 Variances of the sum and difference of the quadratures measured in the two modes: a) unconditionally, b) conditioned on photon subtraction events in both channels $A$ and $B$ in \cite{Kurochkin2014}. The noise level corresponding to the double vacuum state is 1. The minimum variance in (a) is indicated by a dashed line. The curves show theoretical fits. The insets show histograms of the sum and difference of the position quadratures corresponding to each case. The dashed line shows the standard quantum limit. Enhancement of squeezing is present in (b) while the loss of Gaussian character is insignificant.}
	\label{fig:kurochkin}
\end{figure}

The results of this experiment are shown in Fig.~\ref{fig:kurochkin}, where the variances of the sum and difference quadratures $x_\pm(\theta_A,\theta_B)= x_A(\theta_A)\pm x_B(\theta_B)$ are plotted as a function of the sum of the phases $\theta_A+\theta_B$ in the two channels. This way of presenting the results takes advantage of the fact that the two-mode squeezed vacuum is insensitive with respect to equal and opposite shifts of phases $\theta_A$ and $\theta_B$. Indeed, shifting the two modes by $(\delta\theta_A,\delta\theta_B)$ will apply the factor of $e^{ik(\delta\theta_A+\delta\theta_B)}$  to each term of the decomposition (\ref{EqStateEPR}), which vanishes for $\delta\theta_A+\delta\theta_B=0$. The setting $\theta_A+\theta_B=0$ corresponds to the measurement of the position quadratures in both modes, so $x_A(\theta_A)$ and $x_B(\theta_B)$ are correlated and $x_-(\theta_A,\theta_B)$ is squeezed. In contrast, for $\theta_A+\theta_B=\pi$, $x_A(\theta_A)$ and $x_B(\theta_B)$ are anticorrelated and $x_+(\theta_A,\theta_B)$ is squeezed. 
We note also that because the initial squeezing is low, the distillation does not compromise the Gaussian character of these states, as is evident from the insets in Fig.~\ref{fig:kurochkin}.

These experiments show the functionality of the photon subtraction approach to CV entanglement distillation. However, this approach has a significant limitation. As is evident from Eqs.~(\ref{eq:psiin}), the enhancement of entanglement what can arise from single-photon subtraction in each mode cannot exceed a factor of 2, which is not sufficient to compensate for a typical loss
occurring in a communication line. Subtracting higher photon numbers  can improve this factor, but at a cost of exponential loss in the event rate.

A useful alternative protocol  \cite{Ulanov2015} relies on the  noiseless linear amplification discussed in the previous section. To see this, let us consider the EPR state, which we, as previously, truncate at the single-photon term in the Fock decomposition as per Eq.~(\ref{eq:psiin}a), assuming $r\ll 1$. Suppose that mode $B$ of this state propagates through a loss channel of amplitude transmissivity $t$, resulting in the following statistical mixture:
\begin{equation}\label{initstateloss}
\hat\rho=(\ket{00}-rt\ket{11})(\bra{00}-rt\bra{11})+r^2(1-t^2)\ketbra{10}{10}.
\end{equation}
Next, the NLA \emph{\`a la} \textcite{Lvovsky2002} is applied to that channel as per Eq.~(\ref{catalysis}). We obtain 
\begin{align}\nonumber
\hat\rho'&=({\sqrt R}\ket{00}-rt\ket{11})({\sqrt R}\bra{00}-rt\bra{11})
\\&+r^2R(1-t^2)\ketbra{10}{10}.\label{finalstaterho}
\end{align}
We see that the vacuum component in the first term of the mixture, which corresponds to the entangled state, is multiplied by a factor $\sqrt R\ll 1$, corresponding to the effective enhancement of the entanglement by the inverse of that factor. In this way, by choosing the reflectivity $R$ of the beam splitter used in the NLA, one can restore the entanglement in the EPR state after  an arbitrarily high loss.

\begin{figure}[tb]
	\includegraphics[width=0.8\columnwidth]{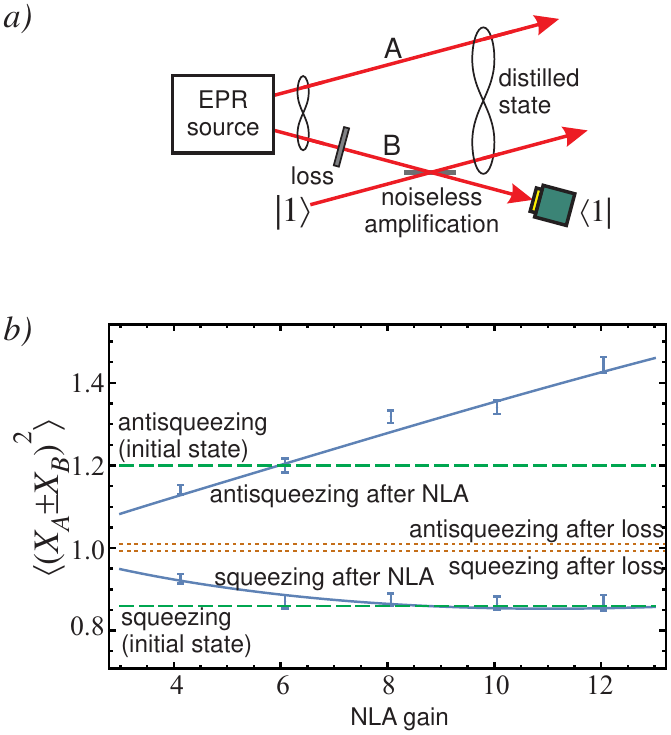}
	\caption{(Color online) 
		Distillation of CV entanglement by means of noiseless linear amplification \cite{Ulanov2015}. a) Experimental setup, b) Results showing two-mode squeezing, measured by the variances of the sum (anti-squeezing) and difference
		(squeezing) of the position quadratures in the two modes of the distilled EPR state. The theoretical curves assume the   detection efficiencies in the undistilled and distilled channels of $\eta_A=0.45$ and $\eta_B = 0.5$, respectively, and a single-photon preparation efficiency of $\eta = 0.65$. 
	}
	\label{fig:CVdistillNLA}
\end{figure}

Figure \ref{fig:CVdistillNLA} shows the setting and results of the experiment by \textcite{Ulanov2015}. Two non-degenerate parametric down-conversion setups are used, one to generate the EPR state (with $r=0.135$), and one to prepare the heralded single photon \cite{Huisman2009}. Mode $B$ of the EPR state propagates through an attenuator with a 5\% transmissivity that models a lossy channel. Subsequently, it is overlapped on a low-reflectivity beam splitter to realize NLA. The initial squeezing of 0.65 dB is degraded by the attenuator to an almost undetectable level, but then fully recovered [Fig.~\ref{fig:CVdistillNLA}(a)].

Both the photon-subtraction and NLA methods of CV entanglement distillation yield a state that approximates the two-mode squeezed vacuum only up to the first order in the photon number decomposition. This means that the amount of squeezing and entanglement attainable through the distillation is limited. However, the fact that the resulting states are non-Gaussian permits further distillation by Gaussian means. In particular, the state (\ref{finalstaterho}) can be further distilled using the Gaussian procedure of \textcite{Eisert2004}, which can be implemented using only four quantum optical memory cells \cite{Datta2012}. By applying this protocol in an iterative manner, an infinitely squeezed EPR state can in principle be obtained.

\subsection{Discrete-continuous interfacing}\label{CVDVIntSec}
Physical systems that can be used for quantum information processing can be divided roughly into two classes. The first class is the systems with non-equidistant energy levels, from which one can select a pair of levels that can serve as a qubit. Examples of the first class include single atoms, quantum dots, superconducting circuits or color centers. The other class includes atomic ensembles, optical or microwave cavities, and optomechanical membranes --- that is, systems whose energy level structure is intrinsically equidistant, and therefore similar to that of the harmonic oscillator. In systems of the second class, it is sometimes more natural to encode quantum information in continuous degrees of freedom.

A technology for coherent and loss-free exchange of quantum information among different quantum systems is essential for constructing functional quantum information processing networks \cite{Kurizki2015}. A natural means for such exchange is the electromagnetic field, as it is capable of carrying quantum information over significant distances. Fortunately, this system, being by nature a harmonic-oscillator-like system, is also capable of coupling efficiently to quantum objects with nonequidistant energy levels. Moreover, it is able to process information in both CV \cite{Braunstein2005} and DV \cite{Kok2007} regimes. However, there is a missing piece: a tool for interconversion of quantum information between these two encodings.

\begin{figure}[tb]
	\includegraphics[width=\columnwidth]{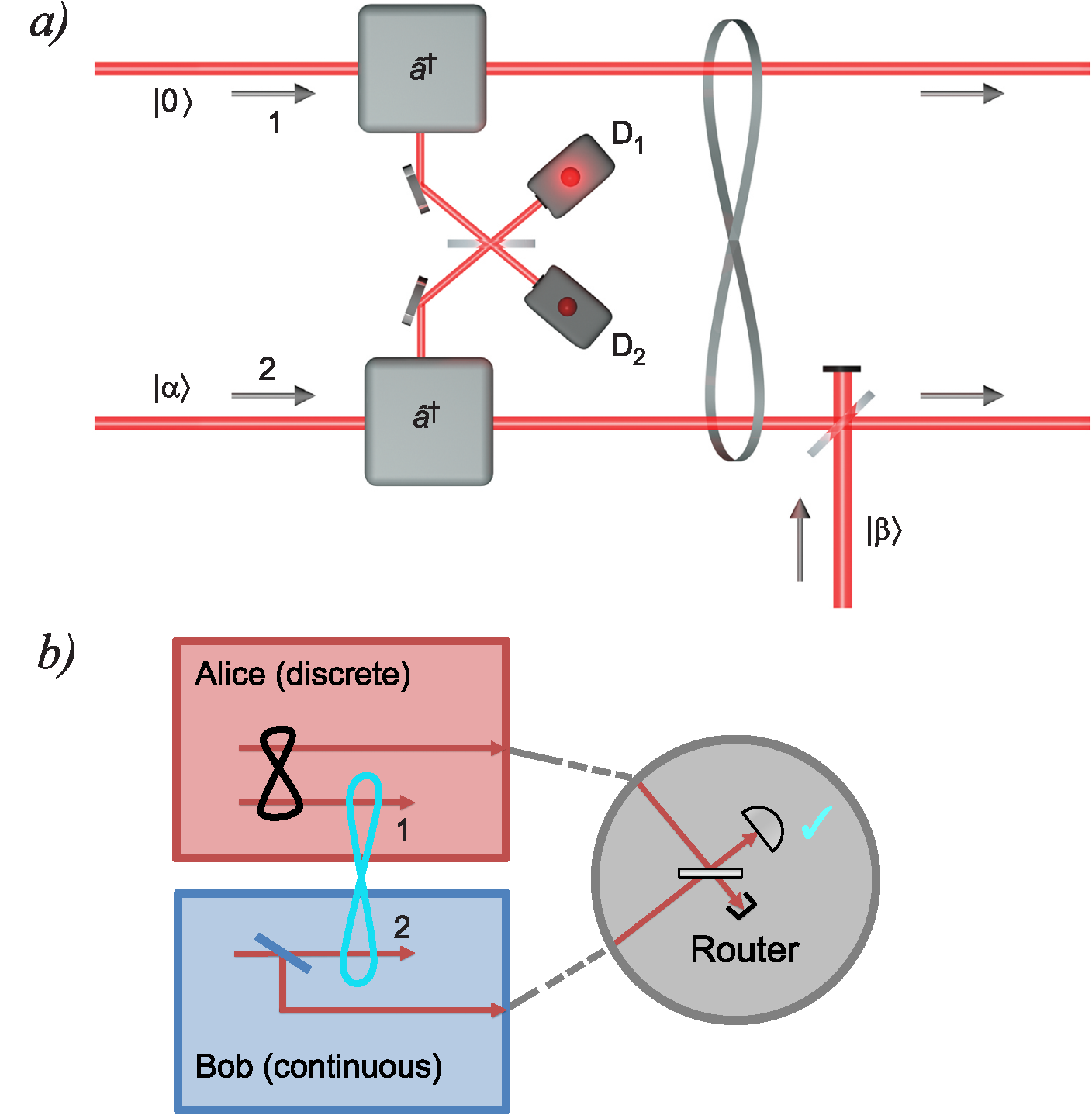}
	\caption{(Color online) 
		Preparation of an entangled state between a CV (cat) qubit and the single-rail DV qubit. a) The scheme of \textcite{Jeong2014}; b) the scheme of \textcite{Morin2014}. See text for details. 
	}
	\label{fig:BelliniLaurat}
\end{figure}

The role of such a tool can be played by an entangled state between qubits encoded in the discrete and continuous bases. Indeed, such a state can be used as the entangled resource for protocols such as teleportation, remote state preparation, and entanglement swapping. In the last few years, there has been a significant effort to develop a technique for generating such a state. In 2014, \textcite{Jeong2014} and \textcite{Morin2014} independently demonstrated an entangled state between a CV  qubit encoded in two coherent states of opposite phases (\emph{cf.}~Sec.~\ref{CatQubitSec}) and the single-rail DV qubit encoded in the Fock states $\ket 0$ and $\ket 1$ using two different schemes.

Figure \ref{fig:BelliniLaurat}(a) shows the scheme of \textcite{Jeong2014}. Modes 1 and 2 are initially prepared in the vacuum and coherent states, respectively, and then subjected to a nonlocal photon addition operator using the setup of Fig.~\ref{fig:AddSub2mode}(b). This produces the state  $a_1^\dag\ket 0\ket\alpha+\ket 0a_2^\dag\ket \alpha$. The effect of the photon creation operator on a coherent state strongly depends on its amplitude. For $\alpha=0$, $\hat a^\dag\alpha=\ket 1$, i.e. the photon-added state is orthogonal to the input. But for a large $\alpha$, the photon-added state $\hat a^\dag\alpha$ is approximated by a coherent state of a slightly larger amplitude $\ket{g\alpha}$, with $g\approx1+1/|\alpha|^2$ \cite{Jeong2014}. In other words, the photon creation operator acts on a high-amplitude coherent state like a noiseless amplifier (Sec.~\ref{section:NLA}). The resulting state in modes 1 and 2 will therefore be approximated by $\ket 1\ket\alpha+\ket 0\ket{g \alpha}$. The remaining step is to apply a phase-space displacement to mode 2 by $-\alpha(g+1)/2$ to transform $\ket\alpha\to\ket{-\alpha(g-1)}$ and $\ket{g\alpha}\to\ket{\alpha(g-1)}$, resulting in the desired DV-CV entangled qubit. This step is implemented by mixing mode 2 with a strong coherent state $\ket\beta$ on a low-reflectivity beam splitter [Fig.~\ref{fig:BelliniLaurat}(a)]

\begin{figure}[tb]
	\includegraphics[width=0.93\columnwidth]{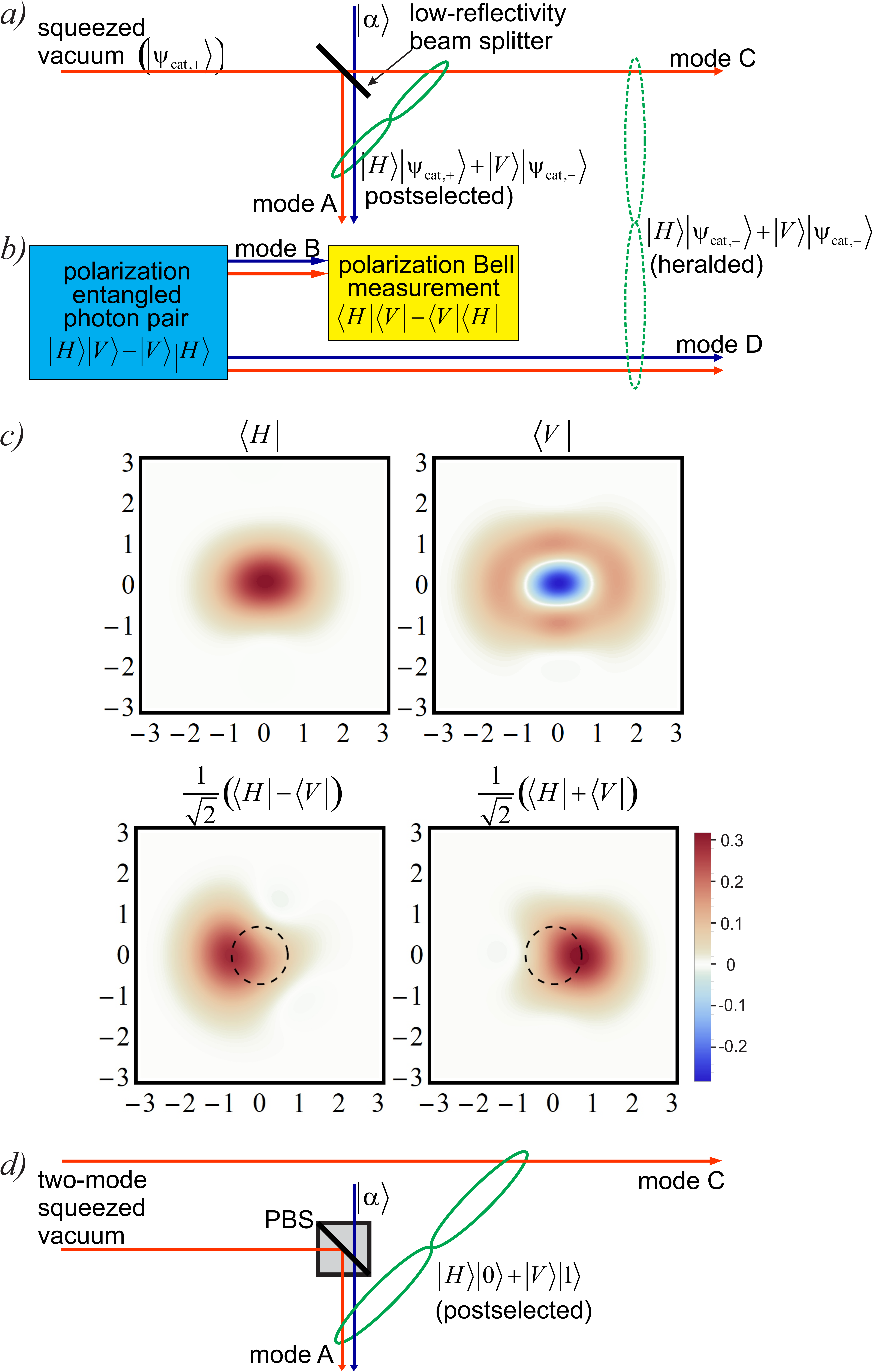}
	\caption{Experiments with the entanglement of the polarization and the cat-like qubits [(a)-(c), \cite{Sychev2017b}] and of the polarization and single-rail qubit [(d), \cite{Drahi2019}]. a) Preparation of the postselected state  (\ref{HVCat}) (conditioned on a single photon present in mode $A$) in modes $A$ and $C$. Red lines denote vertical polarization, blue horizontal. b) An entanglement swapping procedure with a freely propagating polarization-entangled state in modes $B$ and $D$ prepares the same state in a heralded (postselection-free) fashion. c) The state of the CV mode $C$ when the DV mode $A$ is projected onto various polarization states. Top rows: Experimental and corresponding theoretical Wigner functions. Red color indicates positive values, blue --- negative. The dashed circles correspond to the half-height of the vacuum state Wigner function. d) Preparation of the postselected state single-dual-rail entangled state (conditioned on a single photon present in mode $A$) in modes $A$ and $C$ \cite{Drahi2019}. The polarization and notation of the modes are changed with respect to the original work  \cite{Drahi2019} for consistency.}
	\label{fig:HVCat}
\end{figure}

The experiment by \textcite{Morin2014} starts with two OPAs, one operating in the nondegenerate and the other in the degenerate regime. The weekly pumped non-degenerate OPA (marked ``Alice") produces pairs of photons so that, when one photon in the pair is detected, the other emission channel [mode 1 in Fig.~\ref{fig:BelliniLaurat}(b)] will also contain a photon; otherwise it is likely to contain the vacuum state. The degenerate OPO (``Bob") emits squeezed vacuum into mode 2 [Fig.~\ref{fig:BelliniLaurat}(b)], which, as discussed in Sec.~\ref{ScCatTheoSec}, approximates an even cat state. A fraction of that state is tapped using a weakly reflective BS; if a photon in the tapped mode is detected, the state of mode 2 transforms into an odd cat. 

The heralding modes from both parts of the setup are mixed on a beam splitter and directed onto a single-photon detector; the desired state preparation is heralded by an event in this detector. Indeed, such an event can occur due to one of the two situations. If the trigger photon originates from the non-degenerate OPA, the state of modes 1 and 2 becomes $\ket 1\ket{\psi_{{\rm cat},+}}$. If, on the other hand, the photon comes from the degenerate OPA, the state becomes  $\ket 0\ket{\psi_{{\rm cat},-}}$. Provided that the two heralding modes are well-matched, the heralded 2-mode state is the superposition of these two states, i.e.~the entangled DV-CV qubit. The coefficients of this superposition are determined by the relative optical phases of the modes, SPDC amplitudes and the BS parameters.

The entangled state prepared using the method of Ref.~\cite{Morin2014} has been used to implement remote state preparation \cite{LeJeannic2018}. The discrete mode was subjected to homodyne detection. Observation of a specific quadrature projected that mode onto a superposition of the vacuum and single-photon states, and the CV channel was accordingly prepared in the same superposition thanks to the DV-CV entanglement. In a later publication \cite{Cavailles2018}, the same group demonstrated, by more than five standard deviations, that this experiment constitutes Einstein-Podolsky-Rosen steering.

The state of \textcite{Morin2014} has also served as the entangled resource for rudimentary teleportation from a cat qubit onto a qubit spanned by ${\ket0,\ket 1}$ \cite{Ulanov2017a} and for entanglement swapping with a delocalized single photon \cite{Guccione2020}. The Bell measurement in the cat qubit basis was realized in the manner similar to that in \cite{Neergaard-Nielsen2013} (see Sec.~\ref{section:NLA}). For the input state, Ulanov {\it et al.} used a coherent state of a variable phase $\ket{\alpha e^{i\phi}}$, which, as they argued, approximates the cat qubit state $\ket{\psi_{{\rm cat},+}}+e^{i\phi}\ket{\psi_{{\rm cat},-}}$ up to the first term in the Fock decomposition. In fact, this approximation is precise only for $\phi=0$ or $\pi$ (i.e.~the input state being $\ket\alpha$ or $\ket{-\alpha}$); otherwise it fails for large values of $\alpha$. Nevertheless, this method achieves a high non-classical teleportation fidelity averaged over the Bloch sphere of input qubits. A scheme for reverse teleportation --- from the single-rail qubit onto the cat qubit --- has been theoretically analyzed in \cite{Huang2019}.

Both of the above methods of preparing the CV-DV entangled state have their advantages and drawbacks. The disadvantage of the approach used in \cite{Jeong2014} is that the relation $a^\dag\alpha\approx\ket{g\alpha}$ holds only for relatively high values of alpha. However, the amplitudes of the components of the resulting qubit, $\pm(g-1)\alpha\approx \pm1/|\alpha|^2$, diminish with $\alpha$. Therefore one must compromise between the fidelity of the cat and its amplitude: for example, setting $\alpha=2$ results in $\pm(g-1)\alpha\approx 0.25$ and the fidelity of 0.98. On the other hand, the scheme of \cite{Morin2014}, while avoiding this issue, suffers from the amplitudes of the odd and even cat being different by a factor of $\sqrt 3$ (see Sec.~\ref{ScCatTheoSec}), which complicates the practical application of the CV part of the resource.

Another issue with both these approaches is that the single-rail encoding of the qubit is not commonly used in practical quantum optical information processing. This is because of the vulnerability to optical losses and inefficient  detection and the challenges associated with single-qubit operations and measurements \cite{Berry2006,Izumi2020}. Much more common is the dual-rail encoding, where the logical value is assigned to the photon being present in one of two modes --- such as the well-known polarization qubit is a particular case of the dual-rail encoding. Therefore it is desirable to have a technique to produce an entangled qubit of the form 
\begin{equation}\label{HVCat}\ket R=\ket H\ket{\psi_{{\rm cat},+}}+\ket V\ket{\psi_{{\rm cat},-}}, 
\end{equation}where $\ket H$ and $\ket V$ denote horizontally and vertically polarized photons.

This technique has been demonstrated by \textcite{Sychev2017b}. Figure \ref{fig:HVCat}(a) shows its basic principle.  Suppose a weakly squeezed vacuum state, which approximates $\ket{\psi_{{\rm cat},+}}$, is generated in the \emph{vertically} polarized channel of spatial mode $C$ (which we denote as $VC$). This state passes through a low-reflectivity BS, which taps a fraction of the light into spatial mode $A$.  As per our previous arguments, the resulting state can be written as 
\begin{equation}
\ket\psi_{VA,VC}\approx\ket{0}_{VA}\ket{\psi_{{\rm cat},+}}_{VC}+\beta\ket{1}_{VA}\ket{\psi_{{\rm cat},-}}_{VC},
\end{equation}
where $\beta$ depends on the initial squeezing and the beam splitter reflectivity.

Now let us inject a weak \emph{horizontally} polarized coherent state $\ket{\alpha}\approx\ket{0}+\alpha\ket{1}$ into the input mode $A$ of the BS. The state of all the relevant modes becomes  
\begin{align}\label{OmegaAC}
&\ket{\Omega}_{AC}= \ket\alpha_{HA}\ket\psi_{VA,VC}\approx \ket{0}_{HA}\ket{0}_{VA}\ket{\psi_{{\rm cat},+}}_{VC}\\
& +\alpha\ket{1}_{HA}\ket{0}_{VA}\ket{\psi_{{\rm cat},+}}_{VC}+\beta\ket{0}_{HA}\ket{1}_{VA}\ket{\psi_{{\rm cat},-}}_{VC}\nonumber ,
\end{align}
where we approximated to the first order in $\alpha$ and $\beta$. The second line in Eq.~(\ref{OmegaAC}), corresponding to a single photon present in spatial mode $A$, comprises the state \eqref{HVCat} in modes $A$ and $HC$ (because $\ket V_A\equiv \ket{0}_{HA}\ket{1}_{VA}$ and $\ket H_A\equiv\ket{1}_{HA}\ket{0}_{VA}$).  Experimentally, \cite{Sychev2017b} demonstrated this by detecting the photon in mode $A$ in various polarization states. This measurement remotely prepares various superpositions of cat states in mode $HC$, which can be reconstructed by homodyne tomography.

The CV-DV entanglement of the state (\ref{OmegaAC}) becomes manifest only when the photon in mode A is detected and destroyed. This greatly limits the practical utility of this state. However, the issue can be remedied as shown in Fig.~\ref{fig:HVCat}(b). One needs to prepare a polarization-entangled photon pair, e.g. $(\ket H\ket V+\ket V \ket H)/\sqrt 2)$, in a separate set of spatial modes $B$ and $D$ and then subject modes $A$ and $B$ to a measurement in the polarization Bell basis. This will result in the swapping of the CV-DV entanglement \cite{Pan1998} to modes $VC$ and $D$. Because the Bell measurement will only produce an event when two photons are input, the resulting state will be a good approximation to Eq.~(\ref{HVCat}). 

In \textcite{Sychev2017b}, this procedure has been implemented; however, the polarization-entangled pair  used  in that experiment has been of a postselected, rather than freely-propagating, nature. As a result, a freely propagating CV-DV entangled resource could not be produced. However, this goal appears to be within reach of current technology because heralded  preparation of freely polarizating entangled photon pairs has been demonstrated \cite{Barz2010,Wagenknecht2010}. 

\textcite{Drahi2019} used a similar idea to implement entanglement between single- and dual-rail encodings of the qubit. Here, the single-mode squeezed vacuum source is replaced by a weak two-mode squeezed vacuum source operating in the low-gain regime (i.e. a probabilistic photon pair source). One of the modes is mixed on a polarizing beam splitter with a weak coherent state in an orthogonal polarization [Fig.~\ref{fig:HVCat}(d)]. Now, the polarization of the photon in mode A determines whether this photon has originated from the coherent state or from the pair source; accordingly, a twin photon is either absent or present in mode C. The state of modes A and C is therefore given by $(\ket H_A\ket 0_{VC}+\ket V_A\ket 1_{VC})/\sqrt 2$. Again, the postselected nature of this entanglement is addressed using the entanglement swapping scheme of Fig.~\ref{fig:HVCat}(b)].

With the results described above, we now possess tools to interconvert among three primary encodings of the qubit by means of light: single-rail, dual-rail and coherent-state. These tools consolidate the role of light as the medium that enables exchange of quantum information among physical systems of different nature. 

\subsection{Bell's inequalities and device-independent quantum key distribution}

Bell's inequalities (BI) were designed to show incompatibility of remote quantum correlations with local realism \citep{Bell1964}. Their experimental tests are vulnerable to two loopholes: the locality loophole, which arises when the detectors are too close to completely discard the exchange of subluminal signals during the measurements; and the efficiency loophole, which occurs when the detectors are not efficient enough to be described by Bell's model \cite{Pearle1970,Garg1987}.
These loopholes were closed separately \cite{Aspect1982,Weihs1998,Rowe2001} then simultaneously \cite{Hensen2015,Shalm2015,Giustina2015}, and the fundamental debate around quantum nonlocality can be considered settled.

Nevertheless, BI retain a central role in the applied side of quantum science, particularly quantum cryptography. Indeed, loophole-free BI violation constitutes unforgeable evidence of entanglement between Alice and Bob, and hence precludes any eavesdropping as long as the latter is bound by the laws of known physics \citep{Acin2005,Barrett2005,Acin2006}. Hence it would enable device-independent quantum key distribution  \citep{Acin2007}, which remains secure even if an eavesdropper takes control of the detection system  \citep{Lydersen2010,Sauge2011}. 

It is appealing to extend device-independent security to quantum key distribution in the CV domain, which, under some conditions, enables significantly higher communication rates than its DV counterpart \cite{Grosshans2003}.  This would require developing a BI analog that would be compatible with homodyne measurements and allow their experimental loophole-free violation. Unless one trusts the detection system \cite{Thearle2018}, this task cannot be solved without non-Gaussian states with negative Wigner functions, otherwise the quadratures $x$ and $p$ can take the role of hidden variables for which the Wigner function becomes a valid probability distribution. Among the different proposed violation schemes, many require quantum measurements \citep{Banaszek1998,Stobinska2007} or quantum states \citep{Munro1999,Auberson2002,Wenger2003,Cavalcanti2007,Etesse2014} which are still difficult to realize. Others are based on simpler setups but provide only a small violation \citep{Nha2004,Garcia2004}, or can be used only when more than two parties are considered \citep{Acin2009}. Some hybrid schemes, mixing homodyne measurements with photon detection, \citep{Cavalcanti2011,Brask2012}, require more accessible detection efficiencies with simpler states. In general, there is a tradeoff between the required detection efficiency and the complexity of the states used in the protocol \citep{Quintino2012}.

\section{Applications to quantum computing}\label{QCompSec}

\subsection{Linear optical quantum computing}

The challenges of implementing a large-scale, fault-tolerant quantum computer capable of surpassing its classical counterpart for practical computational problems are widely known. However, there exist applications 
where even small scale quantum computing would be useful. Some of these applications are in quantum communications and networking. Because quantum communication is performed by means of light, it would be preferred that the computation is also done by means of optics \citep{OBrien2009}. Used at the nodes of a quantum optical network, such a device would deliver highly secure communications and distributed quantum processing \citep{Kimble2008,Briegel1998,Bennett1993}, associated with 
ultra-high capacity and minimum power in optical communications \citep{Giovannetti2004,Waseda2010}. 

Fault-tolerant optical quantum information processing requires non-Gaussian input states and/or gates capable of turning Gaussian states into non-Gaussian ones \cite{Niset2009}. Their deterministic implementation involves non-linear optical processes at least of the third order \citep{Lloyd1999,Bartlett2002a,Bartlett2002b,Bartlett2002c} which, as already mentioned (section \ref{NonGauss:Principle}) are very difficult to put in practice. On the other hand, they can be realized probabilistically by using linear optics and conditional measurements. For example, \textcite{Costanzo2017} demonstrated a method to enact the transformation 
\begin{equation}\label{Costanzo1}
c_0\ket0+c_1\ket 1+c_2\ket 2 \to c_0\ket 0 +c_1\ket 1-c_2\ket 2,
\end{equation}
which corresponds to self-phase modulation due to the third-order optical nonlinearity, on an arbitrary linear combination of the first three Fock states by using the circuit shown in Fig.~\ref{addsubsc}. As discussed in Sec.~\ref{AssSubSec}, such a circuit implements a linear combination of operators $\hat n=\adag \a$  and $\hat n+\mathbb{I}=\a\adag$, or, equivalently, of operators $\hat n$ and $\mathbb{I}$. By adjusting the parameters of this circuit, one can choose arbitrary coefficients of this linear combinations. In \cite{Costanzo2017}, the operator was $\propto-(2+\sqrt 2)\hat n+\mathbb{I}$, acting as follows:
\begin{equation}\label{Costanzo2}
c_0\ket0+c_1\ket 1+c_2\ket 2 \to c_0\ket 0 +g c_1\ket 1-g^2c_2\ket 2
\end{equation}
with $g=-(\sqrt2+1)$. Subsequently, by applying the noiseless linear amplifier operator $\hat G=(-g)^{-\hat n}$ with the gain $-1/g$ (see Sec.~\ref{section:NLA}), one obtains the transformation (\ref{Costanzo1}).

Applied directly in a computational protocol, such probabilistic gates lead to loss of data. Instead, they can be used off-line to prepare specific entangled resources states, then teleport the computational states using these resources to implement the desired gates. This paradigm is known as linear optical quantum computing (LOQC). In the DV domain, the mainstream scheme of LOQC is that of \textcite{Knill2001}. In the optical CV domain, two approaches are now actively studied. One of them, using two phase-opposite coherent states for qubit encoding \citep{Jeong2002,Ralph2003} is referred to as coherent state quantum computing (CSQC) and discussed in section \ref{section:CSQC}. The other one, relying on phase-space ``grid'' or ``comb'' states known as Gottesman-Kitaev-Preskill (GKP) states \cite{Gottesman2001}, is briefly presented in section \ref{section:GKPQC}. Results on recently proposed ``binomial'' codes for bosonic systems \cite{Michael2016} are so far limited to the microwave domain and remain beyond the scope of this review.


\subsection{Coherent state quantum computing}
\label{section:CSQC}

\textcite{Cochrane1999} introduced logical qubit encoding
using even and odd cat states, and
showed how quantum computing can be implemented by using 
a large enough cross Kerr interaction, 
combined with the displacement operation.
\textcite{Jeong2002} developed this proposal further, using the encoding
with binary coherent states as per Eq.~(\ref{Logical encoding}).
The coherent amplitude should be large enough, so that
the qubit bases can be almost orthogonal.
Assuming that the single-mode Kerr nonlinearity is available online,
they showed that a universal gate set can be constructed
using teleportation-based CNOT gate.
Entangled cat states can also be used as an entanglement resource \citep{vanEnk2001}, but   low-loss Kerr interactions required for these schemes are unfortunately quite hard to realize.

\textcite{Ralph2003} showed a way to construct the universal gate set
only using linear online processes, 
but offline preparation of ancilla states, which are used as resources to be inserted later on into the calculation.
In offline state preparation, two-mode entangled cat states are used to  implement one-bit and CNOT gates.
This scheme can be deterministic in principle,
albeit with a rather demanding requirement on the source of cat states, requiring 
amplitudes $\alpha>2$. 

\textcite{Lund2008} then showed that
by using nondeterministic gates,
based on teleportation with unambiguous state discrimination
combined with error correction, 
the requirement for the amplitudes can be reduced to 
$\alpha>1.2$.
They also studied  tolerance to certain errors,
including those caused by photon loss,
as well as those due to failure events
in the nondeterministic gate teleportation.
It was shown that low enough photon losses (less than
$5\times10^{-4}$), the CSQC scheme can be resource efficient and fault tolerant
in the sense that
approximately $10^4$ cat states are consumed
per error-correction round at the first level coding.
This number is 4 orders of magnitude lower than the number of
Bell pair resource states
consumed under equivalent conditions by the most efficient known
scheme of DV LOQC.

However, when considering practical implementations with currently available
resources,  the remaining challenges are considerable. \textcite{Marek2010}
pointed that if cat states  with amplitudes $\alpha\sim1$ are available,
one must use the one-qubit gate with an incremental phase shift
repeatedly to include a significant angle,
at least ten operations to achieve a $\pi$ phase shift.
So they proposed an alternative way to achieve large phase shift
with currently available resources, 
by using, again, probabilistic operations
that involve single-photon subtractions. They elaborated a universal gate set for CSQC, consisting of the single-mode phase gate, the single-mode Hadamard gate, and the two-mode controlled phase gate.

\begin{figure}[tb]
\centering {\includegraphics[width=\linewidth]{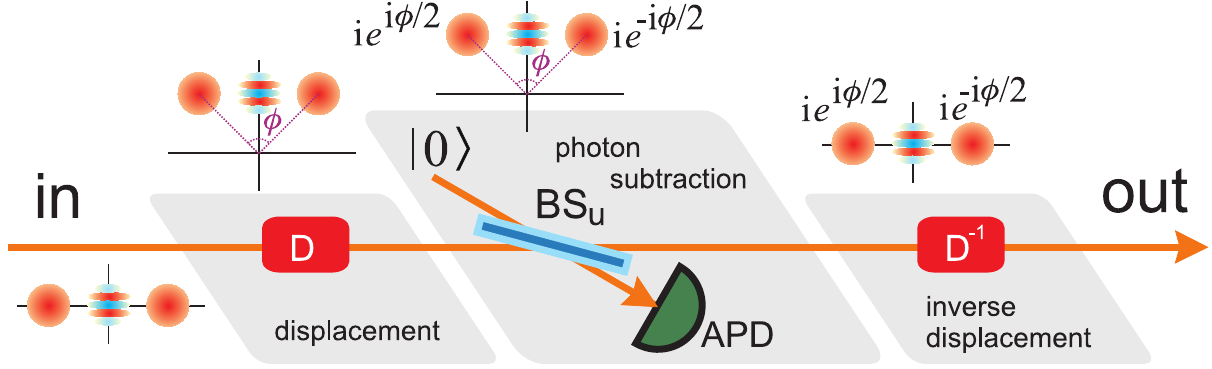}
	\caption{(Color online) Single-mode phase gate in coherent-state linear optical quantum computing \cite{Marek2010}. }
	\label{fig:MarekFiurasek}}
\end{figure}

We discuss the principle of the one-qubit gate from that proposal as an example of a CSQC scheme (Fig.~\ref{fig:MarekFiurasek}). To implement the gate, one first displaces the input qubit $x \ket{\alpha} + y \ket{-\alpha}$ in the phase space along the momentum axis by distance $\beta$, so the qubit components transform according to $\alpha\to\alpha+i\beta,\ -\alpha\to-\alpha+i\beta$ (we assume $\alpha$ to be real). Second, one applies the annihilation operator, which results in a phase shift that is different for the two components: $\a\ket{\alpha+i\beta}=(\alpha+i\beta)\ket{\alpha+i\beta}, \ \a\ket{-\alpha+i\beta}=(-\alpha+i\beta)\ket{-\alpha+i\beta}$. Finally, one displaces the qubit back to its original phase space position, obtaining $x(\alpha+i\beta)\ket{\alpha}+y(-\alpha+i\beta)\ket{-\alpha}$. We see that the two components of the qubit have acquired different quantum phase shifts, with the difference that can be controlled by choosing the value of $\beta$.

At present,
the single-mode phase gate ($\pi$-phase gate) was implemented by
\textcite{Blandino2012b}, and the Hadamard gate by \textcite{Tipsmark2011}. 
Experimental implementation of the two-mode gate remains an open problem.  

Progress in superconducting quantum circuits, where deterministic non-Gaussian operations are feasible, large-amplitude cat states have been realized \cite{Vlastakis2013}, and corresponding error-correction protocols can be implemented \cite{Ofek2016}, stimulated new theoretical proposals for fault-tolerant CSQC using ``multi-legged'' cat states corresponding to superpositions of $d>2$ coherent states lying on a circle in phase space \cite{Mirrahimi2014}. In the optical domain, \textcite{Thekkadath2020} proposed to generate four-legged cats probabilistically using coherent ancillas and photon-number-resolving measurements, while \textcite{Hastrup2020} found an alternative approach which can be deterministic if standard ``two-legged'' cats are available as resources.

\subsection{GKP state quantum computing}
\label{section:GKPQC}

Ideal GKP states correspond to an infinite comb of infinitely squeezed states equally spaced along the $x$ (and, by Fourier transform, also along the $p$) quadrature \cite{Gottesman2001}. In practice, constrains can be relaxed to make these states more physical, but they remain highly non-Gaussian and more difficult to generate than cat states. GKP codes have been experimentally implemented with superconducting qubits \cite{Campagne-Ibarcq2019} and trapped ions \cite{Fluehmann2019}, but in the optical domain they remain at a theoretical stage. 

The protocol proposed by \textcite{Vasconcelos2010} (and independently by \textcite{Etesse2014} in a different  context) would allow conditional generation of such states via a procedure reminiscent of ``Schr\"odinger cat breeding" discussed in Sec.~\ref{BreedingSec}. It begins with two cats mixed on a symmetric beamsplitter, producing an entangled superposition of the form \eqref{eq2}. One of the output modes is then subjected to a homodyne measurement. 
In contrast to ``breeding", the quadrature being measured is momentum rather than position. The resulting conditionally prepared state  in the other mode is then a superposition of $\ket{\psi_{{\rm cat},+}[\sqrt{2}\alpha]}$ and $\ket 0$, which is a momentum-squeezed three-peaked GKP state. This operation can be iteratively repeated to produce GKP states with more peaks. 

\textcite{Weigand2018} have later shown that this protocol can be made more efficient by keeping all states and using the measurement's outcomes for classical feedforward. A similar protocol using photon number resolving counters instead of homodyne detectors for heralding \cite{Eaton2019}, as well as more general but less feasible schemes \cite{Su2019}, have been proposed.

\section{Final outlook}
\label{section:Outlook}

Measurement-induced engineering of non-Gaussian optical states has established itself as an active and mature research field. Its main limitation stems from probabilistic heralding: if the 
probability $p$ of the expected interesting event is small, the probability $p^n$ of getting $n$ times this same event will be 
exponentially smaller. Though  ideas have been proposed to overcome this exponential loss \citep{Knill2001}, their 
implementation is extremely demanding, in particular as they require a very high number of  physical qubits to encode and protect a single logical qubit. Developing on-demand sources of non-Gaussian states, particularly single photons, is hence essential for the further development of the field.


In spite of their probabilistic nature, SPDC sources do possess a promise in this context. An example is the multiplexed photon source \cite{Migdall2002,Kaneda2019}, in which a photon detection in one of the idler modes of a multimode SPDC source triggers an electrooptical switch that directs the counterpart signal photon into a single output mode.
	
Alternatively, a photon (or antother state) prepared in a heralded fashion can be stored in a quantum memory for subsequent on-demand retrieval. 
A good quantum memory should be able to store this state for longer than it takes to generate another one, and the retrieved state should faithfully reproduce the stored one. The latter requirement implies a high combined storage and retrieval efficiency but is not limited to it: in particular, the retrieved state must not exhibit any added quadrature noise with respect to the stored one \cite{Lvovsky2009QM}. 
While these criteria have been satisfied separately \cite{Heinze2013,Vernaz-Gris2018,Radnaev2010,Bouillard2019}, a device that could meet all these requirements at once is still to be developed.

A streamlined variant of the above procedure is the Duan-Lukin-Cirac-Zoller process, in which Raman scattering on an atomic transition creates two-mode squeezing between a long-lived atomic collective state and an optical mode \cite{Duan2001}. Detection of a photon in this mode heralds a single stored atomic excitation, which can be retrieved on demand as another optical photon. In this way, an intermediate step of heralding a photon and storing it in a memory is bypassed, thereby relaxing the efficiency requirements. 
This approach has been tested experimentally in various DV settings \cite{Lvovsky2009QM}. In the context of non-Gaussian optics, it was shown to be efficient enough to observe negative Wigner functions \citep{MacRae2012,Bimbard2014,Brannan2014}.

In parallel, truly deterministic sources of non-Gaussian states made impressive progress. The performance criteria for such sources are (1)  brightness (number of photons emitted per second and per unit bandwidth), (2) antibunching (not more than one photon produced at a time), (3) quality of the mode in which the photon is generated and (4) indistinguishability (if multiple sources run simultaneously). 
Atomic Rydberg \cite{Ornelas-Huerta2020} and cavity-enhanced \cite{Muecke2013} sources, as well as solid-state \cite{Somashi2016,Wang2019} sources are becoming increasingly more efficient according to all these criteria. Atom-based systems largely improved their overall duty cycles and moved beyond the single-photon regime \cite{Hacker2019,Liang2018,Clark2019} while solid-state devices, which were hampered by sample-to-sample variations, became sufficiently tunable to emit indistinguishable photons \cite{Flagg2010}. Many of these systems can not only generate non-Gaussian states but also implement non-Gaussian operations such as entangling photon-photon gates \cite{Gazzano2013,Hacker2016,Tiarks2019}.

Instead of engineering non-Gaussian optical states directly, one could also focus on developing linear interfaces between light and matter-based systems, leaving the complexity of non-Gaussian operations to the latter. Among them, trapped ions and superconducting circuits are currently leading the race in terms of processing capabilities. Efficiently coupling ions to light requires optical cavities: despite additional challenges with respect to neutral atoms, ion-cavity systems are making steady progress \cite{Lee2019}. As for microwave-to-optical photon converters, different implementations can be envisioned \cite{Lambert2019} and some become functional at the quantum level \cite{Forsch2020}.

Scaling quantum optical information processing circuits up to a practically useful size necessarily requires their implementation on an integrated photonic chip. In addition to the scalablity, this platform addresses many challenges associated with free space optics. Integrated interferometers are intrinsically stable, while waveguides provide a transverse confinement enhancing the efficiency of nonlinear processes.  
For atom-based systems, moving from ``bulk'' optical cavities to microresonators enhances the atom-light coupling and the input/output coupling efficiency \citep{Thompson2013,Shomroni2014,Luan2020}. At the same time, taking advantage of these opportunities requires a large range of new technologies to be developed. This includes low-loss waveguides, rapidly controllable linear-optical interferometers, efficient indistinguishable photon and squeezed vacuum sources, as well as on-chip  photon and homodyne detectors. Integrated quantum photonics is a very active field, and we refer to recent reviews for more details \cite{Silverstone2016,Wang2019b}.

We have shown in this review that the tools of non-Gaussian quantum optics achieve results far beyond those that can be obtained using discrete- or continuous-variable methods alone. Hence these tools will undoubtedly be  part of any further progress on manipulating and exploiting quantum light. The main challenge now is to make them deterministic and efficient ---
not an easy challenge, but certainly one which can be met. 


\bibliographystyle{apsrmp4-1}
\bibliography{All_RMP_rev4}
\end{document}